\documentclass[12pt,preprint]{aastex}

\def\kms{km~s$^{-1}$}
\def\degpnt{^{\circ}\kern-1.7mm.\kern+.35mm}
\def\arcpnt{"\kern-1.7mm.\kern+.35mm}
\def\minpnt{'\kern-1.0mm.\kern+.30mm}

\newcommand{\gp}{\ensuremath{g^{\prime}}}
\newcommand{\rp}{\ensuremath{r^{\prime}}}
\newcommand{\ip}{\ensuremath{i^{\prime}}}
\newcommand{\zp}{\ensuremath{z^{\prime}}}

\shorttitle{SNLS SN~Ia Rate}
\shortauthors{Neill et al.}

\received{2006 January 23}
\begin{document}


\title{The Type Ia Supernova Rate at z $\approx$ 0.5 from the Supernova Legacy Survey\footnote{Based on observations obtained with
  MegaPrime/MegaCam, a joint project of CFHT and CEA/DAPNIA, at the
  Canada-France-Hawaii Telescope (CFHT) which is operated by the National
  Research Council (NRC) of Canada, the Institut National des Sciences de
  l'Univers of the Centre National de la Recherche Scientifique (CNRS) of
  France, and the University of Hawaii.  This work is based in part on data
  products produced at the Canadian Astronomy Data Centre as part of the
  Canada-France-Hawaii Telescope Legacy Survey, a collaborative project of
  NRC and CNRS.  Based on observations obtained at the European Southern
  Observatory using the Very Large Telescope on the Cerro Paranal (ESO
  Large Programme 171.A-0486). Based on observations (programs
  GN-2004A-Q-19, GS-2004A-Q-11, GN-2003B-Q-9, and GS-2003B-Q-8) obtained at
  the Gemini Observatory, which is operated by the Association of
  Universities for Research in Astronomy, Inc., under a cooperative agreement
  with the NSF on behalf of the Gemini partnership: the National Science
  Foundation (United States), the Particle Physics and Astronomy Research
  Council (United Kingdom), the National Research Council (Canada), CONICYT
  (Chile), the Australian Research Council (Australia), CNPq (Brazil) and
  CONICET (Argentina).  Based on observations obtained at the W.M. Keck
  Observatory, which is operated as a scientific partnership among the
  California Institute of Technology, the University of California and the
  National Aeronautics and Space Administration.  The Observatory was made
  possible by the generous financial support of the W.M. Keck Foundation.}\\
  \bigskip
	}


\author{J.~D.~Neill\altaffilmark{2}, M.~Sullivan\altaffilmark{3},
D.~Balam\altaffilmark{2}, C.~J.~Pritchet\altaffilmark{2},
D.~A.~Howell\altaffilmark{3}, K.~Perrett\altaffilmark{3},
P.~Astier\altaffilmark{4}, E.~Aubourg\altaffilmark{5,6},
S.~Basa\altaffilmark{7}, R.~G.~Carlberg\altaffilmark{3},
A.~Conley\altaffilmark{3}, S.~Fabbro\altaffilmark{8},
D.~Fouchez\altaffilmark{9}, J.~Guy\altaffilmark{4},
I.~Hook\altaffilmark{10}, R.~Pain\altaffilmark{4},
N.~Palanque-Delabrouille\altaffilmark{6}, N.~Regnault\altaffilmark{4},
J.~Rich\altaffilmark{6}, R.~Taillet\altaffilmark{11,4},
G.~Aldering\altaffilmark{12}, P.~Antilogus\altaffilmark{4},
C.~Balland\altaffilmark{5},
S.~Baumont\altaffilmark{4}, J.~Bronder\altaffilmark{10}, 
R.~S.~Ellis\altaffilmark{13}, M.~Filiol\altaffilmark{7},
A.~C.~Gon\c{c}alves\altaffilmark{14}, M.~Kowalski\altaffilmark{12},
C.~Lidman\altaffilmark{15},
V.~Lusset\altaffilmark{6}, M.~Mouchet\altaffilmark{5},
S.~Perlmutter\altaffilmark{13}, P.~Ripoche\altaffilmark{9},
D.~Schlegel\altaffilmark{12}, C.~Tao\altaffilmark{9}}

\altaffiltext{2}{Department of Physics and Astronomy, University of Victoria, PO Box 3055, Victoria, BC V8W 3P6, Canada}
\altaffiltext{3}{Department of Astronomy and Astrophysics, University of
Toronto, 60 St.\ George Street, Toronto, ON M5S 3H8, Canada}
\altaffiltext{4}{LPNHE, CNRS-IN2P3 and University of Paris VI \& VII, 75005 Paris, France}
\altaffiltext{5}{APC, 11 Pl. M. Berthelot, 75231 Paris Cedex 5, France}
\altaffiltext{6}{DSM/DAPNIA, CEA/Saclay, 91191 Gif-sur-Yvette Cedex, France}
\altaffiltext{7}{LAM CNRS, BP8, Traverse du Siphon, 13376 Marseille Cedex 12, France}
\altaffiltext{8}{CENTRA - Centro Multidisciplinar de Astrof\'{\i}sica, IST, Avenida Rovisco Pais, 1049 Lisbon, Portugal}
\altaffiltext{9}{CPPM, CNRS-IN2P3 and University Aix Marseille II, Case 907, 13288 Marseille Cedex 9, France}
\altaffiltext{10}{University of Oxford Astrophysics, Denys Wilkinson Building, Keble Road, Oxford OX1 3RH, UK}
\altaffiltext{11}{Universit\'{e} de Savoie, 73000 Chamb\'ery, France}
\altaffiltext{12}{LBNL, 1 Cyclotron Rd, Berkeley, CA 94720, USA}
\altaffiltext{13}{California Institute of Technology, E. California Blvd., Pasadena, CA 91125, USA}
\altaffiltext{14}{LUTH,UMR 8102, CNRS and Observatoire de Paris, F-92195 Meudon, France}
\altaffiltext{15}{ESO, Alonzo de Cordova 3107, Vitacura, Casilla 19001, Santiago 19, Chile}


\begin{abstract}

We present a measurement of the distant Type Ia supernova rate derived from the
first two years of the Canada -- France -- Hawaii Telescope Supernova Legacy
Survey.  We observed four one-square degree fields with a typical temporal
frequency of $\langle \Delta t \rangle \sim 4$ observer-frame days over time
spans of from 158 to 211 days per season for each field, with breaks during
full moon.  We used 8-10 meter-class telescopes for spectroscopic followup to
confirm our candidates and determine their redshifts.  Our starting sample
consists of 73 spectroscopically verified Type Ia supernovae in the redshift
range $0.2 < z < 0.6$.  We derive a volumetric SN~Ia rate of $r_V(\langle z
\rangle =0.47) = 0.42^{+0.13}_{-0.09}$ (systematic) $\pm 0.06$ (statistical) $
\times 10^{-4}$ yr$^{-1}$ Mpc$^3$, assuming $h = 0.7, \Omega_m = 0.3$ and a
flat cosmology.  Using recently published galaxy luminosity functions derived
in our redshift range, we derive a SN~Ia rate per unit luminosity of
$r_L(\langle z \rangle =0.47) = 0.154^{+0.048}_{-0.033} $ (systematic) $
^{+0.039}_{-0.031}$ (statistical) SNu.  Using our rate alone, we place an upper
limit on the component of SN~Ia production that tracks the cosmic star
formation history of 1 SN~Ia per 10$^3$ M$_{\odot}$ of stars formed.  Our rate
and other rates from surveys using spectroscopic sample confirmation display
only a modest evolution out to $z=0.55$.

\end{abstract}




\section{Introduction\label{sec_intro}}

Type Ia supernovae (SNe~Ia) have achieved enormous importance as cosmological
distance indicators and have provided the first direct evidence for the dark
energy that is driving the Universe's accelerated expansion
\citep{Riess98AJ,Perlmutter99ApJ}.  In spite of this importance, the physics
that makes them such useful cosmological probes is only partly constrained.
White dwarf physics is the best candidate for producing a standard explosion
due to the well understood Chandrasekhar mass limit \citep{Chandrasekhar31ApJ}.
However, any plausible SN~Ia scenario requires a companion to donate mass and
push a sub-Chandrasekhar C-O white dwarf towards this limit producing some form
of explosion \citep[for a review, see][]{Livio01sgrb}.  The range of possible
companion scenarios needed to accomplish this are currently divided into two
broad categories: the single degenerate scenario, where the companion is a
subgiant or giant star that is donating matter through winds or Roche lobe
overflow \citep{Whelan73ApJ,Nomoto82ApJ,Canal96ApJ,Han04MNRAS}, and the double
degenerate scenario involving the coalescence of two white dwarf stars after
losing orbital angular momentum through gravitational radiation
\citep{Webbink84ApJ,Iben84ApJS,Tornambe86MNRAS,Napiwotzki04ASPC}.

Population synthesis models for these scenarios predict different SN~Ia
production timescales relative to input star formation
\citep[e.g.][]{Greggio05A&A}.  By comparing the global rate of occurrence of
SN~Ia at different redshifts to measurements of the global cosmic star
formation history (SFH), the `delay function', parameterized by its
characteristic timescale, $\tau$, can be derived, which in turn constrains the
companion scenarios.  This comparison requires the calculation of a volumetric
SN~Ia rate and measuring the evolution of this rate with redshift.

\begin{figure}
\includegraphics[scale=0.50,angle=90.]{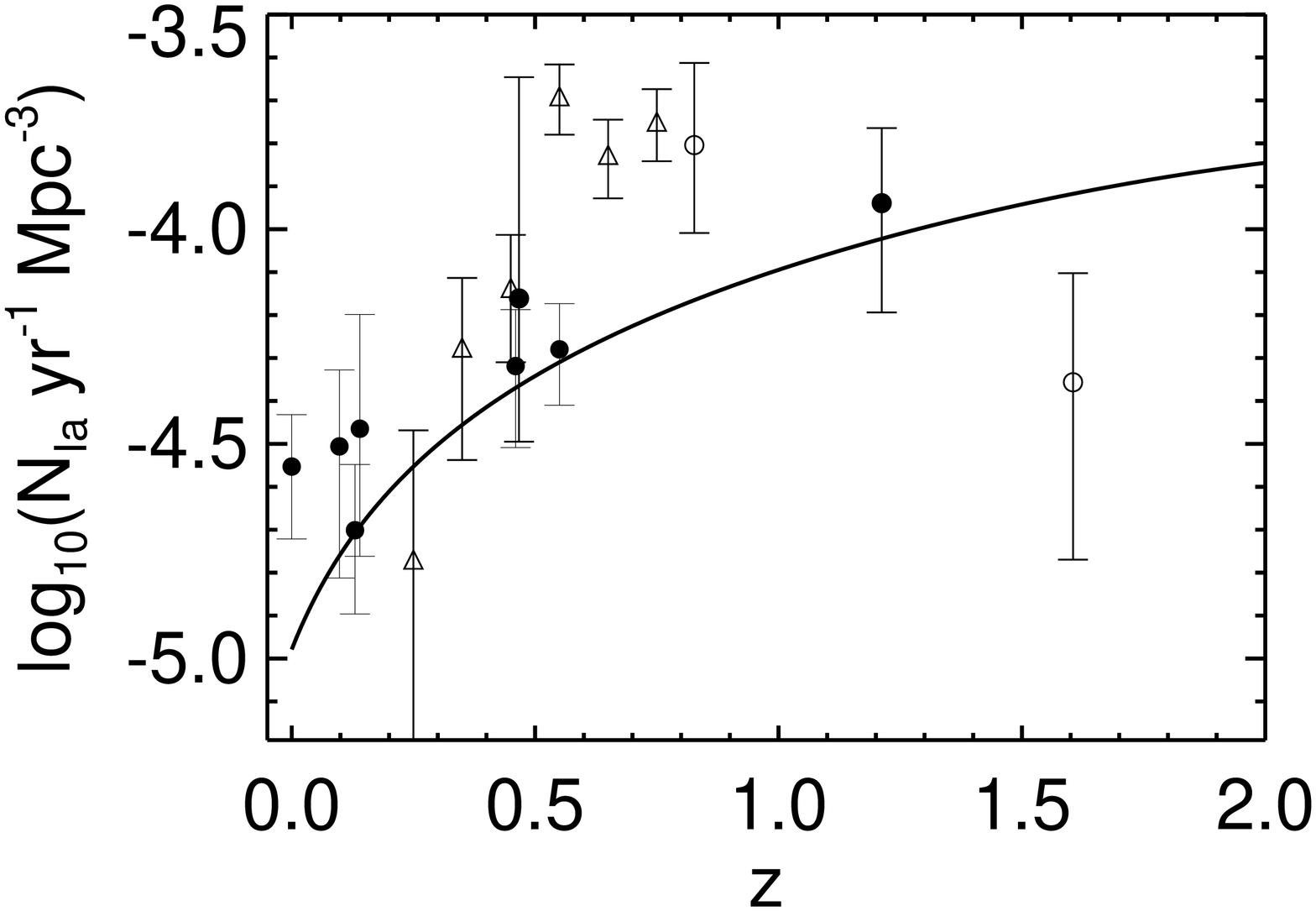}
\caption{Observed SN~Ia rate versus redshift.  The filled circles are
SN Ia rates derived from samples with the majority of objects
confirmed by spectroscopy from the following references (in redshift
order): \citet{Cappellaro99A&A,Madgwick03ApJ,Blanc04A&A,Hardin00A&A,
Tonry03ApJ,Dahlen04ApJ,Pain02ApJ,Dahlen04ApJ}.  The open circles are
the SN~Ia rates from \citet{Dahlen04ApJ}, whose
samples employ only 50\% spectroscopic confirmation.  The open
triangles are the rates from \citet{Barris06ApJ}, whose
samples are confirmed almost entirely with photometric methods.  The
solid curve is a renormalization of the SFH from \citet{Hopkins06}
using a factor of $10^3$, which is can be considered a toy model of
SN~Ia production that assumes 1 SN~Ia is produced instantaneously for
every $10^3$ M$_{\odot}$ of stars formed.
\label{fig_sfh}
}
\end{figure}

Early SN surveys were host targeted \citep[e.g.,][]{Zwicky38ApJ}, and
produced rates per unit blue luminosity that required conversion to
volumetric rates through galaxy luminosity functions.  These surveys suffer
from large systematic uncertainties because of the natural tendency to
sample the brighter end of the host luminosity function.  With the advent
of wide-field imagers on moderately large telescopes, recent surveys have
been able to target specific volumes of space and directly calculate the
volumetric rate.  Examples of volumetric SN~Ia rate calculations at a
variety of redshifts can be found in the following studies (plotted in
Figure~\ref{fig_sfh}): \citet{Cappellaro99A&A,Hardin00A&A,
Pain02ApJ,Madgwick03ApJ,Tonry03ApJ,Blanc04A&A,Dahlen04ApJ,Barris06ApJ}.

We plot the rates from these surveys as a function of redshift in
Figure~\ref{fig_sfh}, along with a recent SFH fit from
\citet{Hopkins06}, renormalized by a factor of $10^3$.  This allows us
to compare the SFH to the observed trend in the SN~Ia rate.  This
trend, compared with the SFH curve, shows some curious properties.
The large gradient just beyond $z=0.5$ observed in \citet{Barris06ApJ}
has no analog in the SFH curve, and neither does the apparent
down-turn beyond $z=1.2$ observed by \citet{Dahlen04ApJ}.  Fits of the
delay function to various subsets of these data have produced no
consensus on $\tau$, or the form of the delay.  Reported values for
$\tau$ range from as short as $\tau \le$~1~Gyr \citep{Barris06ApJ} to
as long as $\tau = 2 - 4$~Gyr \citep{Strolger04ApJ}.  This lack of
consensus and the peculiar features in Figure~\ref{fig_sfh} argue that
systematics are playing a role in the observed SN~Ia rates, especially
at higher redshifts.  It is vital to investigate the sources of
systematic error in deriving SN~Ia rates and to compare the cosmic SFH
with rates that have well characterized systematic errors.

In this paper we take advantage of the high-quality spectroscopy and
well-defined survey properties of the Supernova Legacy Survey
\citep[SNLS,][]{Astier06A&A} to produce a rate that minimizes
systematics, which we then compare with cosmic SFH.  To minimize
contamination, we use only spectroscopically verified SNe~Ia in our
sample.  We examine sources of systematic error in detail and, using
Monte Carlo efficiency experiments, place limits on them.  In
particular, we improve upon previous surveys in the treatment of host
extinction by using the recent dust models of \citet{Riello05MNRAS}.
We also investigate the possibility that SNe~Ia are being missed in
the cores of galaxies with fake SN experiments using real SNLS images.
These experiments allow us to place limits on our own errors and
assess the impact of various sources of systematic error on SN surveys
in general.

In order to avoid large and uncertain completeness corrections, we
employ a simplification in our rate determination that fully exploits
the data set currently available: we restrict our sample to the
redshift range $0.2 < z < 0.6$.  This ensures that the majority of
SNe~Ia peak above our nominal detection limits and thus provides a high
completeness.  This simplification allows us to extract a well-defined
sample of spectroscopically verified SNe~Ia from the survey and
accurately simulate the SNLS survey efficiency, thus producing the
most accurate SN~Ia rate at any redshift.  Our rate alone is
sufficient to constrain some of the SFH delay function models by
placing limits on their parameters.

The structure of the paper is as follows.  In \S\ref{sec_snls} we describe
the survey properties relevant to rate calculation.  In \S\ref{sec_select}
we develop objective selection criteria, derive our SN~Ia sample and
analyze this sample to determine our spectroscopic completeness.  In
\S\ref{sec_eff} we describe our method for calculating the survey
efficiency and present the results of these calculations.  In
\S\ref{sec_res} we present the derived SN~Ia rates per unit volume and per
unit luminosity, an analysis of systematic errors, and a comparison of our
rates with rates in the literature.  In \S\ref{sec_disc} we compare our
volumetric rate and a selection of rates from the literature with two
recent models connecting SFH with SN~Ia production.

For ease of comparison with other rates studies in the literature, we assume 
a flat cosmology throughout with $H_0 = 70$ \kms~Mpc$^{-1}$,
$\Omega_{\Lambda} = 0.7$, and $\Omega_M = 0.3$.

\section{The Supernova Legacy Survey\label{sec_snls}}

The SNLS is a second-generation SN~Ia survey spanning five years,
instigated with the purpose of measuring the accelerated expansion of the
universe and constraining the average pressure-density ratio of the
universe, $\langle w \rangle$, to better than $\pm0.05$ \citep{Astier06A&A}.
In order to achieve this goal, the SNe~Ia plotted on our Hubble diagram
must have well-sampled light curves (LCs) and spectral followup
observations that provide accurate redshifts and solid identifications.
The LC sampling is achieved using MegaCam \citep{Boulade03SPIE}, a 36 CCD
mosaic one-square degree imager, in queued service observing mode on the
3.6 meter Canada-France-Hawaii Telescope (CFHT).  This combination images
four one-square degree fields (D1-4, evenly spaced in right ascension, see
Table~\ref{tab_obs}) in four filters (\gp\rp\ip\zp) with an observer-frame
cadence of $\Delta t \sim 4$ days (rest-frame cadence for a typical SN of
$\Delta t \sim 3$ days) and with a typical limiting magnitude of 24.5 in
\ip.  The queued service mode provides robust protection against bad
weather, as any night lost is re-queued for the following night.

This observing strategy provides dense LC coverage for SNe~Ia out to $z\sim1$
and is ideal for measuring the {\it rate of occurrence} of distant
SNe~Ia.  It also produces high quality SN~Ia candidates identified early enough
so that spectroscopic followup observations can be scheduled near the
candidate's maximum light \citep{Sullivan06AJ}.  This strategy has been very
successful \citep{Howell05ApJ}, and the SNLS has been fortunate to have
consistent access to 8-10 meter class telescopes (Gemini, Keck, VLT) for
spectroscopic followup.  This is critical for providing a high spectroscopic
completeness and the solid spectroscopic type confirmation required to remove
contaminating non-SN~Ia objects from our sample \citep{Howell05ApJ,Basa06}.

\subsection{The Detection Pipeline\label{sec_pipe}}

The imaging data are analyzed by two independent search pipelines in
Canada\footnote{see http://legacy.astro.utoronto.ca/} and
France\footnote{see http://makiki.cfht.hawaii.edu:872/sne/}.  For the rate
calculation in this paper, we use the properties of the Canadian pipeline.

The Canadian SNLS real-time pipeline uses the \ip\ filter images for
detection of SN candidates and images in all filters for object
classification.  Each epoch consists of five to ten exposures which undergo
a preliminary (real-time) reduction which includes a photometric and
astrometric calibration before being combined.  A reference image for each
field is constructed from previously acquired, hand picked, high-quality
images.  The detection pipeline then seeing-matches the reference image to
the (usually lower image quality) new epoch image \citep{Pritchet06}.  The
seeing-matched reference image is then subtracted from the new epoch and
the resulting difference image is analyzed to detect variable objects which
appear as residual (positive) point sources.  A final list of candidate
variable objects is produced from this difference image in two stages:
first, a preliminary candidate list is generated using an automated
detection routine and then, a final candidate list is culled by human
review of the preliminary list.  This visual inspection is conducted by one
of us (D.B.) and is essential for weeding out the large quantity of
non-variable objects (image defects, and PSF matching errors) that remain
after the automated detection stage.

At this stage, all candidate variables are given a preliminary
classification and any object that may possibly be a SN (of any type) has
`SN' in its classification.  The new measurements of variable candidates
are entered into our object database and compared with previously
discovered variable objects.  This comparison weeds out previously
discovered non-SN variables such as AGN and variable stars from the SN
candidate list.  All measurements of the current SN candidates, including
recent non-detections, are then evaluated for spectroscopic followup using
photometric selection criteria.

The details of the photometric selection process for the SNLS are
presented in \citet{Sullivan06AJ}.  In brief, all photometric
observations of the early part of the LC of a SN candidate are fit to
template SN~Ia LCs using a $\chi^2$ minimization in a multi-parameter
space that includes redshift, stretch, time of maximum light, host
extinction, and peak dispersion.  The template LCs are generated from
an updated version of the SN~Ia spectral templates presented in
\citet{Nugent02PASP}.  These spectral templates are multiplied by the
MegaCam filter response functions and integrated, thus accounting for
k-corrections \citep{Sullivan06AJ}.  The results of this fit are used
to measure a photometric redshift, $z_{PHOT}$, for the candidate and
to make a more accurate classification.  If there is any doubt about
the nature of the object, the `SN' classification is retained in the
database.

All SN candidates in the database are available for the observers
doing spectroscopic followup.  The quality of the candidate, deduced
from the template fit and an assessment of usefulness for cosmology,
is used to prioritize the candidates for spectroscopic observation.
Once these observations are taken, they are reduced and compared to
SN~Ia spectral templates \citep{Howell05ApJ,Basa06} to calculate a
spectroscopic redshift, $z_{SPEC}$.  The final typing assessment uses
all available information, both photometric and spectroscopic.  The
photometry provides early epoch colors, which can help identify CC
SNe.  It also provides an accurate phase for the spectroscopic
observation which is also important in discriminating SNe~Ia from CC
SNe.  The galaxy-subtracted candidate spectrum is then checked for the
presence of spectral features peculiar to SNe~Ia.  We then assign a
likelihood statistic for the candidate's membership in the SN~Ia type
\citep{Howell05ApJ}.

\section{Selection Criteria\label{sec_select}}

Selection criteria are used to provide consistency between the observed
sample, the survey efficiency calculation, and the completeness calculation
and thus produce an accurate rate.  In practice, they serve to objectify
the survey goals and properties (which unavoidably include the human
element) such that efficiency simulations are accurate and tractable.  The
criteria we developed consist of the minimum required photometric
observations, expressed in terms of rest-frame epoch and filter, that
guarantee that any real SN~Ia acquires spectroscopic followup.  They were
derived by examining the photometric observations of all our
spectroscopically confirmed SNe~Ia in the redshift range $0.2 < z < 0.6$.
To account for any real SN~Ia that meet these criteria but, for one reason
or another, did not acquire spectroscopic followup, we also apply these
criteria to our entire 'SN' candidate list in a completeness study (see
below).

Since the primary goal of the SNLS is cosmology, when selecting SN
candidates for spectroscopic followup we attempt to eliminate objects, even
SNe Ia, that offer no information for cosmological fitting.  Examples of
these include SNe for which no maximum brightness can be determined, or
for which no stretch or no color information can be measured.  Thus, the
objective criteria that define our sample and survey efficiencies are
expressed by requiring each confirmed SN~Ia to have the following
observations: 

\begin{enumerate}
\item one \ip\ detection at $S/N > 10.0$ between restframe day -15.0 and day -1.5
\item two \ip\ observations between restframe day -15.0 and day -1.5
\item one \rp\ observation between restframe day -15.0 and day -1.5
\item one \gp\ observation between restframe day -15.0 and day +5.0
\item one \ip\ or \rp\ observation between restframe day +11.5 and +30.0
\end{enumerate}

Criterion 1 and 2 implement our need to detect candidate SNe~Ia early
enough to schedule spectroscopic observations near maximum brightness.
Criterion 2 is required to judge if the LC is rising or declining.
Criteria 3 and 4 are required because an early color is important for
photometrically classifying the SN type.  Criterion 5 implements the
requirement that stretch information be available for any cosmologically
useful SN~Ia.  We only require the detection in the pre-max \ip\ because
during the early part of the light curve SNe~Ia are distinguished from
other SN types by having redder colors.  Thus, if we have an early
detection of a candidate in \ip, but can only place a limit on the object
in \rp\ or \gp, then it would have a reasonably high probability of being a
SN~Ia and is likely to be spectroscopically followed up.  This also means
that highly reddened SNe are not selected against.  For this redshift
range, we need not be concerned with criteria based on the \zp\ filter.
Criterion 5 would not logically enter the selection process as a detection,
since these observations could not be taken before the decision to followup
is made.  It is included solely to remove objects that are discovered close
to the end of an observing season, when there is no hope of obtaining the
observations needed to derive a stretch value.

It is important to point out that these criteria are independent of the LC
fitting that is normally done in candidate selection, i.e., there are no
criteria involving the SN~Ia fit $\chi^2$.  This is because we defined
these criteria with spectroscopically confirmed SNe~Ia.  The fitting is
required to derive the type and the redshift of candidate SNe.  In our
sample selection both of these quantities are given by the spectroscopy.
In the efficiency simulations, we are only interested in our detection
efficiency for SNe~Ia, so the type is defined {\it a priori}, and the
redshifts are given by the Monte Carlo simulation (see below).  The LC
fitting does enter into the completeness study, since we are then
interested in objects without spectroscopy.  We describe the LC
fitting criteria used to derive an accurate completeness, given the above
selection criteria, in \S\ref{sec_spec_comp}.

\subsection{The Observed Sample\label{sec_sample}}

In order to define an observed sample consistent with these survey
selection criteria, we must eliminate spectroscopically confirmed SNe Ia in
the initial list that do not meet these criteria.  These `special-case' SNe
Ia acquired spectroscopic followup for two reasons.  First, during some of
our initial runs we attempted to spectrally follow up nearly every
suspected SN to help refine our photometric selection criteria.  Second,
occasionally bad weather can prematurely end a field's observing season
before all the good declining candidates have the required observations to
determine their stretch values.

We derived our starting SN~Ia sample from all spectroscopically confirmed
SNe~Ia with a spectroscopic redshift, $z_{SPEC}$, in the range $0.2 <
z_{SPEC} < 0.6$, discovered in the first two full seasons of each deep
field.  The starting and ending dates and resulting time span in days is
listed for each season of each field in Table~\ref{tab_obs}.
Figure~\ref{fig_iobs} illustrates the field observing seasons for the
sample by plotting the Julian Day of the epochs versus their calculated
limiting \ip\ magnitude (see \S\ref{sec_det}).

Table~\ref{tab_sne} individually lists the 73 spectroscopically
confirmed SNe~Ia from the SNLS that comprise our starting sample.
Column 1 gives the SNLS designation for the SN, columns 2 and 3 give
the J2000.0 coordinates, column 4 gives the redshift, column 5 gives
the MJD of discovery, column 6 indicates if the object was culled from
the initial list by enumerating which of the criteria from
\S\ref{sec_select} it failed, and column 7 lists the references in
which further information about the object is published.  The table is
ordered by field and by time within each field with breaks between the
two seasons of the given field.

Figure~\ref{fig_snfits} plots the nightly averaged photometry for each
of the 73 spectroscopically confirmed SNe~Ia from Table~\ref{tab_sne}
with 1$\sigma$ error bars on a normalized AB magnitude scale.  The
best-fit SN~Ia template is overplotted \citep{Sullivan06AJ}.  The
magnitude scale is normalized such that the brightest tick mark is
always 20 magnitudes.  This procedure preserves the relative magnitude
difference between filters for a given SN.  The day scale on the
bottom of each plot is the observed day relative to maximum light.
The day scale on the top is the restframe day relative to maximum
light.  The designation from Table~\ref{tab_sne} (minus the SNLS
prefix) is given in the upper left corner and the spectroscopic
redshift in the upper right corner of each panel.  If the object was
culled from the initial list, this is indicated under the designation
with the word `Rej' (e.g. SNLS-03D1dj was culled).

Almost every object from season one of each field is included in the
cosmology fit of \citet{Astier06A&A}.  The three exceptions are
SNLS-03D1ar, which had insufficient observations at the time,
SNLS-03D4cj, which was a SN 1991T-like SN~Ia, and SNLS-03D4au, which
was under-luminous, most likely due to extinction.  We point out that
even though these objects were excluded from the cosmology fit, their
identity as SNe~Ia has never been in doubt.  The objects that
were observed with Gemini have their spectra published in
\citet{Howell05ApJ}.  The objects observed with the VLT will have
their spectra published shortly in \citet{Basa06}.  The remaining 10
objects from the first seasons were observed with Keck.

The sample is summarized in Table~\ref{tab_sam}, which lists, for each
season of each field, the total number of spectroscopically confirmed
SNe~Ia and the number after culling the starting list using our
objective selection criteria.

\subsection{Spectroscopic Completeness\label{sec_spec_comp}}

We now calculate the number of objects that passed our selection criteria
but, for one reason or another, were not spectrally followed up.  This
calculation is aided by the high detection completeness of the survey below
$z = 0.6$ (see \S\ref{sec_reseff}), and the classification scheme we use,
where any object remotely consistent with a SN LC, after checking for
long-term variability, retains the `SN' in the classification.  We are also
able to use a final version of the photometry, generated for all objects in
our database from images that have been de-trended with the final
calibration images for each observing run.  This final photometry, which
now covers all phases of the candidate LCs, is fit with SN~Ia templates as
described above to produce a more accurate $z_{PHOT}$ and a $\chi^2_{SNIa}$
for the $\chi^2$ of the SN~Ia template fit to the photometry in all
filters.  We examined all objects with a final photometry $z_{PHOT}$ in the
range $0.2 < z_{PHOT} < 0.6$ discovered within the time spans in
Table~\ref{tab_obs} with the following classifications: `SN', `SN?', `SNI',
`SNII', `SNII?', `SN/AGN', and `SN/var?'.  We measured the offset and
uncertainty in our $z_{PHOT}$ fitting technique by comparing $z_{PHOT}$
with $z_{SPEC}$ and found a mean offset of $\Delta z < 10^{-3}$ and an RMS
scatter of $\sigma_z = 0.08$.  We, therefore, assume that the remaining
error in $z_{PHOT}$ from the the final photometry is small and random such
that as many candidates are scattered out of our redshift range of interest
as are scattered in.  

Of 180 objects from the sample time ranges with `SN' in their type, 50 do not
have the required observations from our object selection criteria listed above
and so, even if they were SNe~Ia, would not be included in our culled sample.
Of the remaining 130 objects, 64 are rejected because their fit to the
templates has a $\chi^2_{SNIa} > 10.0$ and so very unlikely to be SNe~Ia
\citep{Sullivan06b}.  We then apply an upper limit stretch cut, requiring $s <
1.35$, to the remaining 66 objects.  Objects with $s > 1.35$ are also not
SNe~Ia (see Astier et al. 2005, Figure 7 and Sullivan et al. 2006b, Figure 3).
These objects are probably SNe~IIP which have a long plateau in their LCs and
hence produce anomalously high $s$ values when fit with a SN~Ia template.  The
$s < 1.35$ cut removes 33 objects.  We then make a cut by examining the early
colors and remove those that have large residuals in this part of the LC as a
result of being too blue (one signature of a core-collapse SN).  Of the
remaining 33 objects, 14 of these are rejected as too blue in the early colors,
even though the overall $\chi^2_{SNIa}$ is less than 10.0.  We are left with 19
unconfirmed SN~Ia candidates that have a reasonable probability of being
missed, real SNe~Ia.

Table~\ref{tab_cands} lists the 19 unconfirmed SN~Ia candidates, their
coordinates, their $z_{PHOT}$, discovery date, initial type,
$\chi^2_{SNIa}$, and their status.  We group them into those with
$\chi^2_{SNIa} < 5.0$ and those with $\chi^2_{SNIa} > 5.0$ and consider
those in the first group to be probable SNe~Ia, and those in the second
group to be possible SNe~Ia.  We point out that the intrinsic variation in
SN Ia LCs rarely allow template fits with $\chi^2_{SNIa} < 2$, and that
typical fits have $\chi^2_{SNIa}$ in the range 2-3 \citep{Sullivan06b}.  We take
the conservative approach that, aside from the division at $\chi^2_{SNIa} =
5.0$, we must consider each candidate in each group as equal.  This then
determines the range of completeness we consider in calculating our
systematic errors (see below).  Our most likely completeness is defined by
assuming that each `Probable~SN~Ia' in this list is, in fact, a real SN~Ia
and each `Possible~SN~Ia' is not.  The minimum completeness is defined by
the scenario that all 19 are real SN~Ia and the maximum completeness is
defined by the scenario that none of the 19 are real, which amounts to
100\% completeness.  We tabulate the confirmed, probable and possible
SNe~Ia and the minimum and most likely completeness for each field and the
ensemble in Table~\ref{tab_spec_comp}.  We will use this table when we
compute our systematic errors in \S\ref{sec_sys_spec}.

Figure~\ref{fig_snzph} plots the nightly average photometry for the 19
unconfirmed SN~Ia candidates from Table~\ref{tab_cands} using the same
normalized AB magnitude scale and day axes as in
Figure~\ref{fig_snfits}.  The photometric redshift is indicated in the
upper right corner of each panel.  The $\chi^2_{SNIa}$ value from
Table~\ref{tab_cands} for each SN is indicated under its designation
on each panel.

\section{Survey Efficiency\label{sec_eff}}

Since Fritz Zwicky's pioneering efforts to estimate supernova rates
from photographic surveys using the control-time method
\citep{Zwicky38ApJ}, there have been significant improvements in
calculating a given survey's efficiency \citep[for a review
see][Chapter 6]{Wood-Vasey05}.  As a recent example, \citet{Pain96ApJ}
used SN~Ia template LCs to place simulated SNe in CCD survey images to
generate a Monte Carlo simulation that produced a much more accurate
efficiency for their survey.  Most recent surveys using CCDs have
performed some variation of this method to calculate their
efficiencies and from them derive their rates \citep{Hardin00A&A,
Pain02ApJ, Madgwick03ApJ, Blanc04A&A}.

In our particular variation on this method, we do not place artificial SNe on
every image of our survey.  Instead, we characterize how our frame limits vary
with relevant parameters (such a seeing) using a subset of real survey images.
We then use this characterization to observe a Monte Carlo simulation which
uses the updated SN~Ia spectral templates of \citet{Nugent02PASP} and our
survey filter response functions to generate the LCs from a large population of
realistic SNe~Ia.  Thus, to calculate an appropriate survey efficiency, we need
to implement the objective selection criteria defined above in a Monte
Carlo efficiency experiment that simulates the observation of the SN~Ia LCs by
the SNLS.  Criteria 2-5 (see \S\ref{sec_select}) can be implemented simply by
inputting the date and filter of each image in the survey sample time ranges,
and seeing if we have the required observations for each simulated candidate.
Criterion 1 specifies a detection in the \ip\ filter, which requires that we
calculate the SN visibility at each \ip\ epoch in the survey sample time
ranges.

\subsection{\ip\ SN Visibility\label{sec_det}}

The photometric depth reached by a given \ip\ observation depends on the
exposure time ($E_e$), image quality ($IQ_e$), airmass ($X_e$), 
transparency ($T_e$), and the noise in the sky background ($S_e$).  Some of
these data are trivially available from each image header.  The
transparency and the sky background must be derived from the images
themselves.

Our final photometry pipeline includes a photometric calibration process
that calculates a flux scaling parameter, $F_e$, for each image.  We
calculate it by comparing a large number of isolated sources in the object
image with the same objects in a (photometric) reference image.  The
resulting $F_e$ values are applied to each object image to ensure that the
flux measured for a non-variable object is the same in each epoch.  Thus,
$F_e$ accounts for variations in both $T_e$ and $X_e$.  An image with lower
transparency and/or higher airmass will have a larger $F_e$.  During this
process the standard deviation per pixel in the sky is also calculated,
allowing us to account for variations in $S_e$.  The total number of
usable CCD chips, out of the nominal 36, is also tabulated (see
\S\ref{sec_ccdloss}). 

Another factor that determines a spatially localized frame limit is the
galaxy host background light against which the SN must be discerned
($H_{i',gal}$).  This depends on the brightness and light profile of the
host and the brightness and position of the SN within the host.  This
dependence is mitigated somewhat by the subtraction method used in our
detection pipeline (see \S\ref{sec_pipe}), but must still be measured.

We designed a controlled experiment to explore the effects of $IQ_e$, and
$H_{i',gal}$ on SN visibility.  This experiment places many artificial SNe
of varying brightness and host galaxy position (yielding a range of
$H_{i',gal}$) in real SNLS detection pipeline images of varying $IQ_e$.  We
chose a range of $IQ_e$ from $IQ_e = 0\arcpnt60$, close to the median for
the survey, to $IQ_e = 1\arcpnt06$, near the limit of acceptability.  We
used epochs with the canonical exposure time of 3641s and required
that the images were taken under photometric conditions.

Prior to the addition of fake SNe, each image was analyzed with SExtractor
\citep{Bertin96A&AS} to produce a list of potential galaxy hosts over the
entire image.  For a given fake SN, the host was chosen from this list using a
brightness weighted probability, such that brighter galaxies are more likely to
be the host than fainter galaxies.  The location within the host for the fake
SN was also chosen with a brightness weighted probability, such that more SNe
are produced where the galaxy has more light (i.e.  toward the center).  Once
the location within the pipeline image is decided, a nearby isolated, high S/N
star was scaled to have a magnitude in the range $21.0 < \ip < 27.0$ and added
at the chosen position.

There was no correlation of the fake SN magnitude with the host magnitude,
therefore, our simulations were relevant for SNe at all phases of their LC.
This spatial distribution and magnitude range allows us to quantify any
systematic loss of SN visibility near the cores of galaxies in our recovery
experiments (see below).  Once a set of these images was produced it was
put through the same detection pipeline used by the Canadian SNLS for
detecting real SNe \citep{Perrett06,Sullivan06AJ,Astier06A&A}.

Figure~\ref{fig_comp_review} shows the raw recovery percentage of $\sim$
2000 fake SNe after the human review process for two $IQ_e$ values:
$0\arcpnt69$ and $1\arcpnt06$.  The 50\% recovery limits are indicated and
are the most useful for rate calculation since the visible SNe missed below
these limits are gained back by including the invisible SNe above the
limits (see Figure~\ref{fig_comp_review}a).  The loss in visibility going
from automatic detection to human review amounts to a brightening of the
visibility limits of only 0.1 magnitudes at the small $IQ_e$ value.
Figure~\ref{fig_comp_review}b shows no trend with host offset and
Figure~\ref{fig_comp_review}d shows that the cutoff due to background
brightness is 20 mag arcsecond$^{-2}$.  A notable feature of
Figure~\ref{fig_comp_review}a is the maximum recovery percentage of 95\%
for the $IQ_e = 0\arcpnt69$ image.  We examined the spatial distribution of
the fake SNe from this image that were missed above $\ip = 23$ to try to
understand the source of this limit on the recovery.  We saw no correlation
with galaxy host offset, proximity to bright stars, or placement on masked
or edge regions.  This feature appears to be purely statistical in origin
and we account for it when observing the Monte Carlo simulations (see
\S\ref{sec_monte}).

Figure~\ref{fig_det} shows the 50\% recovery limits derived from the fake
SN experiments using nine \ip\ images having a range of $IQ_e$.  These
limits have been corrected for sky noise, transparency, and exposure time
differences.  We plot the histogram of all the $IQ_e$ values for all \ip\
images relevant to this study as a dashed line.  All points are derived at
the automated detection stage unless otherwise indicated.  The corrected
results of the human review experiment from above are plotted as asterisks.
We fit the human review limits with a linear fit (shown as the solid line)
and this fit represents an upper bound on the limits of the images we
sampled.  A constant frame limit of $\ip = 24.5$ is shown as the solid
horizontal line and is a reasonable lower bound.  The two solid lines
encompass all points in Figure~\ref{fig_det}.  We will use the human review
limit fit as our best estimate of the frame limit versus $IQ_e$ function,
with the constant limit as an estimate for the systematic error in our
rates due to the \ip\ frame limits (see \S\ref{sec_sys}).

\subsubsection{\ip\ SN Visibility Equation\label{sec_ddet}}

Our fake SN experiments have provided a way to calculate the visibility
limit in magnitudes, $L_e$, for any \ip\ epoch in our survey sample time
span using the following formula: \begin{equation} L_e = L_{0.5} -
\alpha(IQ_e - 0.5) + 2.5 \log(E_e / E_{ref}) - 2.5 \log(F_e) - 2.5 \log(S_e
/ S_{ref}), \label{eq_flim} \end{equation} where $L_{0.5}$ is reference
visibility limit for an epoch with $IQ_e = 0\arcpnt5$, $F_e =1.0$, exposure
time of $E_{ref}$ seconds and sky noise of $S_{ref}$ counts, $\alpha$ is
the proportionality factor between $IQ_e$ and the visibility limit, $E_e$
is the exposure time of the epoch, $F_e$ is the flux scale factor, and
$S_e$ is the sky noise in counts of the epoch (see Figure~\ref{fig_iobs}).
This formula assumes a linear relationship between $IQ_e$ and $L_{e}$,
which appears to be a reasonable approximation over the range of $IQ_e$
used to discover SNe (see Figure~\ref{fig_det}).

Table~\ref{tab_det} lists the parameters calculated using the 50\% recovery
fraction visibility limits determined from the human review recovery
experiment (see Figure~\ref{fig_comp_review} and \ref{fig_det}).  Columns 1
and 2 of Table~\ref{tab_det} lists the $IQ_e$ for the pair of good and bad
$IQ$ images used in the human review experiment, column 3 lists the
reference exposure time, column 4 lists the reference sky noise in counts,
column 5 lists the visibility limit at $IQ = 0\arcpnt5$, and column 6 lists
the proportionality constant between $IQ_e$ and $L_e$.  For the reference
sky noise, $S_{ref}$, we used the value from the good $IQ$ image and
adjusted the limit from the poor $IQ$ image to correspond to an image with
the same sky noise as the good $IQ$ image.

\subsubsection{Temporary CCD Losses\label{sec_ccdloss}}

Another factor affecting the visibility of SNe in the SNLS must be
accounted for.  Occasionally, a very small subset of the 36 MegaCam CCDs will
malfunction for a short time, usually because of a failure in the readout
electronics.  An even rarer occurrence is the appearance of a condensate of
water on the surface of one of the correctors that covers a localized area
of the field of view rendering that part of the detector temporarily
useless for the detection of SNe.  When we calculate the fluxscale factors
mentioned above, the number of usable CCDs is also recorded.  This number
is used to account for these localized, temporary losses of SN
visibility (see below).

\subsection{Monte Carlo Simulation\label{sec_monte}}

The Monte Carlo technique allows us to determine our survey efficiency to a
much higher precision than permitted by the small number of observed
events.  Using observed SN~Ia LC properties and random number generators,
we simulate a large ($N=10^6$) population of SN~Ia events in the sample
volume occurring over a two year period centered on the observed seasons
for the field.  This large number is sufficient to drive the Poisson errors
down to $\sqrt{N}/N = 0.1$\%.  This population is then observed by using
real SNLS epoch properties and equation~\ref{eq_flim}, combined with our
objective selection criteria, to define the number of simulated spectroscopic
SN~Ia confirmations.  This number is divided by the number of input
simulated SNe~Ia to derive the yearly survey efficiency.

\subsubsection{Generating the Sample Population\label{sec_pop}}

To simulate a realistic population of Type Ia SNe, we use the same LC
templates and software used to determine photometric redshifts for the SNLS
candidate SNe \citep{Sullivan06AJ}.  Figure~\ref{fig_pops} shows the
canonical distributions of the parameters that characterize SN~Ia LCs used
in our efficiency simulations.  The redshifts are chosen with a volume
weighted uniform random number generator, to produce a redshift
distribution over the range $0.2 < z < 0.6$ that is uniform per unit volume
as shown in Figure~\ref{fig_pops}a.  We also calculated the run of $dV(z)$,
given the cosmological parameters from \S\ref{sec_intro}, and over-plotted
this as a dashed line (with an offset of $F = 0.005$ for clarity) to show
that our distribution is indeed constant per unit volume.  The stretch
values are selected using a Gaussian distribution centered on 1 with a
width of $\sigma_{s} = 0.1$ (Figure~\ref{fig_pops}b).  The intrinsic SN
color is determined using the stretch-color relation from
\citet{Knop03ApJ}.  The host color excesses are chosen from the positive
half of a Gaussian distribution, centered on 0.0 with a width of
$\sigma_{E(B-V)_h} = 0.2$ (Figure~\ref{fig_pops}c).  These are converted to
host extinction assuming an extinction law with $R_V = 3.1$
\citep{Cardelli89ApJ}. The peak magnitude offsets (after stretch
correction) shown in Figure~\ref{fig_pops}d are chosen from a Gaussian
distribution centered on 0 with a width of $\sigma_{B_{MAX}} = 0.17$
\citep{Hamuy96AJ}.  A uniform random number generator is used to pick the
day of maximum for each simulated SNe~Ia from a two-year-long interval that
is centered on the middle of the survey range being simulated.  This avoids
problems with edge effects and produces an efficiency per year.  We will
address the systematic uncertainty due to differences between these
distributions and the true distributions in \S\ref{sec_sys}.  

In order to account for the possibility that a given SN can be missed
because of temporary localized losses of SN detectability in the Megacam
array (see above, \S\ref{sec_ccdloss}), we assign a pseudo-pixel position
to each simulated SN.  This is done with a uniform random number generator
that selects one of the 370 million Megacam pixels that are nominally
available as the location of the SN.  The number of real pixels available
on a given epoch is calculated from the number of usable chips, derived
during the fluxscale calculation.  By choosing a random number out of 370
million, we are essentially assigning a probability that the SN will land
on a region of the array that is temporarily unusable.  If all chips are
working, then the number of pixels available equals the nominal number and
no SNe are lost.  If a large number of chips are not working, the the
number of pixels available is much less than the nominal number and a
simulated SN has a higher probability of being missed.

\subsubsection{Observing the Sample Population}

With the input sample population defined and LCs covering the simulated
period generated, we use the data describing the real SNLS survey epochs to
observe the simulation.  First, we use an average Milky Way extinction
appropriate for the field being simulated \citep[Table 1]{Astier06A&A}.
Then we use the epoch properties, equation~\ref{eq_flim}, and
Table~\ref{tab_det} to calculate visibility limits for each \ip\ epoch.
These visibility limits are used to define the signal-to-noise ratio,
$S/N$, for each simulated SN observation using the following formula:
\begin{equation} S/N = 10.0 \times 10^{[-0.4(m_e - L_e)]},\label{eq_sn}
\end{equation} with $m_e$ being the template magnitude of the simulated
SN~Ia in the epoch, and $L_e$ the epoch 50\% visibility limit.  This
formula assumes that an observation at the 50\% visibility limit has a
$S/N$ of 10.  The calculated $S/N$ defines the width of a Gaussian noise
distribution and a Gaussian random number generator is used to pick the
noise offset for the observation.

After the noise offsets are added to the observations, the resulting
magnitudes in each epoch are then compared with the corresponding \ip\
visibility limits, $L_e$, and any magnitude that is brighter than its
corresponding limit is considered an \ip\ detection.  We use a uniform
random number generator to assign a real number ranging from 0.0 to 1.0 for
each \ip epoch.  If this number exceeds 0.95, then the candidate is not
detected in that epoch.  This accounts for the 95\% maximum recovery
fraction observed in Figure~\ref{fig_comp_review}a.  The shape of the
recovery fraction at fainter magnitudes is already accounted for by using
the 50\% recovery magnitudes in the visibility limit calculation (see
Figure~\ref{fig_comp_review}a).  To account for localized visibility
losses, we calculate the number of pixels available on the epoch from the
number of good CCDs available on that epoch.  If the candidate was assigned
a pseudo-pixel number larger than the number of good pixels on the epoch,
then the candidate is not detected on that epoch.

The restframe phases (relative to peak brightness) of all the relevant \ip\
epochs are calculated for each simulated SN~Ia using its given redshift.
If a simulated SN~Ia ends up with a detection in the restframe phase range
from criterion 1 (see \S\ref{sec_select}) then we evaluate it with respect
to the remaining criteria.  We calculate the restframe phase for each
observation in the \gp, and \rp\ epochs and then the remaining
criteria are applied to decide if the simulated SN should be counted as a
spectroscopically confirmed SN~Ia.

For the yearly efficiency, we keep track of the total number of SNe~Ia that
are simulated, since they are generated in yearly intervals.  We also keep
track of the number of SNe~Ia that were simulated during the observing
season for each field (from 158 to 211 days, see Table~\ref{tab_obs}).  This
allows us to compute our on-field detection efficiency and our on-field
spectroscopic confirmation efficiency.

\subsubsection{The Monte Carlo Survey Efficiency\label{sec_reseff}}

The resulting efficiencies for each field are presented in
Table~\ref{tab_mc}.  As we stated above, the statistical errors in these
numbers are $\sim$0.1\%.  We present the on-field \ip\ detection
efficiencies in column 2, which are all within 5\% of 100\%. This is
expected considering the redshift range of our sample and the nominal \ip\
frame limits.  It also bolsters our spectroscopic completeness analysis by
showing that our SN candidate list is not missing a significant population,
in our redshift range.  The on-field spectroscopic efficiency (column 3)
averages close to 60\%, reflecting the spectroscopic followup criteria
applied to the detected SNe~Ia.  The yearly efficiency (column 4) averages
close to 30\% which reflects the half-year observing season for each field.

We can compare Figure~\ref{fig_iobs} with Table~\ref{tab_mc} as a
consistency check.  Starting with the on-field detection efficiencies
(column 2), we notice that D1 has the lowest value.  In
Figure~\ref{fig_iobs} we see that D1 has the largest variation in the
visibility limits with some limits approaching $\ip = 20$.  Going to the
spectroscopic on-field efficiency (column 3), we see that D2 has the lowest
value.  This is due to the large gap in the relatively short first season
of D2.  We also see that D4 has the highest on-field spectroscopic
efficiency and the lowest scatter in the visibility limits in both seasons.
In the last column of Table~\ref{tab_mc} we see that D3 has the highest
yearly spectroscopic efficiency, due to the fact that D3
consistently has the longest seasons of any field (see also
Table~\ref{tab_obs}, column 7).  D2 has the shortest season and
consequently, has the lowest efficiency.

\section{Results\label{sec_res}}

We are now ready to apply our survey efficiencies to the culled, observed
sample of SNe~Ia and thereby derive a rate.  The high detection efficiency
of the survey from column 2 of Table~\ref{tab_mc} illustrates that our
sample for this study constitutes a volume limited sample as opposed to a
magnitude limited sample.  This means that we do not produce a predicted
redshift distribution to define our rate and average redshift as was done
in \citet{Pain02ApJ}, for example.  Instead, we apply our efficiency
uniformly to our sample and our average redshift is the volume weighted
average redshift in the range $0.2 < z < 0.6$.  We apply the appropriate
efficiency to the sample of each field individually, propagating the
Poisson errors of the field's sample, and then take an error-weighted
average to derive our best estimate of the cosmic SN~Ia rate averaged over
our redshift range.  We present the results from these calculations below.
We also derive a rate per unit luminosity, present an analysis of our
systematic errors, and compare our results with rates in the literature in
this section.

\subsection{SN Type Ia Rate Per Unit Comoving Volume\label{sec_vol}}

We first need to calculate the true observed number of SNe~Ia per year in
each field.  We then need to correct for the fact that at higher redshift,
we are observing a shorter restframe interval due to time dilation.  To
derive the final volumetric rate we then calculate the total volume
surveyed in each field and divide this out.  We express these calculations
with the following formula: \begin{equation} r_V =
\frac{N_{Ia}/2}{\epsilon_{yr}\ C_{SPEC}}\ [1+\langle z \rangle_V]\
\{\frac{\Theta}{41253} [V(0.6) - V(0.2)]\}^{-1},\label{eq_rv}\end{equation}
where $N_{Ia}/2$ is the number of confirmed SNe~Ia in the sample (see
column 2 of Table~\ref{tab_spec_comp}) divided by the number of seasons (2),
$\epsilon_{yr}$ is the yearly spectroscopic efficiency from column 4 of
Table~\ref{tab_mc}, $C_{SPEC}$ is the spectroscopic completeness presented
in column 6 of Table~\ref{tab_spec_comp}, $1+\langle z \rangle_V$ is the
time dilation correction using the volume weighted average redshift over
our redshift range, $\Theta$ is the sky coverage in square degrees which is
divided by 41253 (the total number of square degrees on the sky), and
$V(z)$ is the total volume of the universe out to the given redshift.
These volumes are calculated using: \begin{equation} V(z) = \frac{4}{3}\pi
\left [ \frac{c}{H_0}\ \int_0^z\frac{dz'}{\sqrt{\Omega_m\ (1+z')^3 +
\Omega_{\Lambda}}} \right ]^3, \label{eq_vol}\end{equation} with the
parameters listed in \S\ref{sec_intro} and assuming a flat cosmology
($\Omega_k = 0$).

The columns in Table~\ref{tab_rates} give the results at several
stages in applying equation~\ref{eq_rv} along with some of the
parameters used in the calculation.  Column 2 presents the observed
raw rate, $r_{RAW}$, calculated by simply dividing the average yearly
sample for each field by the yearly spectroscopic efficiency,
$\epsilon_{yr}$.  Column 3 shows $r_{obs}$, the true observed yearly
rate of SNe~Ia in each field, which is the result of applying our
spectroscopic completeness corrections ($C_{SPEC}$) to $r_{RAW}$.
Column 4 shows the results of accounting for time dilation by
multiplying the observed rates by $1 + \langle z \rangle_V$, where
$\langle z \rangle_V = 0.467$ is the volume weighted average redshift.
Column 5 lists the areal coverage for each field, after accounting for
unusable regions of the survey images which include masked edge
regions, and regions brighter than 20 mag arcsec$^{-2}$ in \ip\ (see
Figure~\ref{fig_comp_review}d).  The resulting survey volume between
redshift $0.2 < z < 0.6$ is reported in column 6, using
equation~\ref{eq_vol} which gives $1.035 \times 10^6$ Mpc$^3$
Deg$^{-2}$, using the cosmological parameters described in
\S\ref{sec_intro}.  Column 7 is the resulting volumetric rate.  At
each stage in the calculation the results are listed for each field
and for a weighted average for the ensemble.  Our derived rate per
unit comoving volume, $r_V$, is $r_V = 0.42 \pm 0.06 \times 10^{-4} $
yr$^{-1}$ Mpc$^{-3}$ (statistical error only).

\subsection{SN Type Ia Rate Per Unit Luminosity\label{sec_lumf}}

We use the galaxy LF derived from the first epoch data of the
VIMOS-VLT Deep Survey \citep{Ilbert05A&A} to calculate the B-band
galaxy luminosity density.  This recent survey derives the LF in the
redshift range $0.2 < z < 0.6$ from 2,178 galaxies selected at $17.5
\leq I_{AB} \leq 24.0$.  The Schechter parameters for the rest-frame B
band LF are tabulated in their Table 1.  We integrated the Schechter
function \citep{Schechter76ApJ} in the two redshift bins $0.2 < z <
0.4$ and $0.4 < z < 0.6$ and used a volume weighted average to get a
luminosity density in the B-band of $\sigma_B = 2.72 \pm 0.48 \times
10^8 L_{\odot,B}$ Mpc$^{-3}$.

By using parameters derived from galaxies in our redshift range of
interest, we do not need to evolve a local LF.  As long as the slope of the
faint end of the LF is well sampled and hence the $\alpha$ parameter is
well determined, this produces an accurate luminosity density and hence an
accurate SN~Ia rate per unit luminosity.  Figure~4 of \citet{Ilbert05A&A}
shows that the LF in the highest redshift bin ($0.4 < z < 0.6$) is well
sampled to $\sim$3.5 magnitudes fainter than the `knee' of the function.

We now convert our volumetric rate into the commonly used luminosity
specific unit called the ``supernova unit'' (SNu), the number of SNe per
century per 10$^{10}$ solar luminosities in the rest-frame B band.
Dividing the luminosity density in the rest-frame B band by our
volumetric rate from Table~\ref{tab_rates} gives $r_L =
0.154^{+0.039}_{-0.031}$ SNu (statistical error only).

\subsection{Systematic Errors\label{sec_sys}}

The rates for each field in Table~\ref{tab_rates} are all within one
$\sigma_{STAT}$ of each other at each stage of the calculation of $r_V$.
This tells us that there are no statistically significant systematic errors
associated with our individual treatment of the fields.  In the subsequent
analysis, we examine sources of systematic error that affect the survey in
its entirety.  We tabulate the values and sources of statistical and
systematic errors in Table~\ref{tab_errs} for both $r_V$ and $r_L$ and
describe each systematic error below.

\subsubsection{Spectroscopic Incompleteness\label{sec_sys_spec}}

We estimate the systematic error due to spectroscopic incompleteness by
using our detailed examination of the SN candidates from
\S\ref{sec_spec_comp} as tabulated in Table~\ref{tab_spec_comp}, column 4.
Using the extremes of completeness for the ensemble (75\% to 100\%) as
limits on this systematic error, the spectroscopic incompleteness is
responsible for a systematic error on $r_V$ of $(+0.03,-0.08) \times
10^{-4}$ yr$^{-1}$ Mpc$^{-3}$ and on $r_L$ of $(+0.010,-0.031)$ SNu.

\subsubsection{Host Extinction\label{sec_sys_dust}}

For our canonical host extinction, we used a positive valued Gaussian
$E(B-V)_h$ distribution with a width of $\sigma_{E(B-V)_h} = 0.2$ (see
Figure~\ref{fig_pops}c) combined with an extinction law with $R_V = 3.1$
\citep{Cardelli89ApJ}.  We follow the procedure described in
\citet{Sullivan06AJ}, with the exception that our host extinction is allowed
to vary beyond $E(B-V)_h = 0.30$.  Systematics are introduced if our
canonical distribution differs significantly from the real SN~Ia host color
excess distribution, or if there is evolution of dust properties such that
the $R_V = 3.1$ model is significantly inaccurate.  Preliminary results
from submm surveys of SN~Ia host galaxies out to redshift $z = 0.5$ show no
significant evolution in the dust properties when compared to hosts at $z =
0.0$ \citep{Clements05MNRAS}.  We thus concentrate on the distribution of
$E(B-V)_h$ as the major source of systematic errors in our redshift range.

In an effort to quantify the systematic contribution of host extinction to
an underestimation of the SN~Ia rate, we re-ran our Monte Carlo efficiency
experiments setting $E(B-V)_h = 0.0$ for each simulated SN.  We analyzed
the results of this experiment exactly as before (see \S\ref{sec_vol}) and
derived a volumetric rate of $r_V = 0.38 \pm 0.05 \times 10^{-4} $
yr$^{-1}$ Mpc$^{-3}$.  This rate is 10\% ($0.67 \sigma_{STAT}$) lower than the
rate using our canonical distribution (see Table~\ref{tab_rates}) and
quantifies the magnitude of the error possible, if host extinction is
ignored.  We will also use this zero dust rate to calculate rate
corrections due to host extinction (see below).

If we assume that our empirical host color excess distribution is biased by
not including SNe in hosts with extreme $E(B-V)_h$, and hence extreme
$A_V$, then the systematic error on $r_V$ is positive only.  In an attempt to
quantify this error, we compare our $E(B-V)_h$ distribution to models of
SN~Ia host extinction presented in \citet[hereafter RP05]{Riello05MNRAS}.

RP05 improve upon the simple model of \citet{Hatano98ApJ}, motivated by the
findings of \citet{Cappellaro99A&A} that the \citet{Hatano98ApJ} model
over-corrects the SN Ia rate in distant galaxies.  RP05 use a more
sophisticated model of dust distribution in SN~Ia host galaxies and include
the effects of varying the ratio of bulge-to-disk SNe~Ia within the host.
The resulting $A_V$ distributions, binned by inclination, are strongly
peaked at $A_V = 0.0$, have high extinction tails, and do not have a
Gaussian shape (RP05, Figure 3).   The smearing of the large fraction of
objects with $E(B-V)_h \sim 0$ by photometric errors would produce a more
Gaussian shape.

Because the $A_V$ distributions of RP05 and the $A_B$ distributions of
\citet{Hatano98ApJ} produce tails of objects with high extinction that
extend beyond Gaussian wings, we performed two additional experiments using
exponential distributions for $E(B-V)_h$ to simulate these tails.  We
generated these exponential distributions using a uniform random number
generator to produce a set of random real numbers between 0 and 1, which we
will designate as $\Re$, and applied the following equation:
\begin{equation} E(B-V)_h = -\ln \Re / \lambda_{E(B-V)_h}\ ,\end{equation}
where $\lambda_{E(B-V)_h}$ is the exponential distribution scale factor.
The smaller the value of $\lambda_{E(B-V)_h}$ the larger the tail of the
distribution.  Figure~\ref{fig_hex} shows the two exponential distributions
with $\lambda_{E(B-V)_h} = 5$ and $3$, along with the canonical
distribution, converted to $A_V$ using $R_V = 3.1$ and binned using the
same bin size as RP05 ($dA_V = 0.1$).  If we compare these distributions
with Figure 3 of RP05, we see that our canonical distribution is closest to
the form of their model with an inclination range of $45^{\circ} \leq i
\leq 60^{\circ}$.  Specifically, both our distributions have a maximum of
$A_V \sim 2.5$.  The exponential distributions are closer matches to their
highest inclination bin, $75^{\circ} \leq i \leq 90^{\circ}$, showing tails
extending beyond $A_V = 7.0$ (although at very low probability).

Using these distributions, we re-ran our Monte Carlo efficiency experiments
and re-derived the volumetric rates to quantify the effect on our derived
rate of missed SNe due to exponential tails in the host extinction
distribution.  For the $\lambda_{E(B-V)_h} = 5$ distribution, we derived a
rate of $r_V = 0.44 \pm 0.06 \times 10^{-4}$ yr$^{-1}$ Mpc$^{-3}$, which is
only 5\% higher than our canonical value.  The $\lambda_{E(B-V)_h} = 3$ case
produced a rate of $r_V = 0.52 \pm 0.07 \times 10^{-4}$ yr$^{-1}$
Mpc$^{-3}$, which is 24\% or $1.67\sigma_{STAT}$ higher.  This distribution is
appropriate for spiral SN~Ia hosts with high inclination, but will
over-estimate the correction to rates from hosts with a range of
inclinations and host morphologies. We, therefore, regard it as a measure
of the upper limit on the statistical error due to host extinction.

We can also compare the rate correction factors from RP05 with the
correction factor resulting from the exponential dust distributions.  Using
our zero extinction experiment and the $\lambda_{E(B-V)_h} = 3$ dust
distribution, this factor is ${\mathcal R} = 0.52/0.38 = 1.37$.  This value
encompasses the factors reported in RP05 (see their \S5) for their models
with bulge-to-total SNe ratios of $0.0 \leq B/T \leq 0.5$, which are given
as $1.22 \leq {\mathcal R}_{B/T} \leq 1.31$.  It also bounds their
correction factors derived for dust models with $R_V = 3.1$, $4.0$, and
$5.0$ which are given as ${\mathcal R}_{R_V} = 1.27, 1.31, 1.34$.  Finally,
this correction is not exceeded by the corrections derived from the RP05
models with total face-on optical depth, $\tau_V = 0.5$ and $1.0$, which
are given as ${\mathcal R}_{\tau_V} = 1.16,1.27$.

These comparisons demonstrate that it is reasonable to assume that we
encompass the systematic errors due to host extinction if we use the
$\lambda_{E(B-V)_h} = 3.0$ exponential host extinction distributions to
define their upper limit.  This results in a systematic error due to host
extinction on $r_V$ of $(+0.10) \times 10^{-4}$ yr$^{-1}$ Mpc$^{-3}$ and on
$r_L$ of $(+0.037)$ SNu.

\subsubsection{Stretch\label{sec_sys_str}}

We consider the effect of errors in the input stretch distribution on our
derived rates.  We re-ran our efficiency experiment doubling the width of
the stretch distribution to $\sigma_s = 0.2$.  This produced a rate of $r_V
= 0.43 \pm 0.06 \times 10^{-4}$ yr$^{-1}$ Mpc$^{-3}$, which is only 2\%
higher than the rate using $\sigma_s = 0.1$.  The resulting systematic error
due to stretch for $r_V$ is $\pm0.01 \times 10^{-4}$ yr$^{-1}$ Mpc$^{-3}$
and for $r_L$ is $\pm0.040$ SNu.

\subsubsection{Frame Limits\label{sec_sys_flims}}

Figure~\ref{fig_det} shows the distribution of the 50\% frame limits versus
$IQ_e$ for a sample of \ip\ images from the survey.  The slope from
Table~\ref{tab_det}, column 6 is steeper than one would expect from a
simple analysis of the standard CCD $S/N$ equation (e.g., Howell 1999) in
the limit where the noise is dominated by the sky (i.e.\ at the frame
limit).  The expected slope is closer to 1.3, a factor of 1.7 lower than
what we derived from our fake SN experiments.  This could be the result of
the coaddition process, perhaps because the co-alignment accuracy is
sensitive to variations in $IQ_e$.  In order to account for a possible
overestimation of the dependence of frame limit on $IQ_e$ (i.e. too large
an $\alpha$ from equation~\ref{eq_flim}), we re-ran our efficiency
experiment with a constant frame limit of $\ip = 24.5$, which is shown in
Figure~\ref{fig_det} as the solid horizontal line.  The resulting
volumetric rate was $r_V = 0.39 \pm 0.05 \times 10^{-4} $ yr$^{-1}$
Mpc$^{-3}$.  Using this value we estimate that an error in calculating the
frame limits would introduce a systematic error on $r_V$ of $(-0.03) \times
10^{-4}$ yr$^{-1}$ Mpc$^{-3}$ and on $r_L$ of $(-0.011)$ SNu.

\subsubsection{Host Offset\label{sec_sys_hoff}}

One of the factors offered by \citet{Dahlen04ApJ} to account for the
discrepancy between ground-based rates near $z = 0.5$ and the delay-time
models they present is the close proximity of SN candidates to host galaxy
nuclei.  If a candidate is too close to a bright host nucleus, they argue,
it can be mis-classified as an AGN, or it might be passed up for
spectroscopic followup because of the high level of host contamination.
The results of our fake SN experiments (see \S\ref{sec_det},
Figure~\ref{fig_det}b) show that there is no such loss of SN sensitivity
close to the hosts of galaxies in the SNLS in the redshift range $0.2 < z <
0.6$.  Another way to look at these data is shown in Figure~\ref{fig_hoff},
which plots the percentage of fake SN missed as a function of host offset
for two $IQ_e$ values from our fake SN experiments (see \S\ref{sec_det}).
This further illustrates the lack of trend with host offset (see also
Figure~2, Pain et al.  2002).  We specifically used a brightness weighted
probability distribution for placing our fake SNe, shown in
Figure~\ref{fig_hoff} as the dot-dashed histogram, which preferentially
places them in the brightest regions of a galaxy (i.e. near the center), so
that we could detect any such problem.

The study by \citet{Howell00ApJ} also showed no significant loss of
objects at small host offset when comparing a sample of 59 local SN~Ia
discovered with CCD detectors and a sample of 47 higher redshift ($z >
0.3$) CCD-discovered SNe.  We conclude that this effect is not significant,
at least out to $z = 0.6$, for the SNLS.

\subsubsection{Subluminous SN Ia Population\label{sec_sys_slum}}

Subluminous or SN 1991bg-like SNe~Ia are another potential source of
systematic error.  Compared to the so-called normal SNe~Ia these
objects can have peak magnitudes up to two magnitudes fainter, exhibit
different spectral features, have a different stretch color
relationship, yet they still obey the stretch-luminosity relationship
exhibited by the normal SNe~Ia \citep{Garnavich04ApJ}.  As such, they
would be useful on a Hubble diagram, however, none of our
spectroscopically confirmed SNe~Ia fall into the subluminous class.
This has been confirmed independently by equivalent width measurements
of our spectroscopically confirmed sample \citep{Bronder06}.  We
cannot assume, however, that the relative frequency of these objects
decends to zero at higher redshifts.  Their intrinsic faintness and
differing color may produce a bias against these objects in our
followup selection criteria.

It is still likely that these subluminous SNe~Ia are recent phenomena,
since they are strongly associated with older stellar populations
\citep{Howell01ApJ}.  They have yet to be found in significant numbers
at redshifts beyond $z=0.2$, even though many CC SNe of similar peak
magnitude have been found.  It is thus reasonable to assume that the
current fraction of subluminous SNe~Ia forms an upper limit on the
fraction at higher redshifts.  The best estimate of the current
fraction is from \citet{Li01APJ}, who derive a fraction of $16\pm6$\%
for the subluminious class using a volume-limited sample.  Rather than
attempt to evolve this number to the redshift range of interest for
this study, we take the conservative approach and use it to calculate
an upper limit on our systematic error due to missing the subluminous
SNe~Ia, assuming their relative fraction has no evolution.  This
calculation yields a systematic error on $r_V$ of $(+0.08) \times
10^{-4}$ yr$^{-1}$ Mpc$^{-3}$ and on $r_L$ of $(+0.029)$ SNu.

\subsection{Comparison with Rates in the Literature\label{sec_lit}}

Figure~\ref{fig_lit} shows the same rates and SFH as Figure~\ref{fig_sfh},
but now with our rate included as the filled square.  At $z \sim 0.45$ all
four observed rates are in statistical agreement.

The result from \citet{Pain02ApJ} at $z = 0.55$ of $r_V =
0.525^{+0.096}_{-0.086}$(stat)$^{+0.110}_{-0.106}$(syst)$ \times
10^{-4}$ yr$^{-1}$ Mpc$^{-3}$ is higher than our rate, but still in
statistical agreement.  We must consider that \citet{Pain02ApJ} do not
account for host extinction in their rate derivation (see their
\S6.8).  There are two possible explanations.  Either host extinction
has a small effect in calculating SN~Ia rates in this redshift range,
or the lack of host extinction correction was compensated for by an
equal amount of contamination in the result from \citet{Pain02ApJ}.
If we calculate a correction for host extinction from our Monte Carlo
experiment where we set all SNe to have $E(B-V)_h = 0.0$, we can
estimate how their rate would change if they had accounted for host
extinction using our method.  This correction is ${\mathcal R} =
0.42/0.38 = 1.11$ when compared to our canonical host extinction
results.  Applying this 11\% correction to the value from
\citet{Pain02ApJ} produces a rate of $r_V = 0.58 \times 10^{-4}$
yr$^{-1}$ Mpc$^{-3}$, which is only $0.5\sigma$ higher than their
original value.  We, therefore, conclude that the rates derived in
this range are not significantly affected by host extinction.

\subsubsection{Contamination}

Even though all the rates near $z=0.5$ are in statistical agreement, the
rate at $z = 0.55$ from \citet{Barris06ApJ} is within our redshift range
and is nearly five times our value ($> 4\sigma$ greater).  The largest
disagreement between published rates in Figure~\ref{fig_lit} (and in the
literature as far as we know) is that between the rates at $z = 0.55$ of
\citet{Pain02ApJ} and \citet{Barris06ApJ}.  We point out that
\citet{Barris06ApJ} is a re-analysis of the data from \citet{Tonry03ApJ},
which reported a rate that agrees with \citet{Pain02ApJ}. The rate from
\citet{Barris06ApJ} is a factor of 3.9 higher than the rate from
\citet{Pain02ApJ}.  If we add the errors on these rates in quadrature, this
amounts to a 3.8$\sigma$ difference.

We have shown that host extinction cannot be the sole explanation for these
discrepancies.  Our estimate for the host extinction correction factor for
\citet{Pain02ApJ} is ${\mathcal R} = 1.11$, which would not be enough to
resolve it.  A host extinction correction factor of ${\mathcal R} = 2.61$
would be required to bring the rate from \citet{Pain02ApJ} just into
statistical agreement with the rate of $r_V = 2.04 \pm 0.38 \times 10^{-4}$
yr$^{-1}$ Mpc$^{-3}$ at $z = 0.55$ from \citet{Barris06ApJ}.  This
correction is larger than the model from RP05 for hosts with total face-on
optical depth of $\tau_V = 10$ which gives ${\mathcal R}(\tau_V) = 2.35$,
the largest correction listed in RP05.  An even larger correction would be
required to bring our rate into statistical agreement with the rate from
\citet{Barris06ApJ} at $z=0.55$.  We must await deeper submm studies to see
if corrections that large are reasonable for SN~Ia hosts out to redshift $z
= 0.6$.  Indications from \citet{Clements05MNRAS} are that this does not
describe SN~Ia hosts out to redshift $z = 0.5$.

We maintain that contamination is the largest source of systematic error
in SN~Ia rates beyond $z = 0.5$.  \citet{Barris06ApJ} used LCs generated from
relatively sparsely sampled ($\Delta t =$ 2-3 weeks) {\it RIZ} filter
photometry \citep{Barris04ApJa} combined with a training set of 23
spectrally identified SN~Ia to verify their SN~Ia sample.
\citet{Dahlen04ApJ} used low-resolution grism spectroscopy in combination
with photometric methods to identify the majority of their candidate
SNe~Ia.  \citet{Strolger04ApJ} state that luminous SNe Ib/c can occupy
nearly the same magnitude-color space as SNe~Ia.  \citet{Johnson06} also
conclude that SNe Ib/c are the biggest challenge in phototyping SNe~Ia.
\citet{Strolger04ApJ} point out that the bright SNe Ib/c make up only $\sim
20\%$ of all SNe Ib/c and that SNe Ib/c make up only $\sim 30\%$ of all
core-collapse (CC) SNe according to \citet{Cappellaro99A&A}.  One obvious
caution is the fact that these ratios are based on a small sample ($< 15$)
from the local universe.  Star-formation increases with redshift and it is
plausible that the relative frequency of SNe~Ib/c may increase as well.

We also assert that CC~SNe at lower redshifts can masquerade as SNe~Ia
at higher redshifts.  We support this assertion by pointing out
Figure~9 in \citet{Sullivan06AJ}.  This figure plots photometrically
determined redshifts (using a SN~Ia template) against
spectroscopically determined redshifts for SNe~Ia and several types of
CC~SNe.  The redshifts of the CC~SNe are systematically over-estimated
by as much as $\Delta z = +0.5$.  Contamination of this kind is not
addressed by using the typical magnitude difference between CC~SNe and
SNe~Ia to cull out CC~SNe \citep{Richardson02AJ}, since the CC~SNe
appear at the wrong redshift.  This problem is strongly mitigated when
host redshifts are available to cross-check SN photo-redshifts, as
long as the host redshifts are reasonably accurate.

Another problem for photometric typing is reddening.  As mentioned above,
SNe~Ia are distinguished from CC~SNe in the early part of their LCs by
having redder colors.  Thus, a highly reddened CC~SNe can appear to be a
SN~Ia and weeding these objects out of a SN~Ia sample requires good epoch
coverage of the later epochs.  The SNLS has the benefit of 4 filter
photometry which helps distinguish even highly reddened CC~SNe early on.
Even with this advantage, $\sim 10$\% of our candidates promoted for
spectroscopic followup turn out to be an identifiable type of CC~SN
\citep{Howell05ApJ}.

The diversity of CC~SNe, as compared to SNe~Ia, is another challenge for
photometric identification of SNe.  Neither \citet{Dahlen04ApJ} nor
\citet{Barris06ApJ} include a large database of spectrally identified
CC~SNe light curves in their training or test data sets.  Until photometric
methods can prove themselves convincingly against the full diversity of
CC~SNe, spectroscopy is the most reliable way to identify SNe.  The pay-off
for developing an accurate classifier based only on photometry is huge,
however, given the expense of obtaining spectroscopy of SNe at high
redshifts.  \citet{Johnson06} point out that having good spectral-energy
distribution coverage and having dense time-sampling will improve the
accuracy of this method, a statement that agrees with our experience with
the SNLS \citep{Sullivan06AJ,Guy05A&A}.

We have emphasized the importance of verifying the majority of the
SN~Ia sample with spectroscopy in all of our figures comparing rates
and SFH by plotting all the rates from samples that fulfill this
criterion as filled symbols.  Assuming these surveys have carefully
characterized their spectroscopic completeness, the trend they display
is the one that should be compared to SN~Ia production models.

If we look in detail at the \citet{Dahlen04ApJ} sample, we point out
that the two redshift bins in which the majority of objects are
spectroscopically confirmed \citep[the `Gold' objects,
see][]{Strolger04ApJ} at $z=0.5$ (2 out of 3) and $z=1.2$ (5 out of 6)
are included in our comparison of SFH and SN~Ia rate evolution (see
Figure~\ref{fig_lit}).  The other two bins have only 50\% of their
sample in the `Gold' category: 7 out of 14 at $z=0.8$, and 1 out of 2
at $z = 1.6$, and are therefore not included in this comparison.

\section{Discussion\label{sec_disc}}

Our rest-frame SN~Ia rate per unit comoving volume in the redshift range $
0.2 < z < 0.6$ ($\langle z \rangle_V \simeq 0.47$) and using the
cosmological parameters $\Omega_m = 0.3$, $\Omega_{\Lambda} = 0.7$ and
$H_0 = 70$ km s$^{-1}$ is \begin{equation}
r_V(\langle z \rangle_V = 0.47) = 0.42^{+0.13}_{-0.09}(syst) \pm 0.06(stat)
\times 10^{-4} yr^{-1} Mpc^{-3},\end{equation}
and we also report our SN~Ia rate in SNu, for comparison with previously
determined rates at lower redshift: \begin{equation} r_L(\langle z
\rangle_V = 0.47) = 0.154^{+0.048}_{-0.033}(syst) ^{+0.039}_{-0.031}(stat)
SNu.\end{equation}

\subsection{Comparison with Star Formation History\label{sec_sfh}}

The place to begin investigating the relationship between SFH and SN~Ia
production is where the systematic errors are minimized.  Volumetric rates
are the most appropriate to use in exploring this relationship.  The SNu,
defined using a blue host luminosity, is less ideal especially for SNe~Ia
for a number of reasons.  Galaxy luminosity evolution makes interpreting
trends in SNu with redshift difficult.  Also, using a blue luminosity is
not good for SNe~Ia, since they also appear in galaxies with older and
redder populations than CC~SNe.

The region of minimal systematic uncertainty in the volumetric rate
evolution is delineated by the trade-off between survey sensitivity and
volume sampling.  At low redshifts most searches are galaxy-targeted,
requiring the conversion of the luminosity specific rate (SNu) to a
volumetric rate through the local galaxy luminosity function.  The volume
sampled is low, increasing the influence of cosmic variance on the derived
volumetric rate and hence, increasing the systematic errors.  At high
redshifts, survey sensitivity dominates the systematics since high redshift
SNe are close to the detection limits, must be spectrally confirmed with
lower S/N spectra, or photometrically identified with lower S/N LCs, and
have projected distances that are fewer pixels from host nuclei.

For the SNLS, the redshift range $0.2 < z < 0.6$ is low enough to reduce
systematic errors associated with detection limits, spectral confirmation,
and host offset.  It also samples 4 large, independent volumes of the
universe minimizing the effects of cosmic variance on the errors.  In our
subsequent analysis, we examine two currently popular models of SN~Ia
production and compare them to our SN~Ia rate at $\langle z \rangle_V =
0.467$, and to the other spectroscopically confirmed rates from the
literature. 

\subsubsection{The Two-Component Model\label{sec_sfh2}}

This recent model, first put forth by \citet{Mannucci05A&A} and
applied to a sample of rates from the literature by
\citet{Scannapieco05ApJ}, proposes a delay function with two
components.  One component, called the prompt component, tracks SFH
with a fairly short delay time ($<~1$~Gyr).  The other component,
called the extended component, is proportional to total stellar mass
and has a much longer delay time.  This model arose as a way to
account for the high SN~Ia rate in actively star-forming galaxies,
relative to less active galaxies, and yet still produce the non-zero
SN~Ia rate in galaxies with no active star formation
\citep{Oemler79AJ, vandenBergh90PASP, Cappellaro99A&A, Mannucci05A&A,
Sullivan06b}.  \citet{Gal-Yam04MNRAS} and \citet{Maoz04MNRAS} observed
the SN Ia rates in galaxy clusters and indicated the possibility that
a long delay-time may be inconsistent with their observations, given
the cluster Fe abundances.  \citet{Scannapieco05ApJ} demonstrate that
the Fe content of the gas in galaxy clusters can be explained by the
prompt component of the two-component model.  They also demonstrate
that the two-component model reproduces the observed stellar [O/Fe]
abundance ratios within the Galaxy \citep[Figure 3]{Scannapieco05ApJ}.

The two-component model is described by equation 1 from
\citet{Scannapieco05ApJ}, which gives the relationship between the cosmic
star formation rate, the cosmic stellar mass and the volumetric
SN~Ia rate as a function of time.  We re-write their equation in a more
general form here: \begin{equation} r_V(t) = A M_*(t) + B \dot{M}_*(t),
\label{eq_ab}\end{equation} which gives $A$ in terms of SNe~Ia per year per
unit mass and B in terms of SNe~Ia per year per unit star formation.  The
component scaled by $A$ is the extended component, while $B$ scales the
prompt component.

By comparing this model to our rate, and other spectroscopically confirmed
rates from the literature, we can estimate the relative contributions of
the extended and the prompt components.  Figure~\ref{fig_lit} already
places an upper limit of $B \lesssim 10^{-3}\ yr^{-1} (M_{\odot}\
yr^{-1})^{-1}$ because the SFH curve normalized to go through our rate
represents a pure prompt component model ($A = 0.0$).

Figure~\ref{fig_sfh2} shows a non-linear least squares fit of the
spectroscopically confirmed SN~Ia rates from this study (filled
square) and the literature (filled circles) to the rates predicted by
the two-component model using the SFH from \citet{Hopkins06}.  This
fit produces a reduced $\chi^2$ goodness-of-fit statistic of
$\chi^2_{\nu} = 0.361$ using the published error bars, produced by
adding the statistical and systematic errors in quadrature.  With 7
degrees of freedom (9 data points minus two parameters), this
corresponds to a probability of 93\% that the null hypothesis is
correct, i.e., that the data represent a random sampling from the
parent distribution described by the fit.  The resulting fit
parameters are $A = 1.4\pm1.0 \times 10^{-14}\ yr^{-1} M_{\odot}^{-1}$
and $B = 8.0\pm2.6 \times10^{-4}\ yr^{-1} (M_{\odot}\ yr^{-1})^{-1}$
(see Figure~\ref{fig_sfh2}).

This is the first time this model has been fit directly to volumetric
rate data from the literature.  \citet{Scannapieco05ApJ} normalized
each component separately (using a different SFH) and then compared
the resulting rate evolution to the observed rates.  For the extended
($A$) component, they used the rate per unit mass for E/S0 galaxies
from \citet[Table 2]{Mannucci05A&A} which gives a value of $A =
3.8^{+1.4}_{-1.2} \times 10^{-14}\ yr^{-1} M_{\odot}^{-1}$, corrected
to our cosmology.  An alternative value for the $A$ component can be
derived from Table 3 in \citet{Mannucci05A&A} which gives the rate in
bins of $B-K$ color, independent of morphology.  The reddest bin,
having $B-K > 4.1$ gives $A = 2.4^{+1.5}_{-1.1} \times 10^{-14}\
yr^{-1} M_{\odot}^{-1}$ (again adjusted for our cosmology), which is
more consistent with the value from our fit.  A possibly more
important difference stems from the definition of mass.  Our mass is
derived from the integration of the SFH from high redshift to the
epoch in question and, therefore, includes the mass from stars that
have died.  The mass used in \citet{Mannucci05A&A} was derived from
the $K$-band luminosity of individual galaxies and is more
representative of the mass currently in stars.  Since our method tends
to over-estimate the mass, our $A$ value is correspondingly lower.

For the prompt ($B$) component, \citet{Scannapieco05ApJ} first normalize
the CC rate to the SFH and then use an assumed CC/SN~Ia ratio, a method
that they admit is highly uncertain.  This produces a prompt component with
$B = 23\pm10 \times10^{-4}\ yr^{-1} (M_{\odot}\ yr^{-1})^{-1}$, which is
$1.45\sigma$ higher than our value.  They also mention an alternate
normalization using the SN~Ia rate in actively star-forming galaxies as
indicated by having $B-K \leq 2.6$.  This method produces the value of $B =
10^{+6}_{-5}\times10^{-4}\ yr^{-1} (M_{\odot}\ yr^{-1})^{-1}$, which is
consistent with our value.

We are comparing a method that normalizes the $A$ and $B$ components
separately \citep{Scannapieco05ApJ}, with a method which directly ties the
component values to the rate evolution, as delineated by the rates derived
from spectrally confirmed samples.  Given the differences in SFH, mass
definition, and method, the level of agreement is encouraging for this
model.  Caveats remain, however, including systematics in the lower
redshift rates due to cosmic variance and systematics in the SFH.  Further
tests of this model will come as the SN~Ia rate evolution is more precisely
measured (at low and high redshifts) and as the SN~Ia rate per unit mass
and per unit star-formation is measured more accurately for a larger set of
individual galaxies \citep{Sullivan06b}.

The success of this model implies that it is reasonable to describe
SN~Ia production in terms of two populations with two different delay
times \citep[see also][]{Mannucci06}.  If these two populations
represent two separate channels for SN~Ia production, they may also
exhibit different intrinsic properties.  The SNLS is carefully
examining this \citep{Sullivan06b} to avoid biases in our cosmological
parameters.  At higher redshifts the component tied to SFH will tend
to dominate, while lower redshift samples will contain more of the
extended component SNe~Ia.  Cosmological parameters determined with
SNe~Ia spanning a large range of redshifts may be subject to
systematics, if unaccounted-for intrinsic differences exist.

\subsubsection{Gaussian Delay Time Model}

Figure~\ref{fig_sfhdt} shows a comparison of observed rates with the delay
time model delineated in \citet{Strolger04ApJ}, which is plotted as a
dashed line.  This model convolves the SFH with a Gaussian delay-time
distribution with a characteristic delay time, $\tau$, and a width that is
some fraction of the delay time: $\sigma = 0.2\tau$ in this case.  We
updated the SFH model, using the fit from \citet{Hopkins06}.  We find that
the delay time model still fits the \citet{Dahlen04ApJ} data with a
delay time of $\tau = 3$ Gyr, which is statistically consistent with the
most likely value from \citet{Strolger04ApJ}.  In contrast,
\citet{Mannucci06} used the data from \citet{Dahlen04ApJ} and combined with
host galaxy colors and host radio properties found a bi-modal delay
distribution to be more consistent.  We find that the single Gaussian model
fit to the \citet{Dahlen04ApJ} data consistently over-predicts the rates
near and below $z = 0.5$.  In particular, the rate from \citet{Barris06ApJ}
at $z = 0.25$ is more than $1\sigma$ below this model.

We also show a Gaussian delay time model normalized to our rate in
Figure~\ref{fig_sfhdt} as the solid line.  It is statistically
consistent with the spectrally confirmed SN~Ia rates, except the
highest one at $z=1.2$.  It has the following parameters: $\tau = 4.0$
Gyr and $\sigma = 0.7\tau$.  This model predicts a very low SN~Ia rate
at higher redshifts in contrast to the photometrically typed rates
near a redshift of $z \sim 0.7$.

While the observations can be fit with this model, the favoured delay
times tend to be longer than 3 Gyr.  This is inconsistent with the
finding that the SN~Ia rate is much higher in galaxies with recent
star formation \citep{Oemler79AJ,vandenBergh90PASP,Cappellaro99A&A,
Mannucci05A&A,Sullivan06b}.  Although we use a different set of
observed rates, \citet{Mannucci06} find that a single Gaussian delay
time does not fit observed rates as a function of redshift, host color
and radio loudness as well as a bi-modal delay-time distribution.  If
there is a strong correlation in the SN~Ia rate with host galaxy star
formation rate, then the overall SN~Ia rate evolution should track the
SFH reasonably closely, especially at higher redshifts.

The real test of the Gaussian delay model will come with rates beyond
$z = 1.4$ where the predicted down-turn compared to SFH will become
pronounced.  Our current best estimate in this range is from
\citet{Dahlen04ApJ} and is based on a sample of two SNe, only one of
which was a member of the `Gold' set from \citet{Strolger04ApJ}.  A
larger, spectroscopically confirmed sample from a much deeper survey
is needed to confirm or refute this result from \citet{Dahlen04ApJ} and
therby either support or discredit the Gaussian delay-time model.

This model faces another challenge.  If there is a large component of
the SN~Ia rate that is closely tied to the SFH, then at higher
redshifts the majority of SNe~Ia will arise closer to star-forming
regions in their hosts.  This implies that the effect of host
extinction (dust) on the sensitivity of SN~Ia surveys will grow with
redshift.  It must be shown that the downturn in the SN~Ia rate
measured by \citet{Dahlen04ApJ} and predicted by this model is not the
result of these factors.  To do this requires the use of updated host
extinction models such as RP05 combined with dust evolution models,
derived from the deepest IR and submm surveys.

\section{Summary}

We have produced the most accurate SN~Ia rate to date by using a
spectroscopically confirmed sample and detection efficiencies derived from
a well characterized survey.  We investigated known sources of systematic
errors using recent models of host extinction from RP05, and fake SN
experiments to test host contamination losses.  Our derived volumetric rate
from a culled sample of 58 SNe~Ia in the redshift range $0.2 < z < 0.6$ is
$r_V = 0.42^{+0.13}_{-0.09}$ (syst) $\pm0.06$ (stat) $\times 10^{-4}$
yr$^{-1}$ Mpc$^{-3}$.  We conclude from our experiments and a comparison of
other rates in the literature, that contamination may be the largest source
of systematic error for rates up to redshift $z = 1$, in particular, for those
rates based on samples that are photometrically typed.

Using the recent SFH fit from \citet{Hopkins06}, we compare our rate with
the two-component model from \citet{Scannapieco05ApJ} and place an upper
limit on the contribution from the component of SN~Ia production that is
closely tied to star formation of $B \lesssim 10^{-3}\ yr^{-1} (M_{\odot}\
yr^{-1})^{-1}$.  By fitting this model to our rate and the spectrally
confirmed rates in the literature we make an estimate of both
components directly and find $A = 1.4\pm0.9 \times 10^{-14}\ yr^{-1}
M_{\odot}^{-1}$ and $B = 8.0\pm2.4 \times10^{-4}\ yr^{-1} (M_{\odot}\
yr^{-1})^{-1}$, with the caveat that our mass definition is an
over-estimate (is the integral of SFH and, therefore, includes dead stars).



\acknowledgments

The authors wish to recognize and acknowledge the very significant
cultural role and reverence that the summit of Mauna Kea has always
had within the indigenous Hawaiian community.  We are grateful for our
opportunity to conduct observations on this mountain.  We acknowledge
invaluable assistance from the CFHT Queued Service Observations team,
led by P. Martin (CFHT).  Our research would not be possible without
the assistance of the support staff at CFHT, especially
J.-C. Cuillandre.  The real-time pipelines for supernovae detection
run on computers integrated in the CFHT computing system, and are very
efficiently installed, maintained and monitored by K.  Withington
(CFHT).  We also heavily rely on the real-time Elixir pipeline which
is operated and monitored by J.-C. Cuillandre, E.  Magnier and K.
Withington.  We are grateful to L. Simard (CADC) for setting up the
image delivery system and his kind and efficient responses to our
suggestions for improvements.  The Canadian collaboration members
acknowledge support from NSERC and CIAR; French collaboration members
from CNRS/IN2P3, CNRS/INSU, PNC and CEA.  This work was supported in
part by the Director, Office of Science, Office of High Energy and
Nuclear Physics, of the US Department of Energy.  The France-Berkeley
Fund provided additional collaboration support.  We are indebted to
A. Hopkins and J. Beacom for providing us with a draft of their work
on SFH prior to its publication.  The views expressed in this article
are those of the author and do not reflect the official policy or
position of the United States Air Force, Department of Defense, or the
U.S. Government.

\clearpage

\begin{figure}
\includegraphics[scale=1.0]{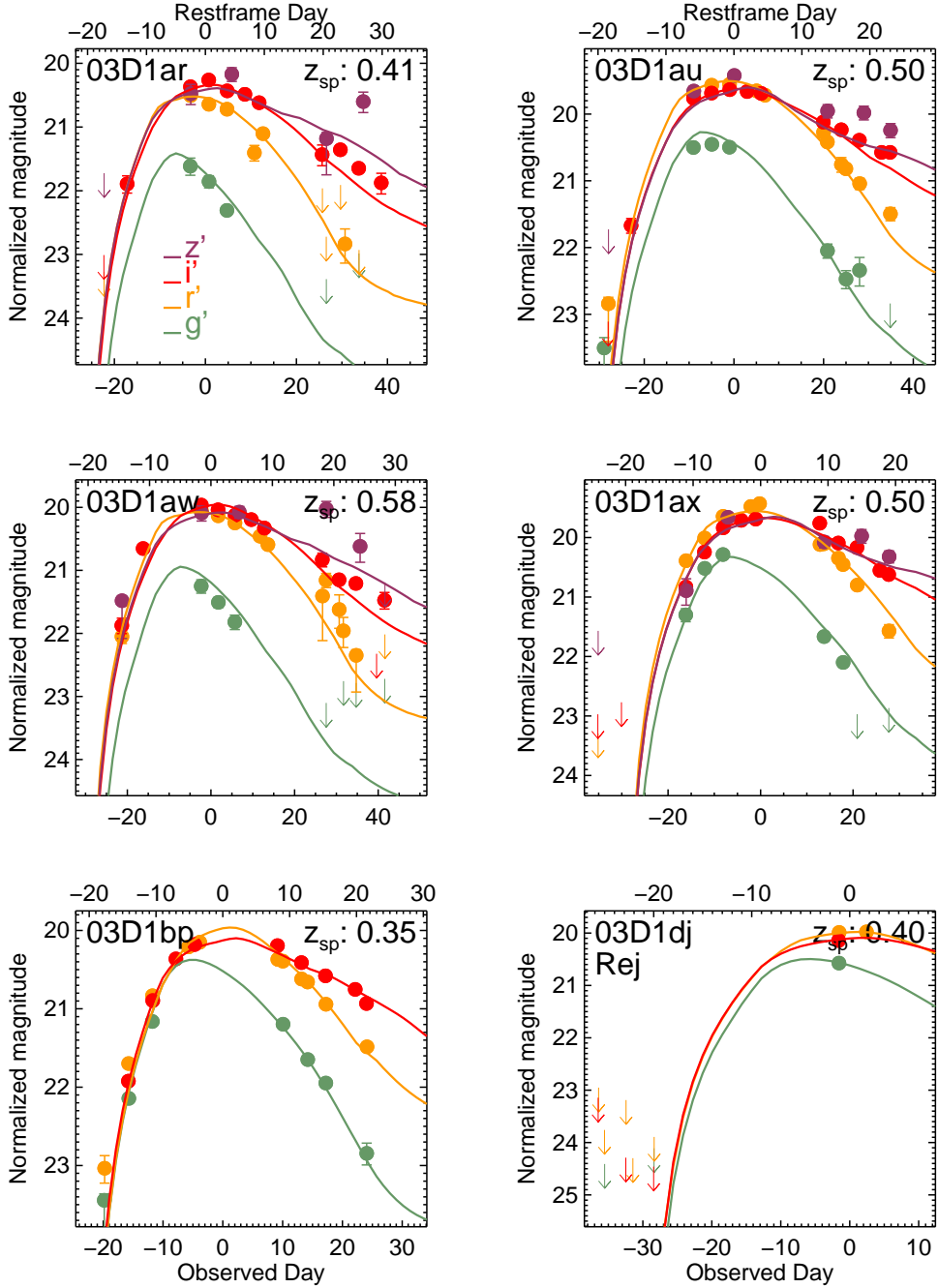}
\caption{Nighly averaged photometry for the spectroscopically confirmed
sample of SNLS SNe~Ia.  Each filled circle represents the nightly average
magnitude for the \gp\ (green), \rp\ (orange), \ip\ (red), and \zp\ (purple)
filters on a normalized AB magnitude scale.  The template fits for each
filter are indicated by the solid lines of the corresponding color.  The
designation from Table~\ref{tab_sne} (minus the SNLS prefix) is indicated in
the upper left corner and the spectroscopic redshift is indicated in the
upper right corner of each panel.  If the SN was culled from the sample,
this is indicated by the word `Rej' under the designation.  The remaining
SNe are presented in the following pages.
\label{fig_snfits}
}
\end{figure}

\begin{figure}
\includegraphics[scale=1.0]{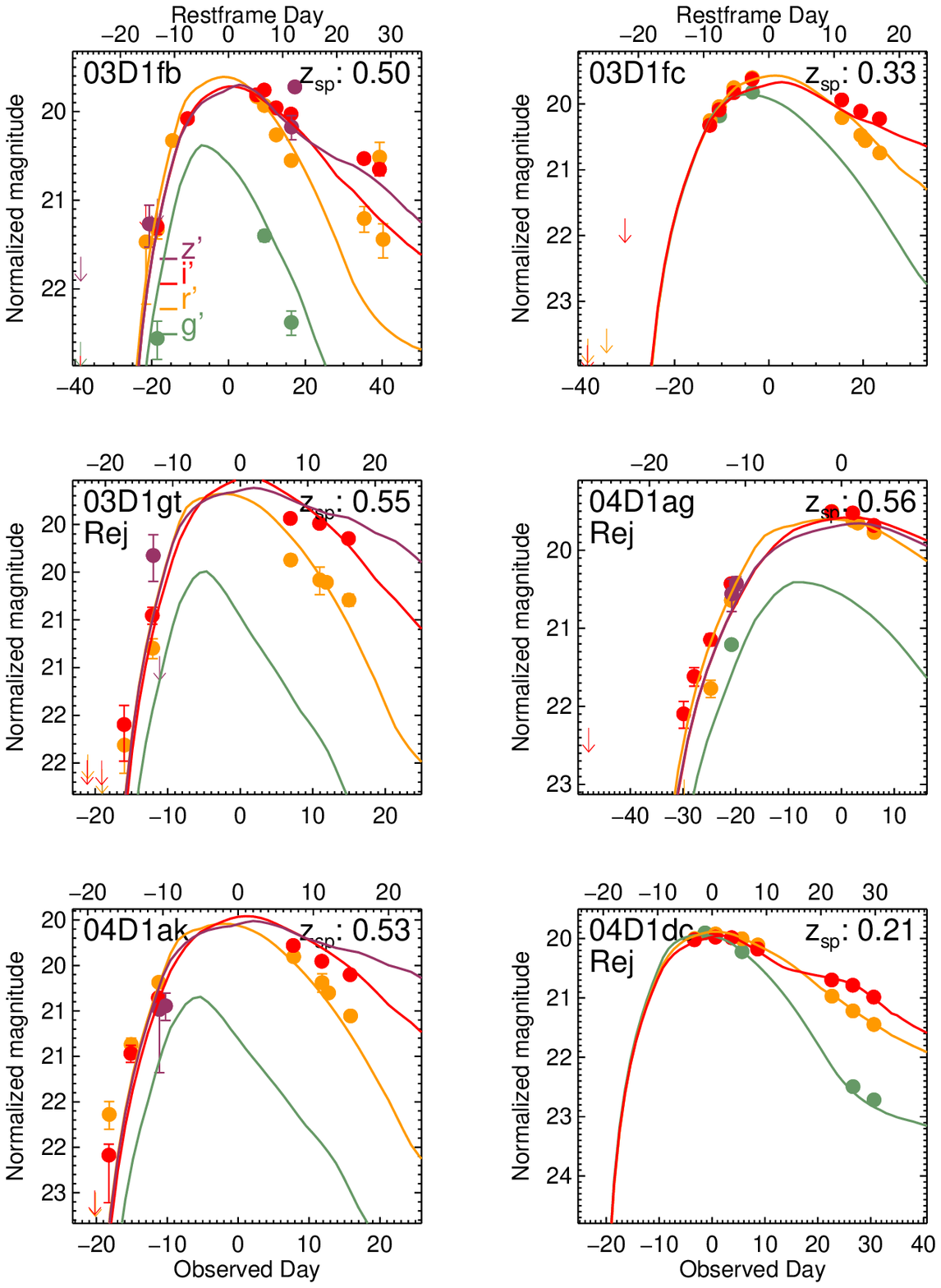}
\figurenum{2}
\caption{continued.}
\end{figure}

\clearpage

\begin{figure}
\includegraphics[scale=1.0]{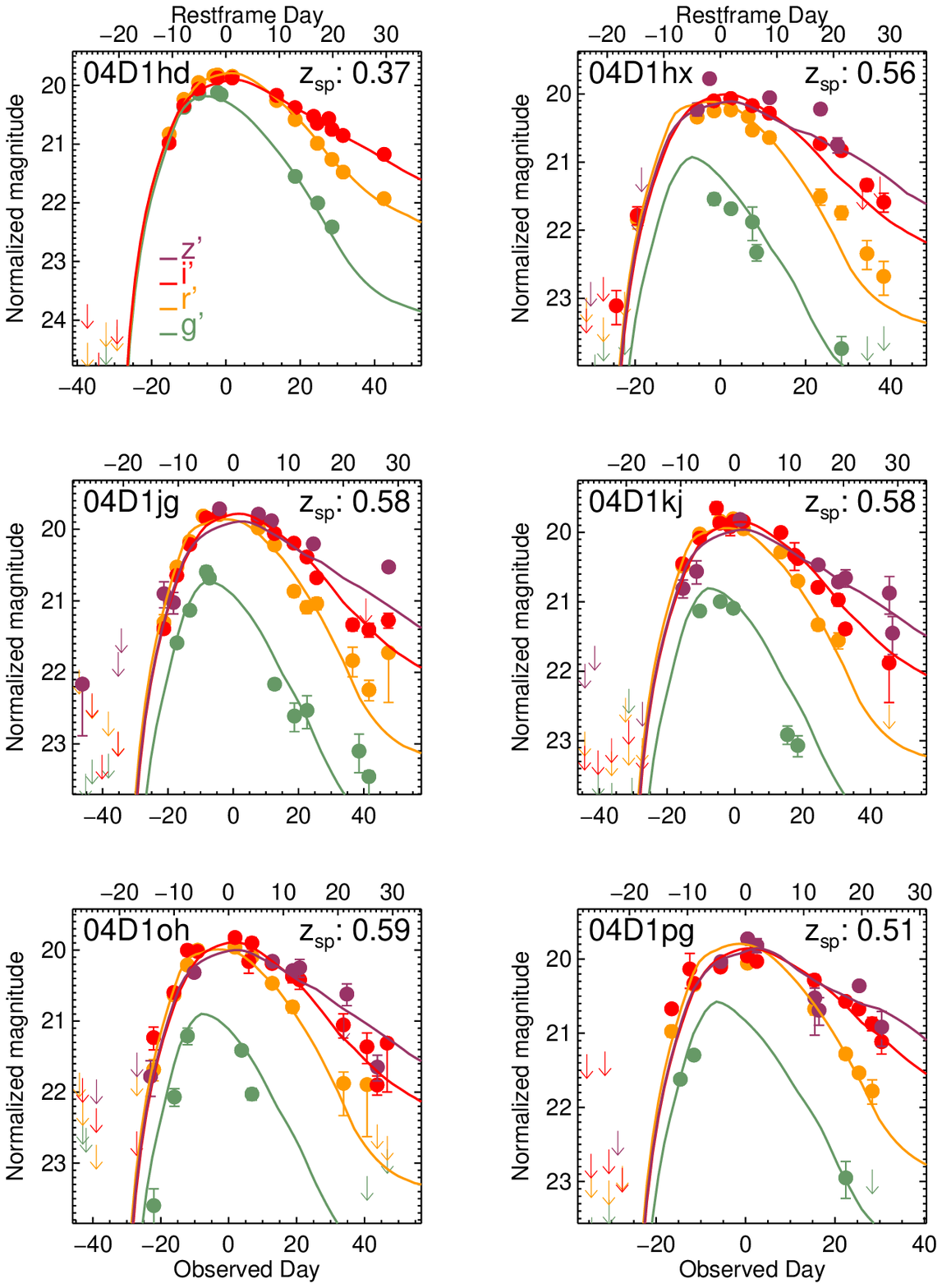}
\figurenum{2}
\caption{continued.}
\end{figure}

\clearpage

\begin{figure}
\includegraphics[scale=1.0]{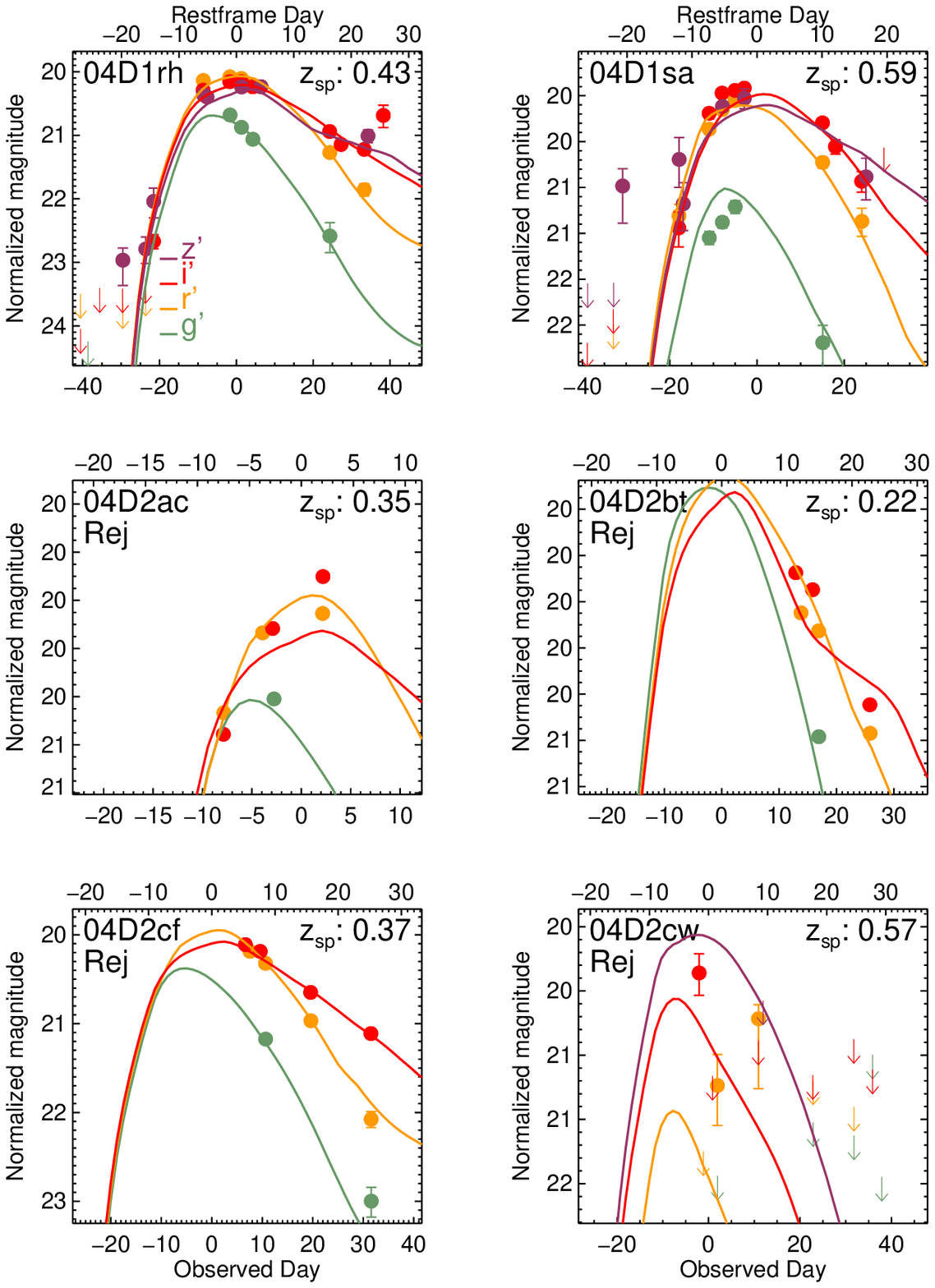}
\figurenum{2}
\caption{continued.}
\end{figure}

\clearpage

\begin{figure}
\includegraphics[scale=1.0]{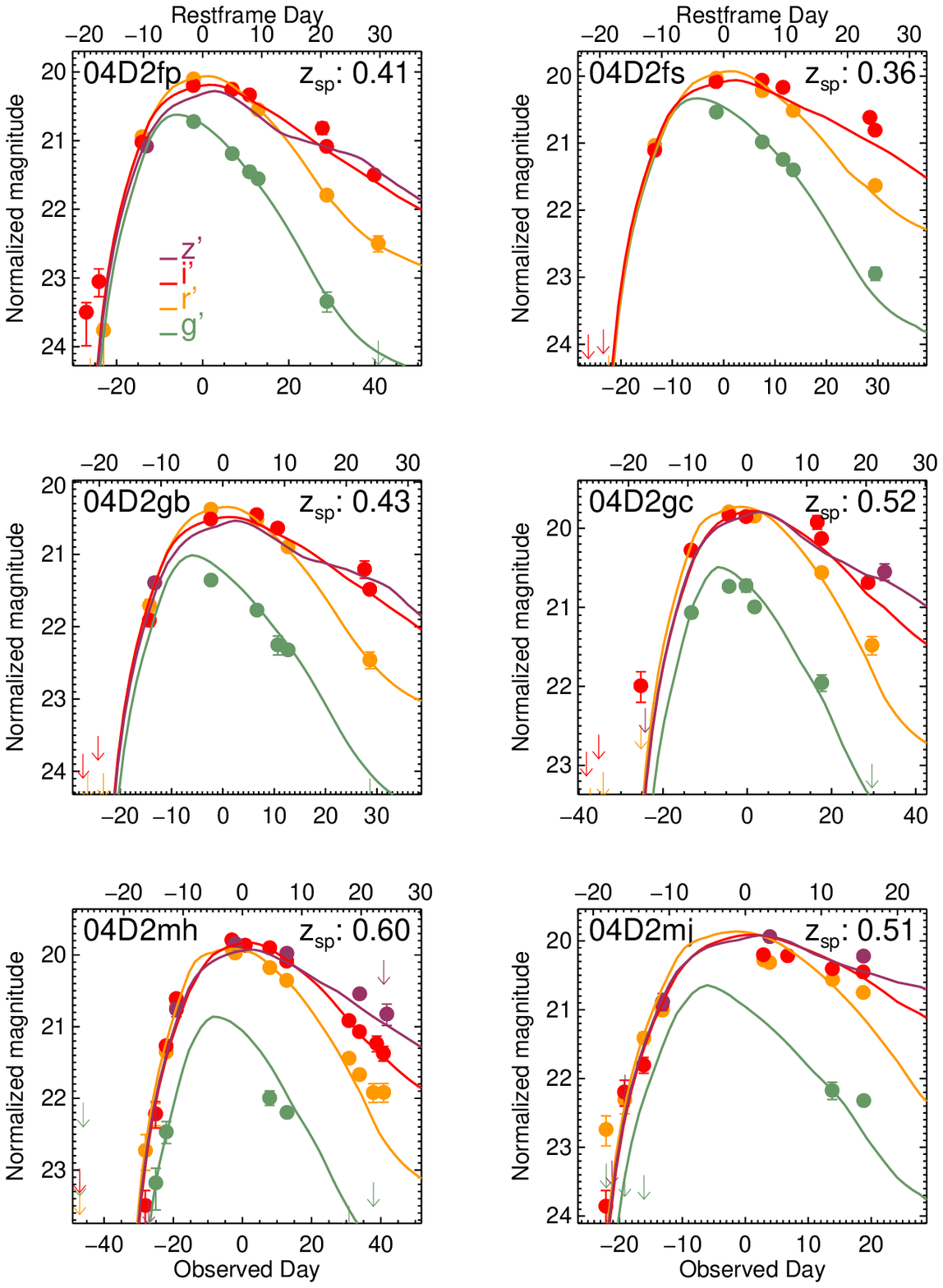}
\figurenum{2}
\caption{continued.}
\end{figure}

\clearpage

\begin{figure}
\includegraphics[scale=1.0]{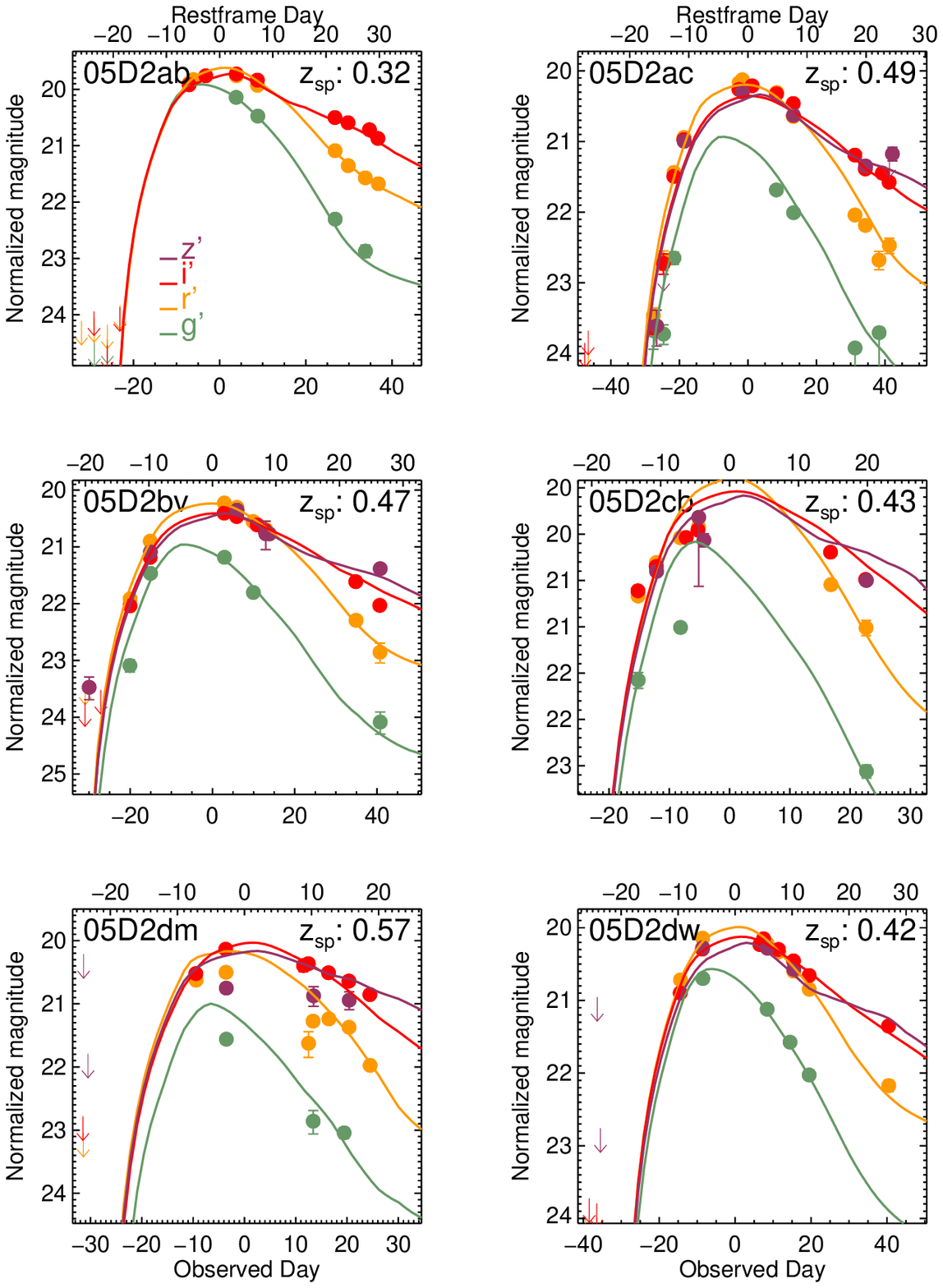}
\figurenum{2}
\caption{continued.}
\end{figure}

\clearpage

\begin{figure}
\includegraphics[scale=1.0]{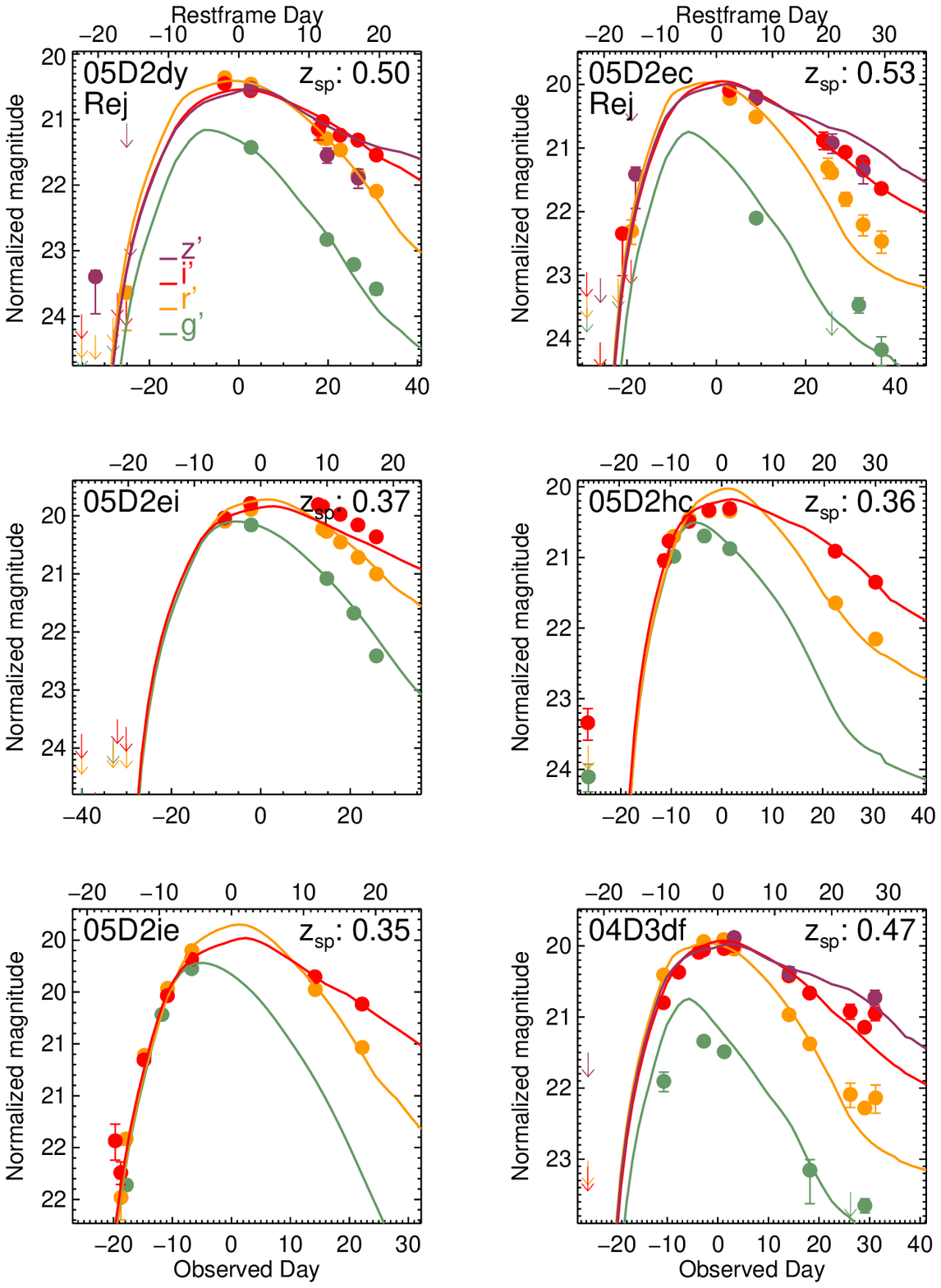}
\figurenum{2}
\caption{continued.}
\end{figure}

\clearpage

\begin{figure}
\includegraphics[scale=1.0]{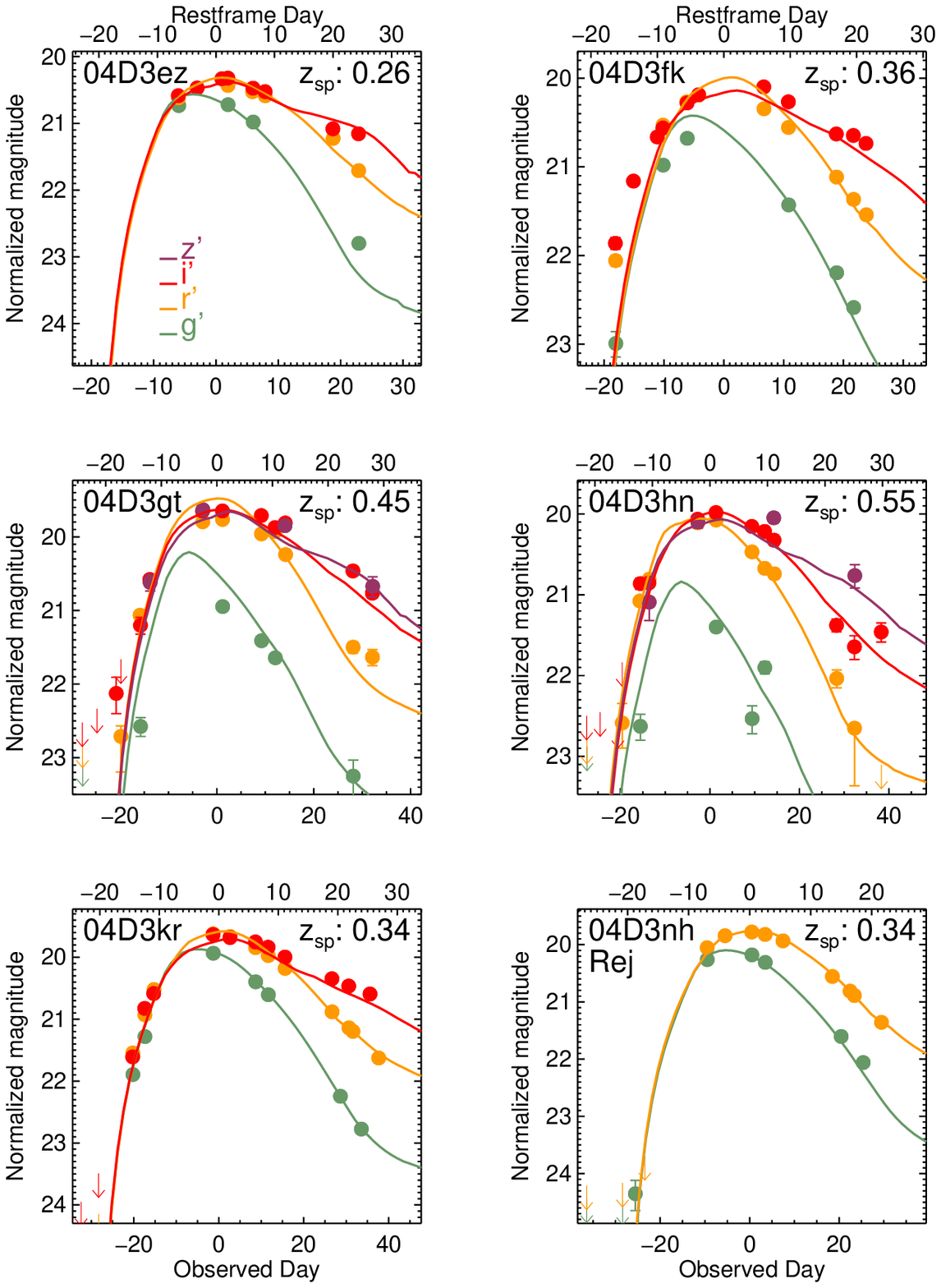}
\figurenum{2}
\caption{continued.}
\end{figure}

\clearpage

\begin{figure}
\includegraphics[scale=1.0]{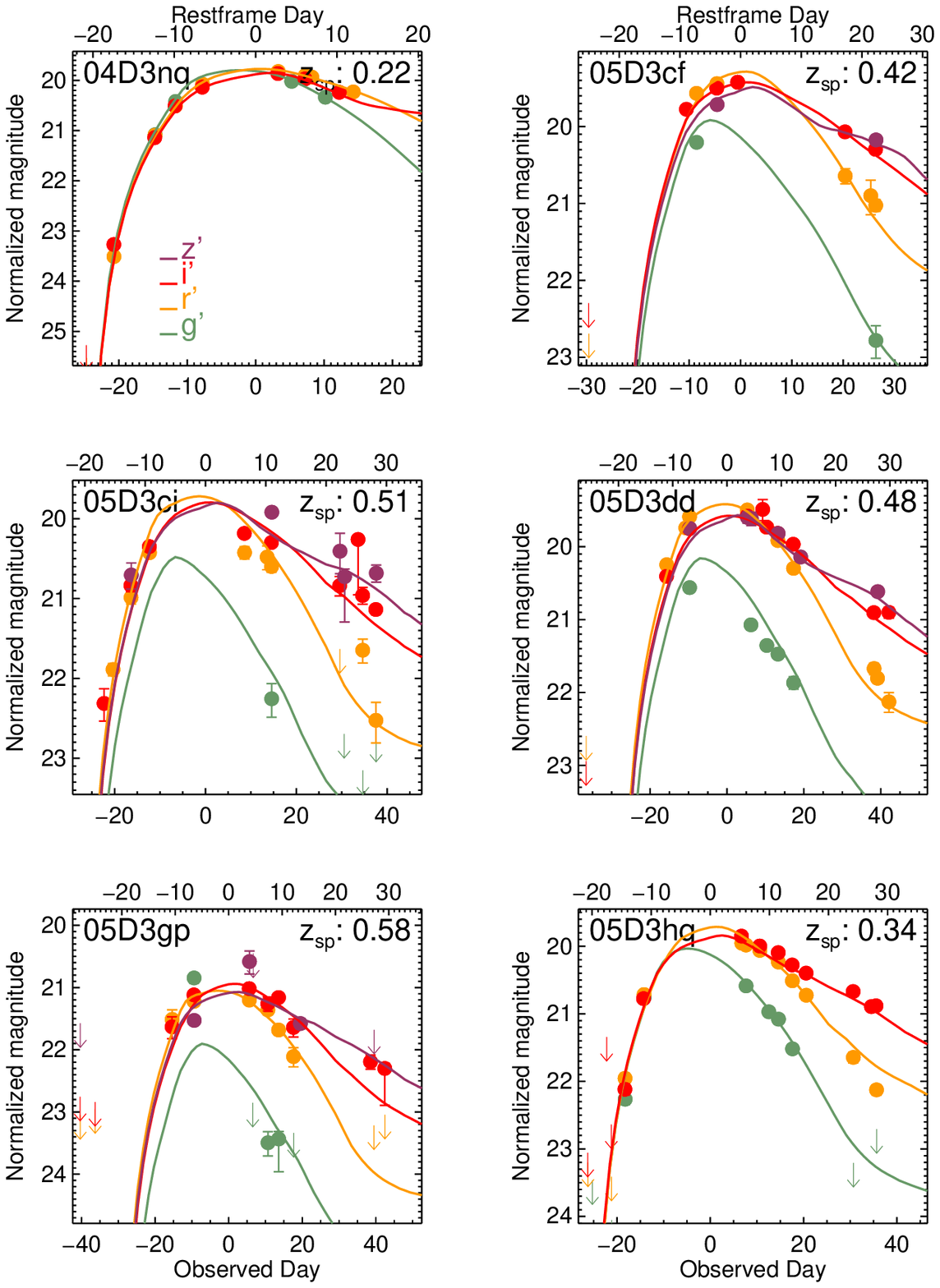}
\figurenum{2}
\caption{continued.}
\end{figure}

\clearpage

\begin{figure}
\includegraphics[scale=1.0]{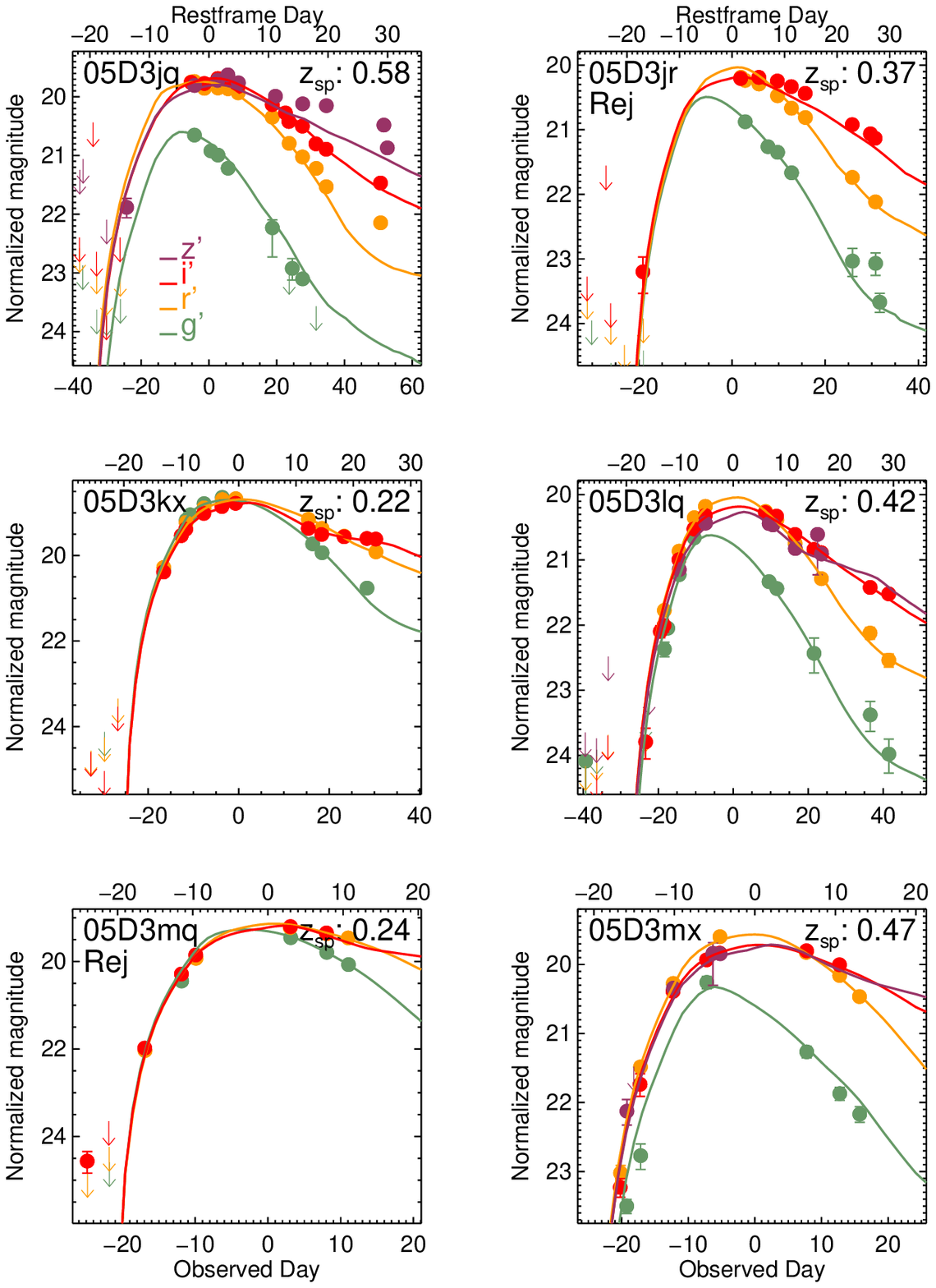}
\figurenum{2}
\caption{continued.}
\end{figure}

\clearpage

\begin{figure}
\includegraphics[scale=1.0]{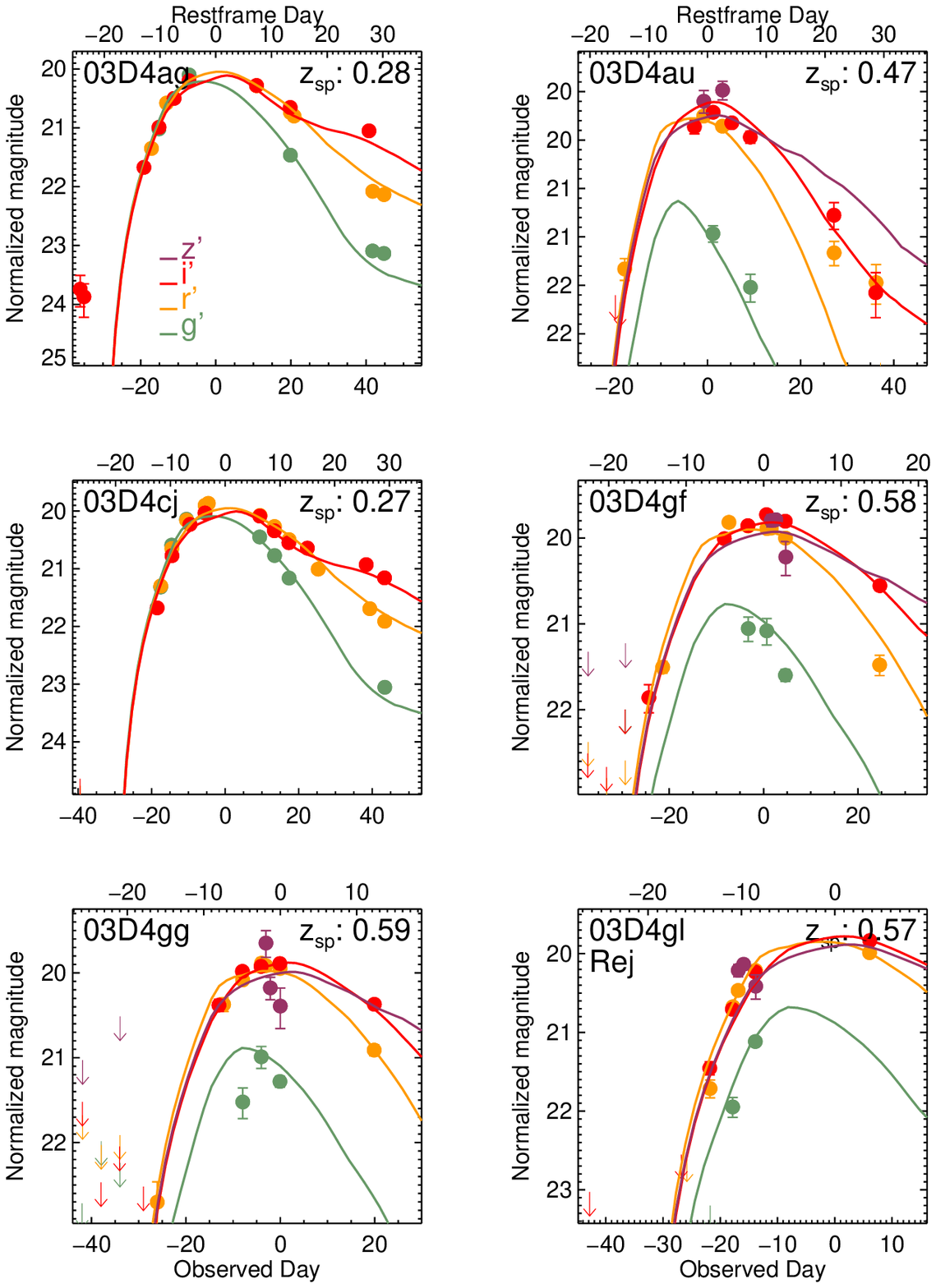}
\figurenum{2}
\caption{continued.}
\end{figure}

\clearpage

\begin{figure}
\includegraphics[scale=1.0]{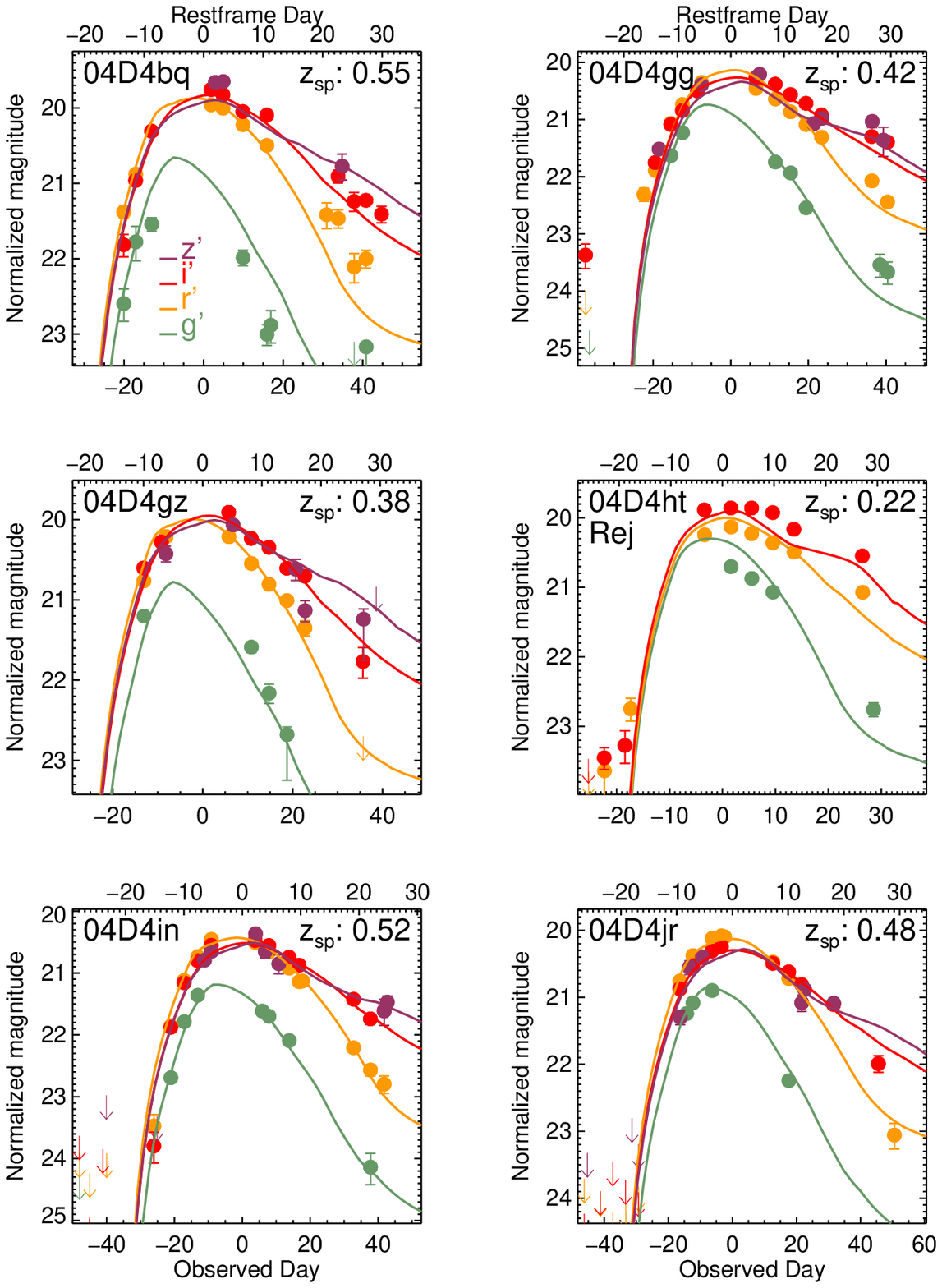}
\figurenum{2}
\caption{continued.}
\end{figure}

\clearpage

\begin{figure}
\includegraphics[scale=1.0]{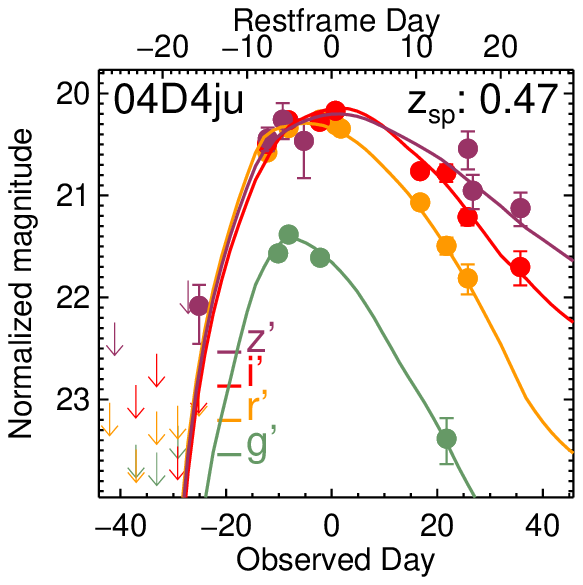}
\figurenum{2}
\caption{continued.}
\end{figure}

\clearpage

\begin{figure}
\includegraphics[angle=90.,scale=0.50]{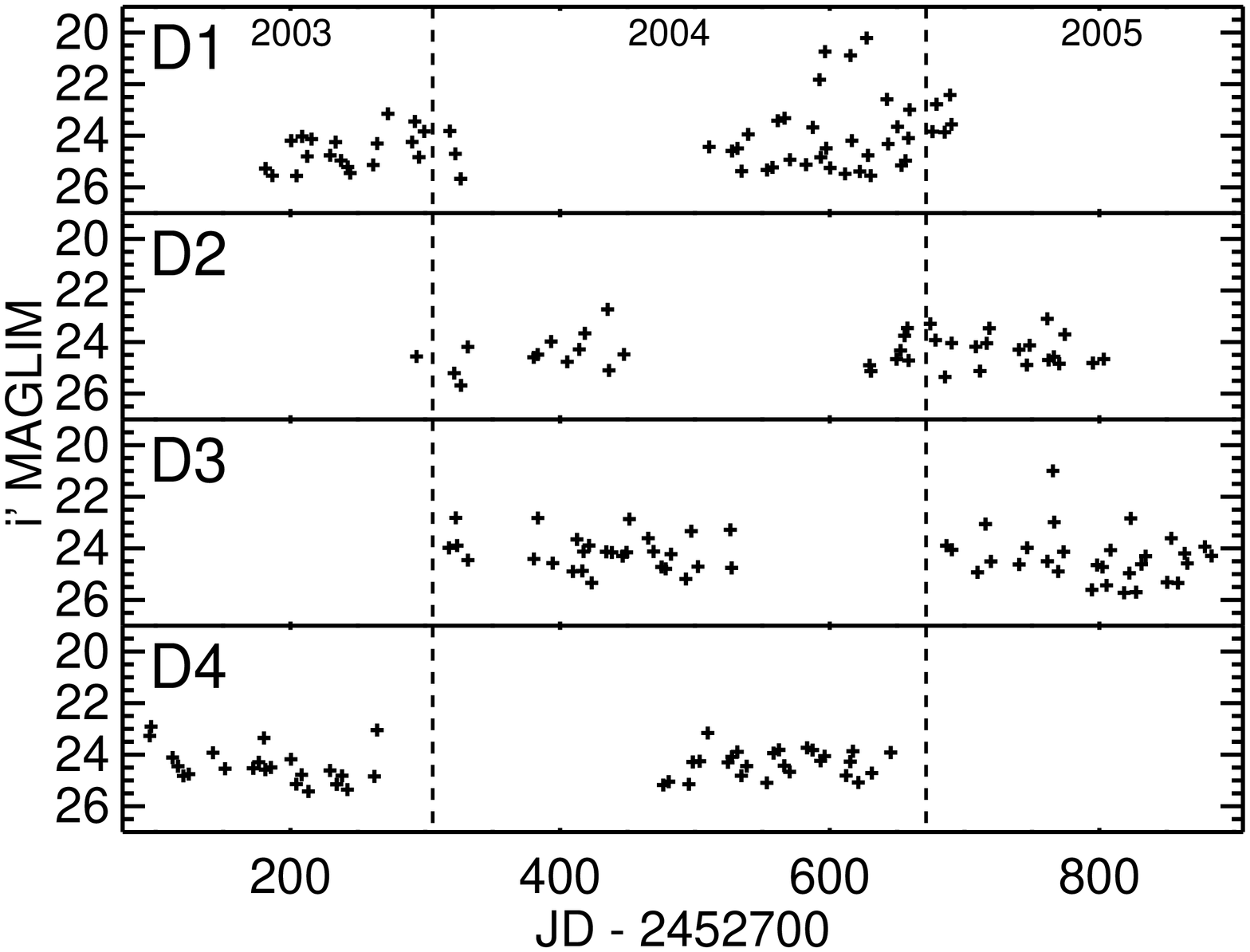}
\caption{Julian Date versus magnitude limit for the \ip\ epochs in each of
the four deep fields used for SN Ia detection in this study.  The calendar 
year transitions are indicated by the vertical dashed lines.  The \ip\
magnitude limits plotted are described in \S\ref{sec_det}.
\label{fig_iobs}
}
\end{figure}

\begin{figure}
\includegraphics[scale=1.0]{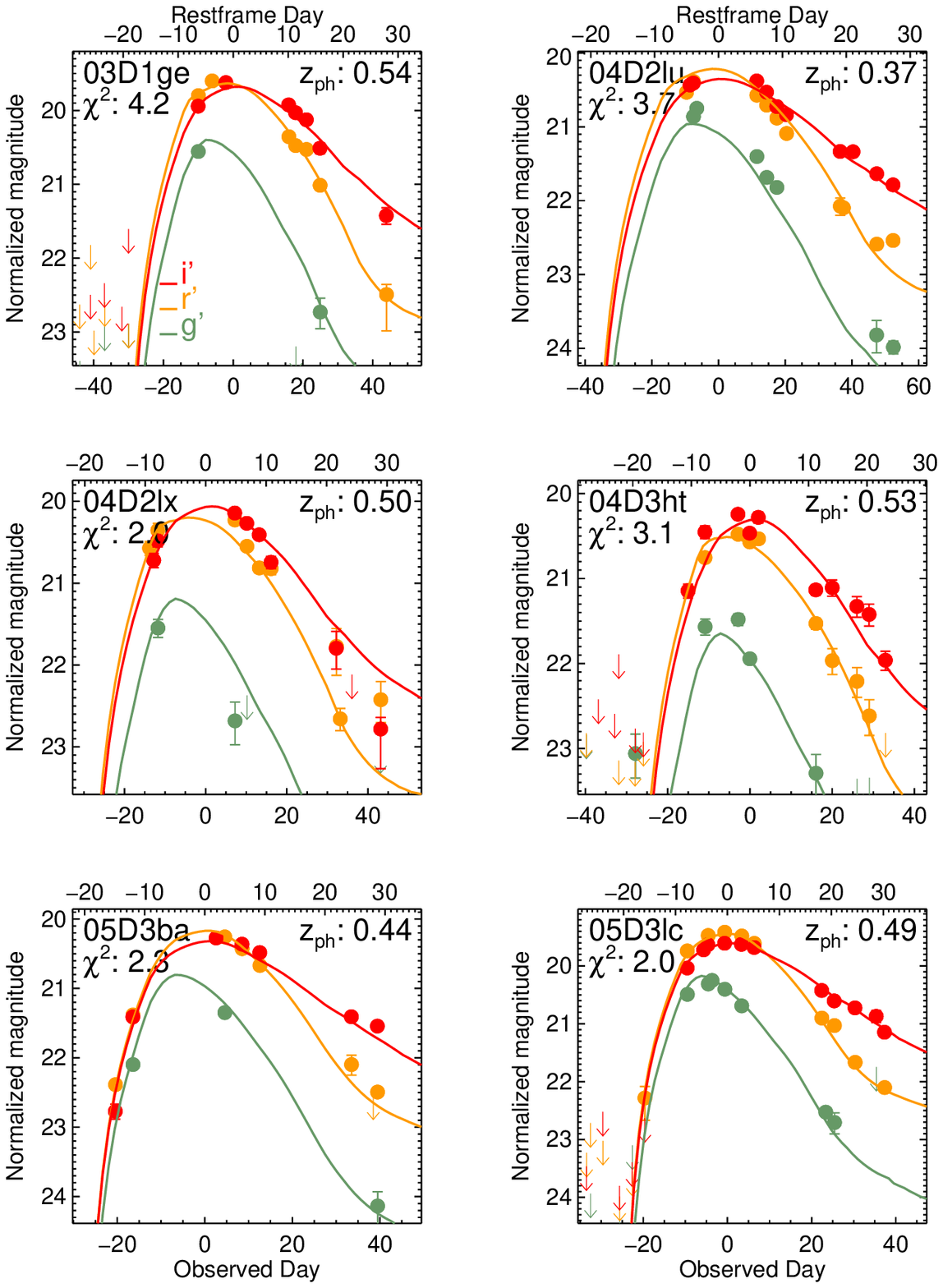}
\caption{Nighly averaged photometry (as in Figure~\ref{fig_snfits}) for
the unconfirmed SNLS SNe~Ia from Table~\ref{tab_cands}.  The designation
is in the upper left corner and the photometric redshift is in the upper
right corner of each panel.  The $\chi_{SNIa}^2$ values from
Table~\ref{tab_cands} are printed under the designation for each SN.
The first six objects are presented here.  The full set is available from
the online version of the article.
\label{fig_snzph}
}
\end{figure}

\begin{figure}
\includegraphics[scale=1.0]{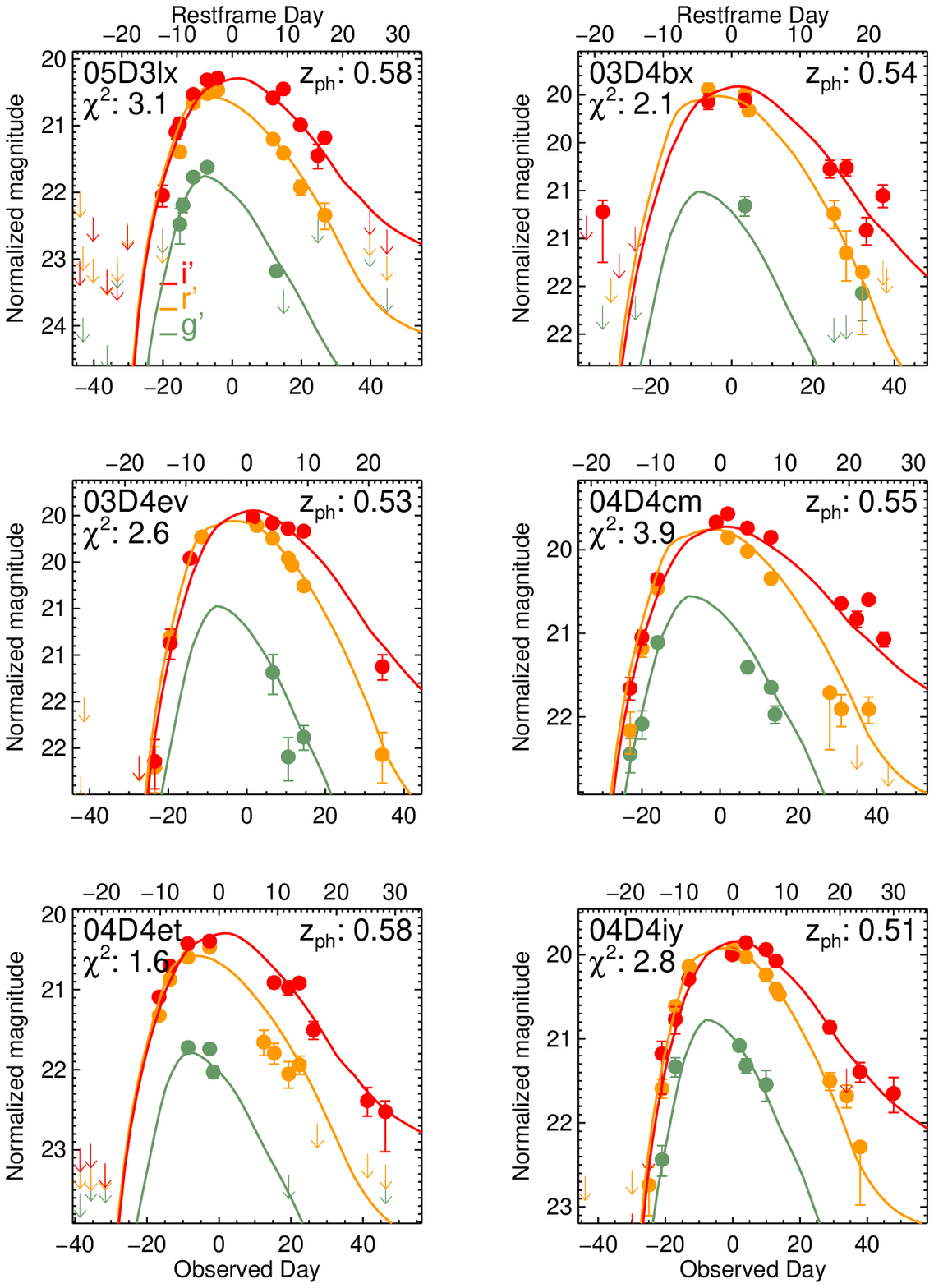}
\figurenum{4}
\caption{continued.}
\end{figure}

\clearpage

\begin{figure}
\includegraphics[scale=1.0]{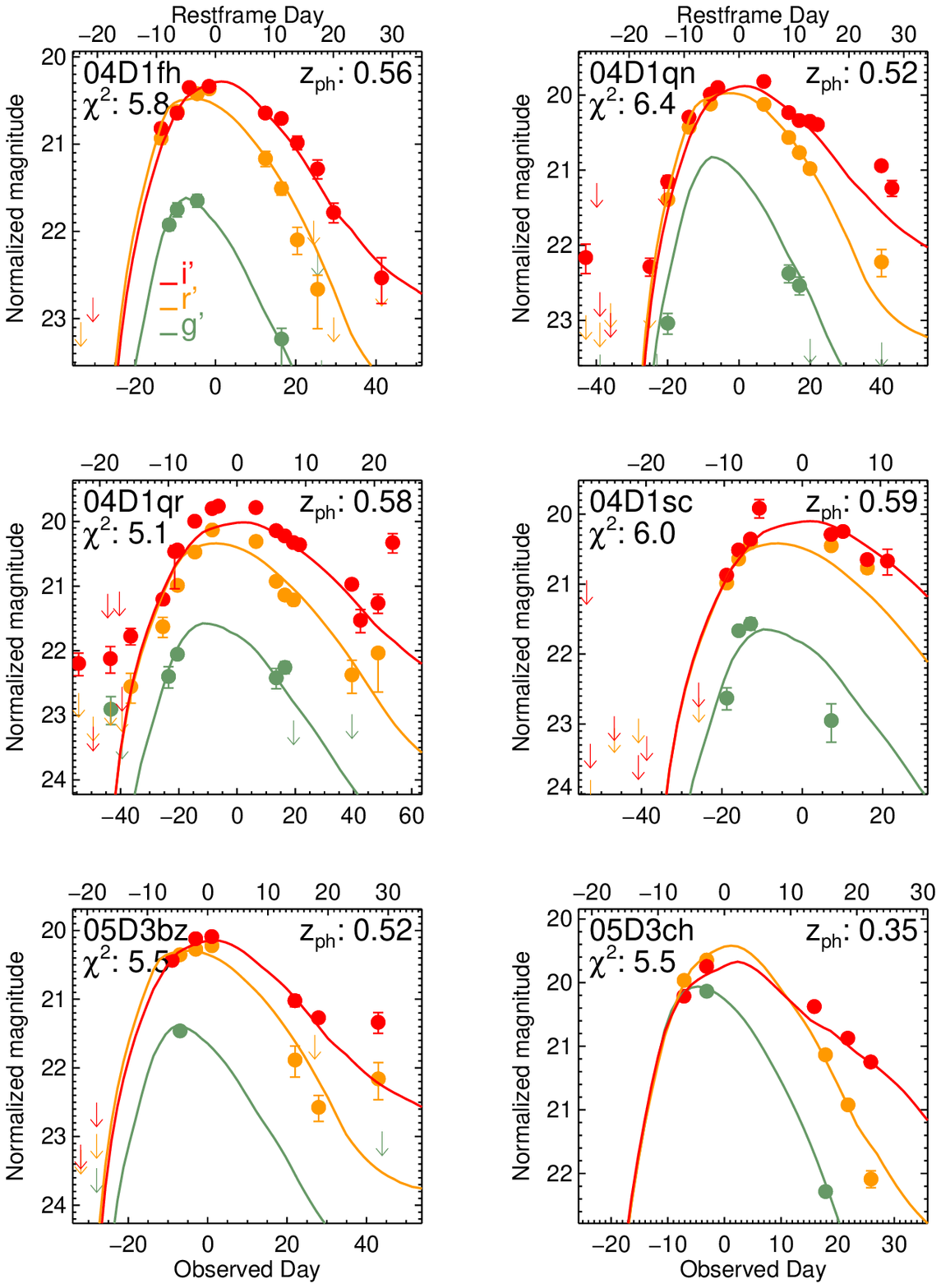}
\figurenum{4}
\caption{continued.}
\end{figure}

\clearpage

\begin{figure}
\includegraphics[scale=1.0]{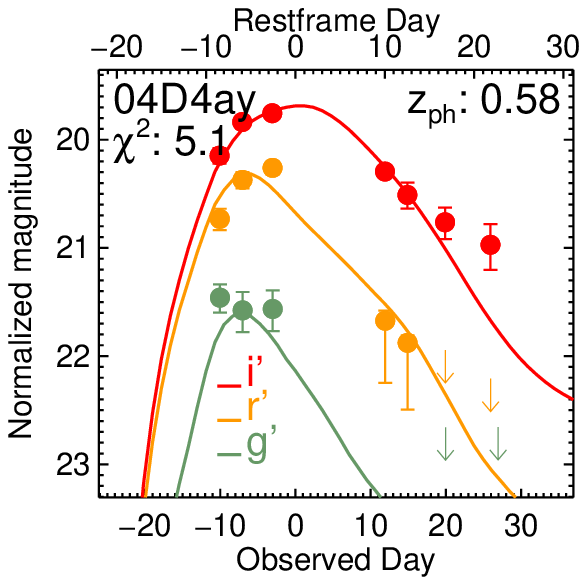}
\figurenum{4}
\caption{continued.}
\end{figure}

\clearpage

\begin{figure}
\includegraphics[angle=90.,scale=0.50]{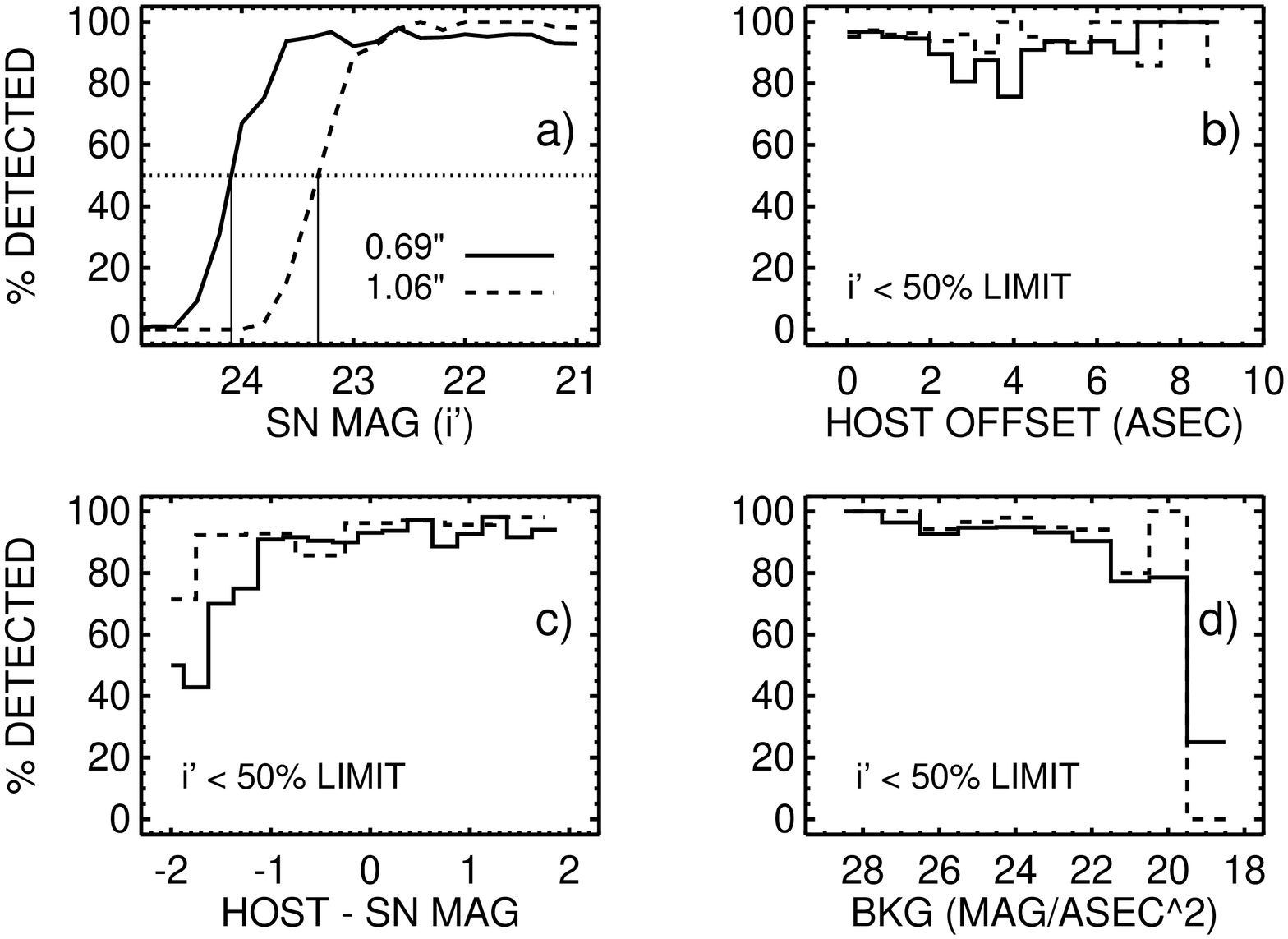}
\caption{Recovery percentage after human review for the two \ip\ images
having $IQ_e = 0\arcpnt69$ (solid line) and $IQ_e = 1\arcpnt06$ 
(dashed line): {\it a)} shows
the total recovered percentage as a function of fake SN \ip\ magnitude with
the 50\% recovery limits for each IQ shown as the vertical lines, {\it b) -
d)} show the recovery fraction versus various parameters for the fake SN
above the 50\% recovery fractions: {\it b)} shows the recovered percentage
as a function of host offset in arcseconds {\it c)} shows the total
recovered percentage as a function of host minus fake SN \ip\ magnitude,
and {\it d)} shows the recovered percentage as a function of \ip\
background measured in \ip\ magnitudes per square arcsecond.
\label{fig_comp_review}
}
\end{figure}

\begin{figure}
\includegraphics[angle=90.,scale=0.50]{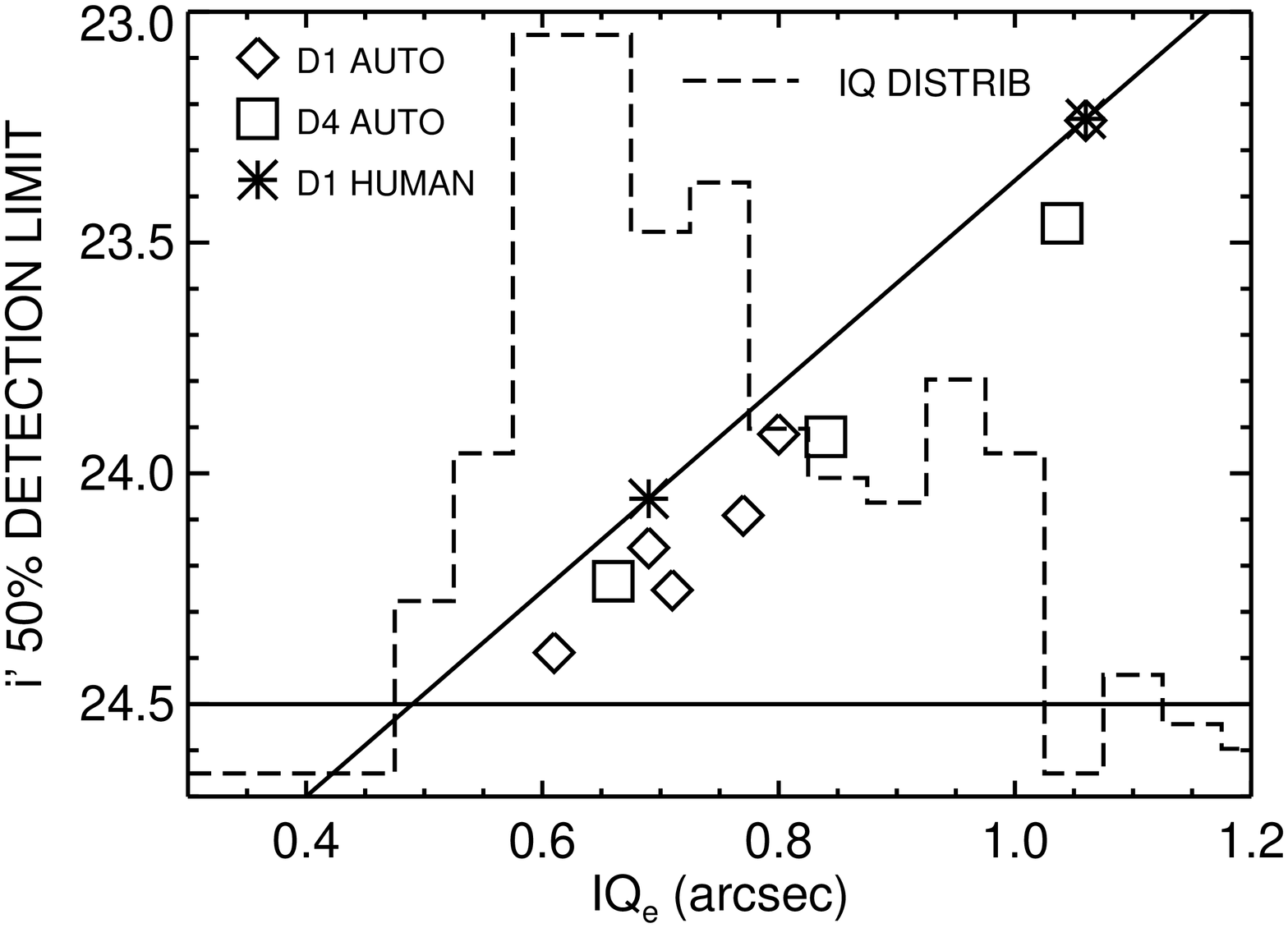}
\caption{50\% detection limit as a function of $IQ_e$ for a range of SNLS
\ip\ images with a range of $IQ_e$.  The $IQ_e$ distribution for all \ip\
images in the survey sample time range is shown as the dashed histogram.
Automated detection limits for field D1 are indicated by the open diamonds,
and for field D4 by the open squares.  The 50\% recovery limits
after human review for two D1 images at $IQ_e = 0\arcpnt69$ and
$1\arcpnt06$ are indicated by the asterisks.  A linear fit to the human
review limits is indicated by the bold solid line.  A constant
frame limit, independent of $IQ_e$, is indicated by the horizontal thin
line at $\ip = 24.5$.  The human review fit and the constant limit at $\ip
= 24.5$ encompass all the limits shown and define the range used in our
calculation of the systematic errors associated with SN~Ia detection in our
simulations (see \S\ref{sec_sys}).
\label{fig_det}
}
\end{figure}

\begin{figure}
\includegraphics[angle=90.,scale=0.60]{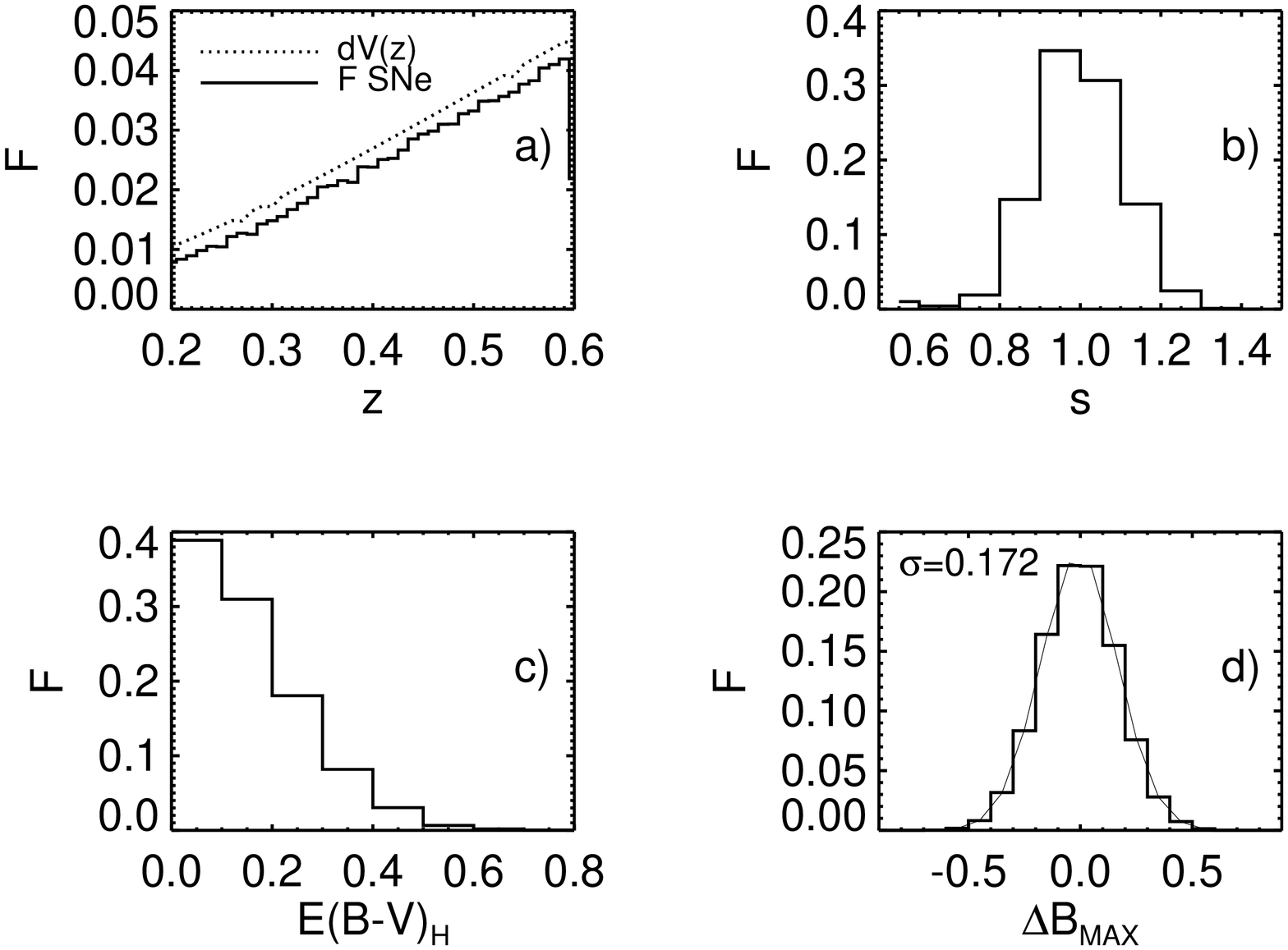}
\caption{Canonical distributions of properties of the simulated SNe~Ia used 
in the Monte Carlo efficiency experiments:  {\it a)} shows the volume-weighted
redshift distribution as the solid histogram and the run of $dV(z)$ as the
dotted line,  {\it b)} shows the Gaussian stretch distribution
with $\sigma=0.1$,  {\it c)} shows the positive-valued
Gaussian host extinction distribution with $\sigma_{E(B-V)_h} = 0.2$,  and {\it
d)} shows the Gaussian $\Delta B_{MAX}$ distribution with $\sigma_{B_{MAX}}
= 0.17$ with a Gaussian fit overplotted as a thin solid line.  The fitted
$\sigma$ of the Gaussian is annotated on the plot and matches the
distribution of real SNe~Ia from \citet{Hamuy96AJ}.
\label{fig_pops}
}
\end{figure}

\begin{figure}
\includegraphics[scale=0.50,angle=90.]{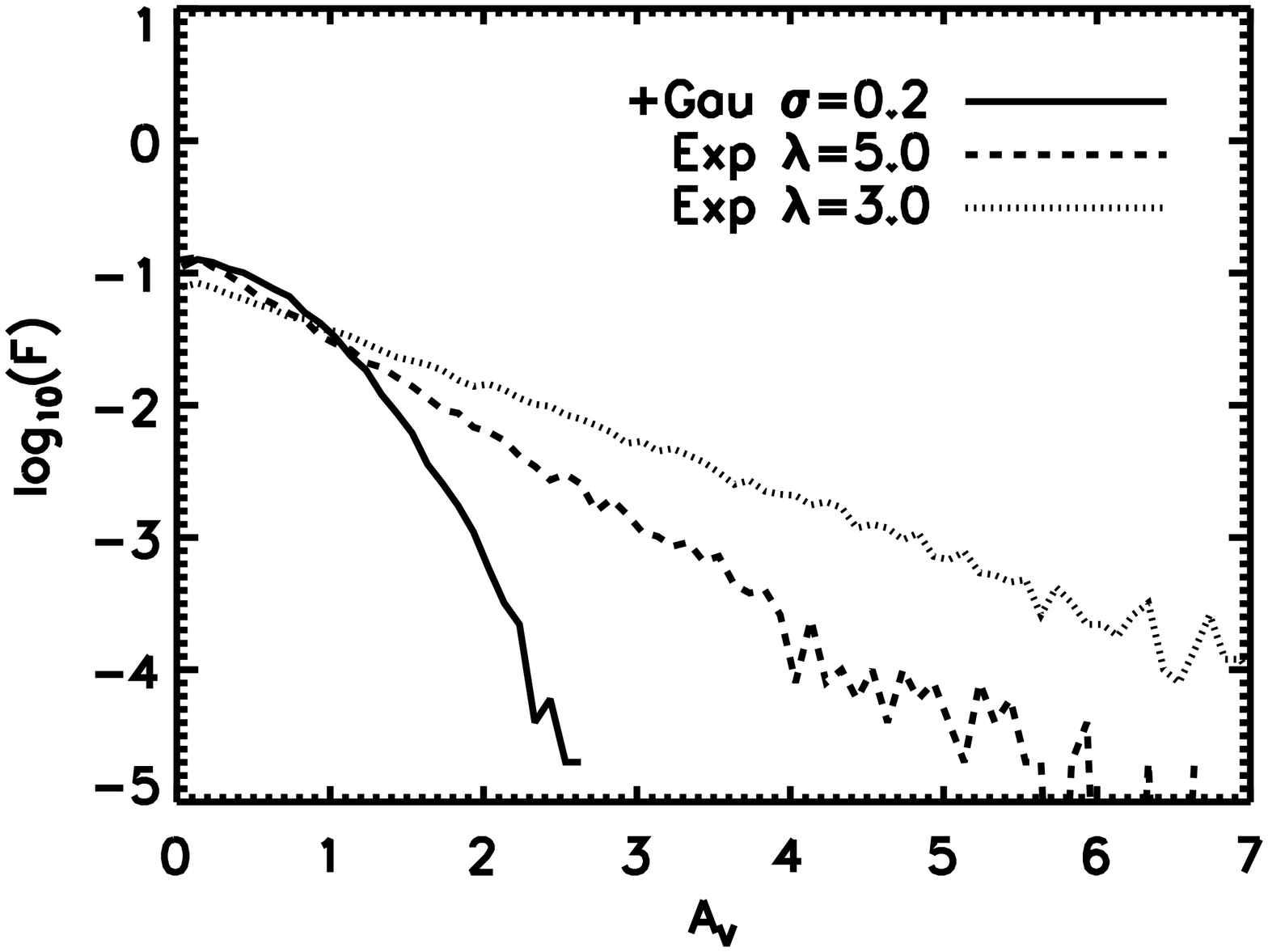}
\caption{Distributions of total V-band extinction, $A_V$, for three models of
SN~Ia host extinction.  The solid line represents the positive-valued
Gaussian with a width of $\sigma_{E(B-V)_h} = 0.2$ that was used for the
canonical Monte Carlo efficiency experiments.  The dashed line represents
an exponential distribution of $E(B-V)_h$ with a scale parameter of 
$\lambda_{E(B-V)_h} = 5.0$.  The dotted line shows an exponential
distribution with a scale parameter of $\lambda_{E(B-V)_h} = 3.0$.  When
comparing these distributions to those in Figure 3 of \citet{Riello05MNRAS}, we see
that our canonical host extinction model is appropriate for an intermediate
host inclination model ($45^{\circ} \leq i \leq 60^{\circ}$), while the
exponential distributions are closer to the extreme host inclination model
($75^{\circ} \leq i \leq 90^{\circ}$).
\label{fig_hex}
}
\end{figure}

\begin{figure}
\includegraphics[scale=0.50,angle=90.]{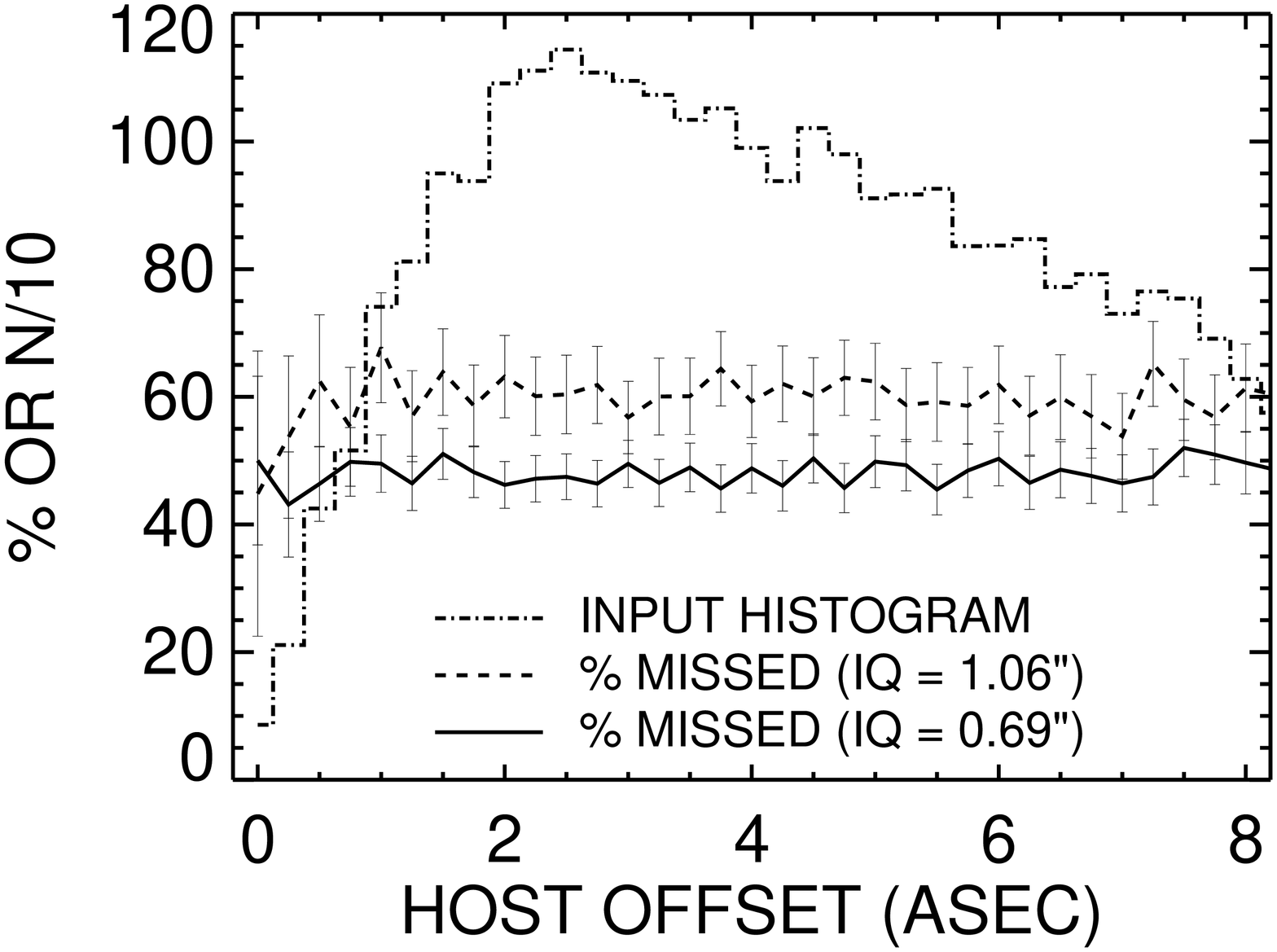}
\caption{Percent missed versus host offset in arcseconds for $IQ_e =
0\arcpnt69$ (solid line) and $IQ_e = 1\arcpnt06$ (dashed line) from the
fake SN experiments described in \S\ref{sec_det}.  The error bars are the
Poisson errors in each bin.  The histogram of the
input host offsets (divided by 10) is plotted as the dot-dashed line.
All data were binned with $0\arcpnt25$ bins.  The percentage missed does
not turn upwards at low host offset as would be expected if there was a
loss of SN visibility near the host galaxy nuclei.
\label{fig_hoff}
}
\end{figure}

\begin{figure}
\includegraphics[scale=0.50,angle=90.]{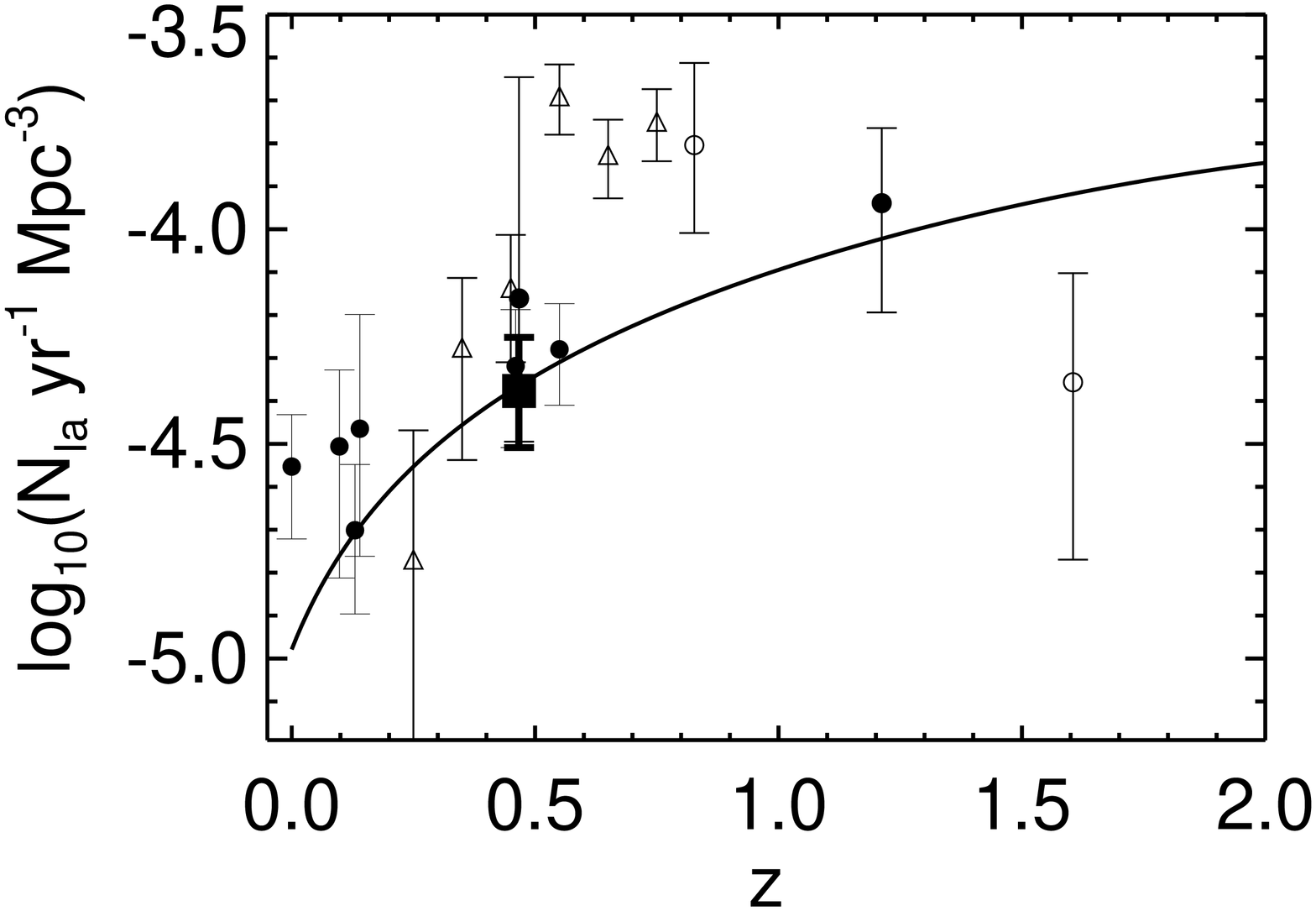}
\caption{Same as Figure~\ref{fig_sfh}, but with the rate from this study
plotted as a filled square.  Because the renormalization of the SFH from
\citet{Hopkins06} using a factor of $10^3$ fits our rate, we can
immediately place an upper limit on any component of SN~Ia rate production
that is tied directly to star formation of $r_{SFH} \lesssim 1$ SN~Ia /
$10^3$ M$_{\odot}$ ($B \lesssim 10^{-3}\ yr^{-1} (M_{\odot}\ yr^{-1})^{-1}$, see \S\ref{sec_sfh2}).
\label{fig_lit}
}
\end{figure}

\begin{figure}
\includegraphics[scale=0.50,angle=90.]{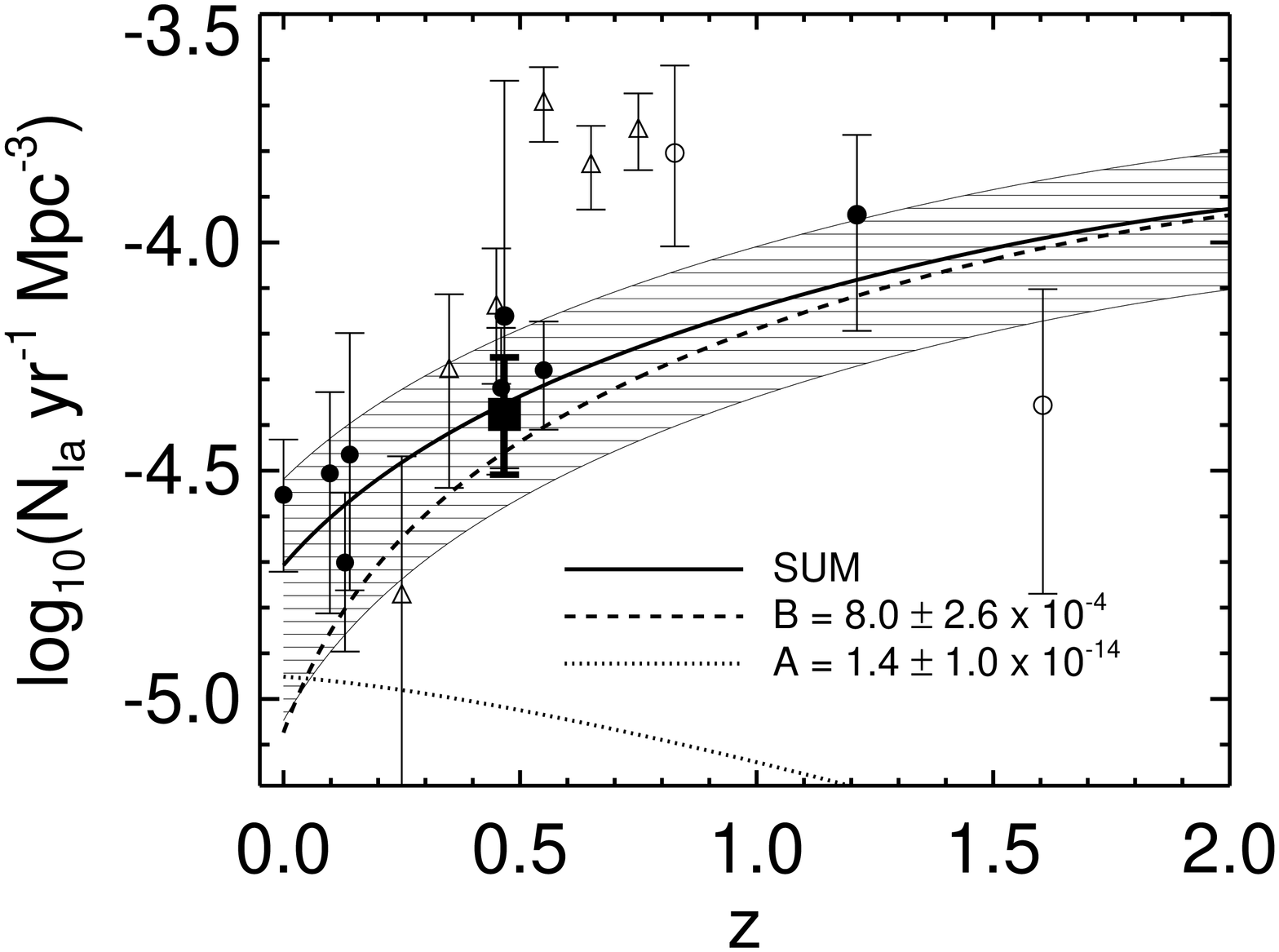}
\caption{Fit of observed, spectrally confirmed SN~Ia rates (see
Figures~\ref{fig_sfh} and \ref{fig_lit}) to the two component model
from equation~\ref{eq_ab} shown as the solid line, with the one-sigma
errors defining the shaded region.  The extended ($A$) component is
proportional to the current mass density, defined by integrating the
SFH of \citet{Hopkins06}, and is shown by the dotted line, while the
the prompt ($B$) component is proportional to the instantaneous SFH
and is shown by the dashed line.  The non-linear least-squares fit to
the spectroscopically confirmed rates has $\chi^2_{\nu} = 0.361$, and
produces an extended component with $A = 1.4 \pm 1.0 \times 10^{-14}\
yr^{-1} M_{\odot}^{-1}$ and a prompt component with $B = 8.0 \pm 2.6
\times 10^{-4}\ yr^{-1} (M_{\odot}\ yr^{-1})^{-1}$.  The errors quoted
are statistical only and do not include systematics due to errors in
the SFH or the mass definition.
\label{fig_sfh2}
}
\end{figure}

\begin{figure}
\includegraphics[scale=0.50,angle=90.]{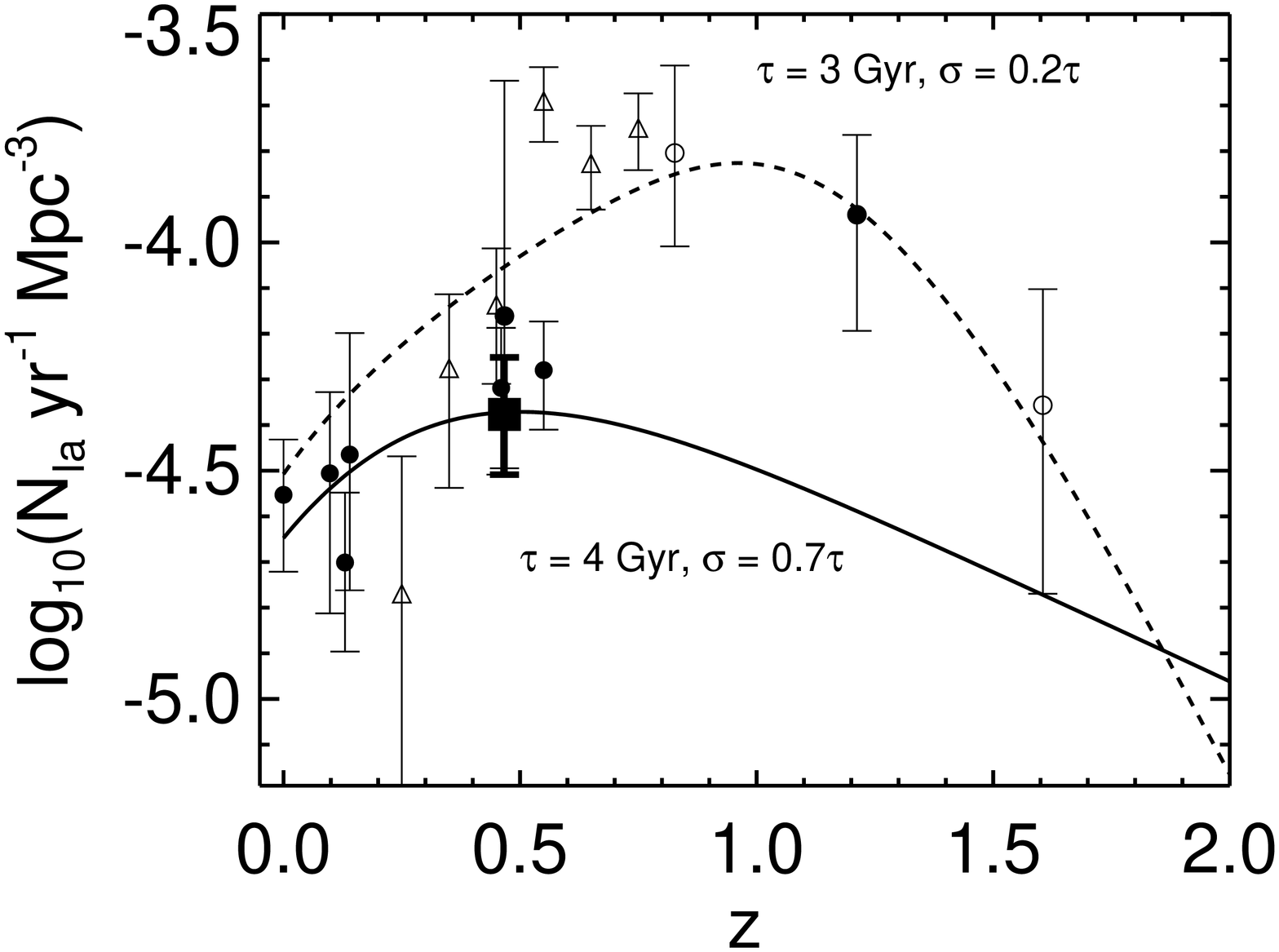}
\caption{Comparison of observed SN~Ia rates with the delay-time model
reported in \citet{Strolger04ApJ}, but using the SFH from
\citet{Hopkins06}.  The filled and open symbols are the same as in
Figure~\ref{fig_sfh2}.  The dashed line is a fit to the data from
\citet{Dahlen04ApJ} with parameters of $\tau = 3$ Gyr, and $\sigma =
0.2\tau$.  The solid line is a fit that is normalized at our rate having
parameters of $\tau = 4$ Gyr, and $\sigma=0.7\tau$.  The fit to the data
from \citet{Dahlen04ApJ} systematically over-predicts the rates at lower
redshift.  Both fits predict low SN~Ia rates beyond $z=1.5$.
\label{fig_sfhdt}
}
\end{figure}

\clearpage

%
%
\begin{deluxetable}{ccccccccccc}
\tabletypesize{\scriptsize}
\tablecaption{SNLS Observations\label{tab_obs}}
\tablewidth{0pt}
\tablehead{
\colhead{ } & \multicolumn{2}{c}{Position(J2000)} & \colhead{ } & 
	\multicolumn{2}{c}{Start Date} & \colhead{ } &
	\multicolumn{2}{c}{End Date} & \colhead{Span} \\
\cline{2-3} \cline{5-6} \cline{8-9} \\
\colhead{Field} & \colhead{RA} & \colhead{Dec} & \colhead{Season} &
	\colhead{(MJD)} & \colhead{(YYYY-MM-DD)} & \colhead{ } &
	\colhead{(MJD)} & \colhead{(YYYY-MM-DD)} & \colhead{(days)} \\
}

\startdata
D1  & 02:26:00.00 & -04:30:00.0 &  1 & 52852 & 2003-08-01 & & 53026 & 2004-01-22 & 174 \\
    & & &  2 & 53207 & 2004-07-21 & & 53390 & 2005-01-20 & 183 \\
\\
D2  & 10:00:28.60 & +02:12:21.0 &  1 & 52993 & 2003-12-20 & & 53151 & 2004-05-26 & 158 \\
    & & &  2 & 53328 & 2004-11-19 & & 53503 & 2005-05-13 & 175 \\
\\
D3  & 14:19:28.01 & +52:40:41.0 &  1 & 53017 & 2004-01-13 & & 53228 & 2004-08-11 & 211 \\
    & & &  2 & 53386 & 2005-01-16 & & 53586 & 2005-08-04 & 200 \\
\\
D4  & 22:15:31.67 & -17:44:05.0 &  1 & 52795 & 2003-06-05 & & 52964 & 2003-11-21 & 169 \\
    & & &  2 & 53173 & 2004-06-17 & & 53350 & 2004-12-11 & 177 \\
\enddata
\end{deluxetable}

%
%
\begin{deluxetable}{llllccc}
\tabletypesize{\scriptsize}
\tablecaption{Spectroscopically Confirmed Type Ia SNe: $0.2 < z < 0.6$\label{tab_sne}}
\tablewidth{0pt}
\tablehead{
\colhead{Name} & \colhead{$\alpha_{J2000.0}$} & \colhead{$\delta_{J2000.0}$} &
\colhead{$z_{SPEC}$} & \colhead{Discovery (MJD)} &
\colhead{Culling Status\tablenotemark{a}} & \colhead{Refs\tablenotemark{b}}\\
}

\startdata
SNLS-03D1ar & 02:27:14.680 & -04:19:05.05 & 0.41 & 52901 & \\
SNLS-03D1au & 02:24:10.380 & -04:02:14.96 & 0.50 & 52901 & & A06\\
SNLS-03D1aw & 02:24:14.780 & -04:31:01.61 & 0.58 & 52901 & & A06\\
SNLS-03D1ax & 02:24:23.320 & -04:43:14.41 & 0.50 & 52901 & & A06,H05\\
SNLS-03D1bp & 02:26:37.720 & -04:50:19.34 & 0.35 & 52913 & & A06,B06\\
SNLS-03D1dj & 02:26:19.082 & -04:07:09.38 & 0.40 & 52962 & 1-3,5 & A06\\
SNLS-03D1fb & 02:27:12.855 & -04:07:16.40 & 0.50 & 52991 & & A06,B06\\
SNLS-03D1fc & 02:25:43.602 & -04:08:38.77 & 0.33 & 52991 & & A06,B06\\
SNLS-03D1gt & 02:24:56.012 & -04:07:37.08 & 0.55 & 53000 & 2,4 & A06,B06\\
SNLS-04D1ag & 02:24:41.108 & -04:17:19.69 & 0.56 & 53019 & 5 & A06,B06\\
SNLS-04D1ak & 02:27:33.399 & -04:19:38.73 & 0.53 & 53019 & & A06,B06\\
\\
SNLS-04D1dc & 02:26:18.477 & -04:18:43.28 & 0.21 & 53228 & 2\\
SNLS-04D1hd & 02:26:08.850 & -04:06:35.22 & 0.37 & 53254 & & H05\\
SNLS-04D1hx & 02:24:42.485 & -04:47:25.38 & 0.56 & 53254 & \\
SNLS-04D1jg & 02:26:12.567 & -04:08:05.34 & 0.58 & 53267 & \\
SNLS-04D1kj & 02:27:52.669 & -04:10:49.29 & 0.58 & 53283 & \\
SNLS-04D1oh & 02:25:02.372 & -04:14:10.52 & 0.59 & 53294 & \\
SNLS-04D1pg & 02:27:04.162 & -04:10:31.35 & 0.51 & 53312 & \\
SNLS-04D1rh & 02:27:47.160 & -04:15:13.60 & 0.43 & 53344 & \\
SNLS-04D1sa & 02:27:56.161 & -04:10:34.31 & 0.59 & 53351 & \\
\\
SNLS-04D2ac & 10:00:18.924 & +02:41:21.45 & 0.35 & 53022 & 5 & A06,B06\\
SNLS-04D2bt & 09:59:32.739 & +02:14:53.22 & 0.22 & 53081 & 1-4 & A06,B06\\
SNLS-04D2cf & 10:01:56.048 & +01:52:45.90 & 0.37 & 53081 & 1-4 & A06,B06\\
SNLS-04D2cw & 10:01:22.821 & +02:11:55.66 & 0.57 & 53081 & 2-5 & A06,B06\\
SNLS-04D2fp & 09:59:28.183 & +02:19:15.20 & 0.41 & 53094 & & A06,B06\\
SNLS-04D2fs & 10:00:22.110 & +01:45:55.64 & 0.36 & 53094 & & A06,B06\\
SNLS-04D2gb & 10:02:22.712 & +01:53:39.16 & 0.43 & 53094 & & A06\\
SNLS-04D2gc & 10:01:39.267 & +01:52:59.52 & 0.52 & 53106 & & A06\\
\\
SNLS-04D2mh & 09:59:45.872 & +02:08:27.94 & 0.60 & 53356 & \\
SNLS-04D2mj & 10:00:36.535 & +02:34:37.44 & 0.51 & 53356 & \\
SNLS-05D2ab & 10:01:50.833 & +02:06:23.02 & 0.32 & 53375 & \\
SNLS-05D2ac & 09:58:59.244 & +02:29:22.22 & 0.49 & 53375 & \\
SNLS-05D2bv & 10:02:17.008 & +02:14:26.05 & 0.47 & 53391 & \\
SNLS-05D2cb & 09:59:24.592 & +02:19:41.34 & 0.43 & 53409 & \\
SNLS-05D2dm & 10:02:07.611 & +02:03:17.35 & 0.57 & 53441 & \\
SNLS-05D2dw & 09:58:32.058 & +02:01:56.36 & 0.42 & 53441 & \\
SNLS-05D2dy & 10:00:58.083 & +02:10:59.52 & 0.50 & 53441 & 2 \\
SNLS-05D2ec & 09:59:26.170 & +02:00:49.36 & 0.53 & 53441 & 1,2\\
SNLS-05D2ei & 10:01:39.103 & +01:49:12.02 & 0.37 & 53441 & \\
SNLS-05D2hc & 10:00:04.574 & +01:53:09.94 & 0.36 & 53463 & \\
SNLS-05D2ie & 10:01:02.907 & +02:39:28.90 & 0.35 & 53467 & \\
\\
SNLS-04D3df & 14:18:10.020 & +52:16:40.13 & 0.47 & 53110 & & A06\\
SNLS-04D3ez & 14:19:07.916 & +53:04:18.88 & 0.26 & 53110 & & A06\\
SNLS-04D3fk & 14:18:26.212 & +52:31:42.74 & 0.36 & 53117 & & A06\\
SNLS-04D3gt & 14:22:32.594 & +52:38:49.52 & 0.45 & 53124 & & A06\\
SNLS-04D3hn & 14:22:06.878 & +52:13:43.46 & 0.55 & 53124 & & A06,H05\\
SNLS-04D3kr & 14:16:35.937 & +52:28:44.20 & 0.34 & 53147 & & A06,H05\\
SNLS-04D3nh & 14:22:26.729 & +52:20:00.92 & 0.34 & 53166 & 1,2 & A06,H05\\
SNLS-04D3nq & 14:20:19.193 & +53:09:15.90 & 0.22 & 53176 & & A06,H05\\
\\
SNLS-05D3cf & 14:16:53.369 & +52:20:42.47 & 0.42 & 53410 & \\
SNLS-05D3ci & 14:21:48.085 & +52:26:43.33 & 0.51 & 53416 & \\
SNLS-05D3dd & 14:22:30.410 & +52:36:24.76 & 0.48 & 53441 & \\
SNLS-05D3gp & 14:22:42.338 & +52:43:28.71 & 0.58 & 53462 & \\
SNLS-05D3hq & 14:17:43.058 & +52:11:22.67 & 0.34 & 53474 & \\
SNLS-05D3jq & 14:21:45.462 & +53:01:47.53 & 0.58 & 53474 & \\
SNLS-05D3jr & 14:19:28.768 & +52:51:53.34 & 0.37 & 53474 & 1,2 \\
SNLS-05D3kx & 14:21:50.020 & +53:08:13.49 & 0.22 & 53519 & \\
SNLS-05D3lq & 14:21:18.449 & +52:32:08.29 & 0.42 & 53528 & \\
SNLS-05D3mq & 14:19:00.398 & +52:23:06.81 & 0.24 & 53559 & 5 \\
SNLS-05D3mx & 14:22:09.078 & +52:13:09.35 & 0.47 & 53559 & \\
\\
SNLS-03D4ag & 22:14:45.790 & -17:44:23.00 & 0.28 & 52813 & & A06\\
SNLS-03D4au & 22:16:09.920 & -18:04:39.37 & 0.47 & 52815 & & A06,B06\\
SNLS-03D4cj & 22:16:06.660 & -17:42:16.72 & 0.27 & 52873 & & A06,H05\\
SNLS-03D4gf & 22:14:22.907 & -17:44:02.49 & 0.58 & 52930 & & A06,B06\\
SNLS-03D4gg & 22:16:40.185 & -18:09:51.82 & 0.59 & 52930 & & A06,B06\\
SNLS-03D4gl & 22:14:44.177 & -17:31:44.47 & 0.57 & 52935 & 5 & A06,H05\\
\\
SNLS-04D4bq & 22:14:49.391 & -17:49:39.37 & 0.55 & 53174 & & A06,B06\\
SNLS-04D4gg & 22:16:09.268 & -17:17:39.98 & 0.42 & 53228 & & H05\\
SNLS-04D4gz & 22:16:59.018 & -17:37:19.02 & 0.38 & 53235 & \\
SNLS-04D4ht & 22:14:33.289 & -17:21:31.33 & 0.22 & 53254 & 2 \\
SNLS-04D4in & 22:15:08.585 & -17:15:39.85 & 0.52 & 53267 & \\
SNLS-04D4jr & 22:14:14.335 & -17:21:00.93 & 0.48 & 53284 & \\
SNLS-04D4ju & 22:17:02.733 & -17:19:58.34 & 0.47 & 53284 & \\
\enddata
\tablenotetext{a}{Numbers indicate the criteria from
  \S\ref{sec_select} that caused the SN to be rejected from the sample}
\tablenotetext{b}{References indicated as follows: A06 -
  \citet{Astier06A&A}, H05 - \citet{Howell05ApJ}, B06 - \citet{Basa06}}
\end{deluxetable}

%
%
\begin{deluxetable}{cccc}
\tabletypesize{\small}
\tablecaption{SN~Ia Sample\label{tab_sam}}
\tablewidth{0pt}
\tablehead{
\colhead{ } & \colhead{ } & \colhead{Total} & \colhead{Culled} \\
\colhead{Field} & \colhead{Season} & \colhead{($N_{SN}$)} & 
\colhead{($N_{SN}$)}\\
}

\startdata
D1  &  1 & 11 &  8 \\
    &  2 & 9 &  8 \\
\\
D2  &  1 & 8 &  4 \\
    &  2 & 13 & 11 \\
\\
D3  &  1 & 8 &  7 \\
    &  2 & 11 &  9 \\
\\
D4  &  1 & 6 &  5 \\
    &  2 & 7 &  6 \\
\\
ALL &    & 73 & 58 \\
\enddata
\end{deluxetable}

%
%
\begin{deluxetable}{lrrccccc}
\tabletypesize{\scriptsize}
\tablecaption{Unconfirmed SN Ia Candidates\label{tab_cands}}
\tablewidth{0pt}
\tablehead{
\colhead{Name} & \colhead{$\alpha_{J2000.0}$} & \colhead{$\delta_{J2000.0}$} &
\colhead{$z_{PHOT}$} & \colhead{Discovery (MJD)} & \colhead{Type} &
\colhead{$\chi^2_{SNIa}$} & \colhead{Status}\\
}

\startdata
SNLS-03D1ge & 02:24:06.043 & -04:23:19.14 & 0.54 & 52993 & SN? & 4.208 & Probable SN Ia\\
SNLS-04D2lu & 10:01:09.465 & +02:32:14.52 & 0.37 & 53353 & SN? & 3.741 & Probable SN Ia\\
SNLS-04D2lx & 10:01:17.159 & +01:42:50.97 & 0.50 & 53353 & SN? & 2.031 & Probable SN Ia\\
SNLS-04D3ht & 14:16:17.101 & +52:19:28.40 & 0.53 & 53135 & SN? & 3.129 & Probable SN Ia\\
SNLS-05D3ba & 14:18:26.790 & +52:41:50.56 & 0.44 & 53387 & SN? & 2.323 & Probable SN Ia\\
SNLS-05D3lc & 14:22:22.902 & +52:28:44.11 & 0.49 & 53519 & SN  & 2.027 & Probable SN Ia\\
SNLS-05D3lx & 14:17:56.809 & +52:20:23.26 & 0.58 & 53532 & SN? & 3.102 & Probable SN Ia\\
SNLS-03D4bx & 22:14:48.602 & -17:31:17.58 & 0.54 & 52843 & SN? & 2.089 & Probable SN Ia\\
SNLS-03D4ev & 22:16:51.395 & -17:20:02.37 & 0.53 & 52914 & SN? & 2.591 & Probable SN Ia\\
SNLS-04D4cm & 22:13:28.782 & -18:03:40.56 & 0.55 & 53177 & SN? & 3.928 & Probable SN Ia\\
SNLS-04D4et & 22:14:51.788 & -17:47:22.86 & 0.58 & 53204 & SN? & 1.558 & Probable SN Ia\\
SNLS-04D4iy & 22:17:07.977 & -18:07:07.18 & 0.51 & 53267 & SN? & 2.801 & Probable SN Ia\\
\\
SNLS-04D1fh & 02:26:59.401 & -04:29:42.41 & 0.56 & 53235 & SN? & 5.762 & Possible SN Ia\\
SNLS-04D1qn & 02:27:28.186 & -04:20:35.78 & 0.52 & 53323 & SN? & 6.400 & Possible SN Ia\\
SNLS-04D1qr & 02:25:49.083 & -04:29:00.23 & 0.58 & 53323 & SN? & 5.135 & Possible SN Ia\\
SNLS-04D1sc & 02:26:34.371 & -04:02:45.60 & 0.59 & 53351 & SN? & 6.038 & Possible SN Ia\\
SNLS-05D3bz & 14:17:50.119 & +52:51:24.16 & 0.52 & 53410 & SN? & 5.499 & Possible SN Ia\\
SNLS-05D3ch & 14:19:09.668 & +52:47:35.93 & 0.35 & 53416 & SN? & 5.553 & Possible SN Ia\\
SNLS-04D4ay & 22:15:54.038 & -18:02:48.95 & 0.58 & 53174 & SN? & 5.064 & Possible SN Ia\\
\enddata
\end{deluxetable}

%
%
\begin{deluxetable}{ccccccc}
\tablecaption{Spectroscopic Completeness\label{tab_spec_comp}}
\tablewidth{0pt}
\tablehead{
\colhead{} & \multicolumn{3}{c}{SNe Ia} & &
\multicolumn{2}{c}{Completeness Fraction\tablenotemark{a}}\\
\cline{2-4} \cline{6-7} \\
\colhead{Field} & \colhead{Confirmed} & \colhead{Probable} &
\colhead{Possible} & & \colhead{Minimum} & \colhead{Most Likely ($C_{SPEC}$)}\\
}

\startdata
D1  &  16 & 1 & 4 & & 0.76 & 0.94 \\
D2  &  15 & 2 & 0 & & 0.88 & 0.88 \\
D3  &  16 & 4 & 2 & & 0.73 & 0.80 \\
D4  &  11 & 5 & 1 & & 0.65 & 0.69 \\
\hline\\
ALL & 58 & 12 & 7 & & 0.75 & 0.83 \\
\enddata
\tablenotetext{a}{The maximum completeness is 1.00}
\end{deluxetable}

\begin{deluxetable}{cccccc}
\tablecaption{\ip\ Frame Limit Equation Parameters\label{tab_det}}
\tablewidth{0pt}
\tablehead{
\colhead{$IQ_{GOOD}$} & \colhead{$IQ_{BAD}$} & 
\colhead{$E_{ref}$} & \colhead{$S_{ref}$} & \colhead{$L_{0.5}$} &
\colhead{$\alpha$} \\
\colhead{(arcsec)} & \colhead{(arcsec)} &
\colhead{(s)} & \colhead{(DN)} & \colhead{(mag)}\\
}

\startdata
0.69 & 1.06 & 3641 & 29.18 & 24.5 & 2.22\\
\enddata
\end{deluxetable}

\begin{deluxetable}{cccccc}
\tablecaption{Monte Carlo Efficiencies\label{tab_mc}}
\tablewidth{0pt}
\tablehead{
\colhead{} & \multicolumn{2}{c}{On-Field} & \colhead{} & \colhead{Yearly} \\
\cline{2-3} \cline{5-5} \\
\colhead{Field} & \colhead{\ip\ Detection} & \colhead{Spec} & \colhead{} &
\colhead{Spec ($\epsilon_{yr}$)} \\
}

\startdata
D1 & 0.948 & 0.612 & & 0.299 \\
D2 & 0.981 & 0.528 & & 0.217 \\
D3 & 0.971 & 0.629 & & 0.313 \\
D4 & 0.979 & 0.654 & & 0.310 \\
\enddata
\end{deluxetable}

\begin{deluxetable}{llllccc}
\tablecaption{SNLS Type Ia SN Volumetric Rates\label{tab_rates}}
\tablewidth{0pt}
\tablehead{
\colhead{} & \colhead{$r_{RAW}$} &
\colhead{$r_{obs}$\tablenotemark{a}} &
\colhead{$r_{1+z}$\tablenotemark{b}} & 
\colhead{$\Theta$} & \colhead{$V_{\Theta,0.2 < z < 0.6}$} & \colhead{$r_V$} \\
\colhead{Field} & \colhead{(yr$^{-1}$)} & \colhead{(yr$^{-1}$)} &
\colhead{(yr$^{-1}$)} & \colhead{(degrees$^2$)} & 
\colhead{($\times 10^4$ Mpc$^3$)} &
\colhead{($\times 10^{-4}$ yr$^{-1}$ Mpc$^{-3}$)} \\
}

\startdata
 D1 &   26.7 $\pm$    6.7 &   28.4 $\pm$    7.1 &   41.7 $\pm$   10.4 &   1.024 &   106.0 &    0.39 $\pm$    0.10 \\
 D2 &   34.6 $\pm$    8.9 &   39.4 $\pm$   10.2 &   57.7 $\pm$   14.9 &   1.026 &   106.2 &    0.54 $\pm$    0.14 \\
 D3 &   25.5 $\pm$    6.4 &   31.9 $\pm$    8.0 &   46.8 $\pm$   11.7 &   1.029 &   106.5 &    0.44 $\pm$    0.11 \\
 D4 &   17.7 $\pm$    5.3 &   25.7 $\pm$    7.8 &   37.7 $\pm$   11.4 &   1.027 &   106.3 &    0.35 $\pm$    0.11 \\
 \\
AVG\tablenotemark{c} &   24.1 $\pm$    3.3 &   30.3 $\pm$    4.0 &   44.4 $\pm$    5.9 &   1.026 &   106.2 &    0.42 $\pm$    0.06\tablenotemark{d} \\
\enddata
\tablenotetext{a}{rates after correcting for spectroscopic incompleteness}
\tablenotetext{b}{rates after correcting for time dilation}
\tablenotetext{c}{Poisson error weighted averages}
\tablenotetext{d}{statistical error only}
\end{deluxetable}

\begin{deluxetable}{lrr}
\tablecaption{Summary of Uncertainties\label{tab_errs}}
\tablewidth{0pt}
\tablehead{
\colhead{Source} & \colhead{$\delta r_V$\tablenotemark{a}} & 
\colhead{$\delta r_L$\tablenotemark{b}} \\
}

\startdata
Poisson               & $\pm0.06$ & $\pm0.020$ \\
Luminosity Estimate   & \nodata & $^{+0.033}_{-0.023} $ \\
\\
Spec. Completeness    & $^{+0.03}_{-0.08}$ & $^{+0.010}_{-0.031}$\\
Host Extinction       & $+0.10$ & $ +0.037$ \\
Frame Limits          & $-0.03$ & $-0.011$ \\
Stretch               & $\pm0.01$ & $\pm0.004$ \\
Subluminous SNe~Ia    & $+0.08$ & $+0.029$ \\
\\
\hline \\
Total Statistical     & $\pm0.06$  & $^{+0.039}_{-0.031}$ \\
\\
Total Systematic      & $^{+0.13}_{-0.09} $  & $^{+0.048}_{-0.033}$\\
\enddata
\tablenotetext{a}{Volumetric uncertainty in units of $10^{-4}$ yr$^{-1}$ Mpc$^{-3}$}
\tablenotetext{b}{{Luminosity specific uncertainty in SNu}}
\end{deluxetable}


\end{document}